\documentclass[10pt]{article}

\usepackage{amsmath}
\usepackage{array}
\usepackage{appendix}
\usepackage{graphicx}
\usepackage{amsfonts}
\usepackage{amssymb}
\usepackage{mathrsfs}
\usepackage{yfonts}
\usepackage{euscript}
\usepackage{upgreek}
\usepackage{slantsc}
\usepackage{calligra}
\usepackage[T1]{fontenc}
\usepackage{epsf}
\usepackage{latexsym}

\usepackage{tipa}

\def\be{\begin{equation}}
\def\ee{\end{equation}}
\def\beq{\begin{equation}}
\def\eeq{\end{equation}}
\def\bea{\begin{eqnarray}}
\def\eea{\end{eqnarray}}

\def\ni{\noindent}
\def\foo{\footnote}

\def\hat{\widehat}

\def\!{\hspace{-1.6667em}}

\def\mD{\mbox{D}}

\def\mF{\mbox{F}}
\def\mG{\mbox{G}}

\def\mK{\mbox{K}}
\def\mL{\mbox{L}}
\def\mM{\mbox{M}}
\def\mN{\mbox{N}} 

\def\mP{\mbox{P}}

\def\mR{\mbox{R}}
\def\mS{\mbox{S}}
\def\mT{\mbox{T}}

\def\mX{\mbox{X}}

\def\md{\mbox{d}} 
\def\me{\mbox{e}}

\def\mg{\mbox{g}}
\def\mh{\mbox{h}}

\def\mn{\mbox{n}}   

\def\mp{\mbox{p}}

\def\ms{\mbox{s}}
\def\mt{\mbox{t}}
\def\muu{\mbox{u}}
\def\mv{\mbox{v}}

\def\uupbeta{{\underline{\upbeta}}}
\def\suupbeta{\mbox{\scriptsize$\underline{\upbeta}$}}

\def\BigupRho{\mbox{\Large $\rho$}}                 

\def\uupbeta{\mbox{\underline{$\upbeta$}}}

\def\bupSigma{\mbox{\boldmath$\Sigma$}}                 
\def\bupPhi{\mbox{\boldmath$\Phi$}}                     
\def\sbSigma{\mbox{\scriptsize\boldmath$\Sigma$}}

\def\fa{\mbox{\sffamily a}}

\def\fA{\mbox{\sffamily A}}

\def\fF{\mbox{\sffamily F}}

\def\fH{\mbox{\sffamily H}}

\def\fQ{\mbox{\sffamily Q}}

\def\fW{\mbox{\sffamily W}}

\def\ux{\underline{{x}}}

\def\bh{\underline{\underline{\mbox{h}}}  }            

\def\bp{\mbox{\bf p}}

\def\bK{\mbox{\bf K}}

\def\bM{\mbox{\bf M}}

\def\bg{\mbox{{\bf g}}}
\def\bM{\mbox{{\bf M}}}
\def\bM{\mbox{{\bf M}}}

\def\bh{\mbox{{\bf h}}}

\def\scA{\mbox{\scriptsize ${\cal A}$}}

\def\scC{\mbox{\scriptsize ${\cal C}$}}          
\def\scD{\mbox{\scriptsize ${\cal D}$}}          
\def\scE{\mbox{\scriptsize ${\cal E}$}}          
\def\scF{\mbox{\scriptsize ${\cal F}$}}
\def\scG{\mbox{\scriptsize ${\cal G}$}}          
\def\scH{\mbox{\scriptsize ${\cal H}$}}          
\def\scI{\mbox{\scriptsize ${\cal I}$}}

\def\scL{\mbox{\scriptsize ${\cal L}$}}          
\def\scM{\mbox{\scriptsize ${\cal M}$}}          
\def\scN{\mbox{\scriptsize ${\cal N}$}}
\def\scO{\mbox{\scriptsize ${\cal O}$}}
\def\scP{\mbox{\scriptsize ${\cal P}$}}
\def\scQ{\mbox{\scriptsize ${\cal Q}$}}          
\def\scR{\mbox{\scriptsize ${\cal R}$}}          
\def\scS{\mbox{\scriptsize ${\cal S}$}}

\def\scU{\mbox{\scriptsize ${\cal U}$}}          

\def\iB{\mbox{\scriptsize$B$}}   
\def\iD{\mbox{\scriptsize$D$}}   
\def\iK{\mbox{\scriptsize$K$}}   
\def\iS{\mbox{\scriptsize$K$}}   
\def\iX{\mbox{\scriptsize$X$}}   

\def\FrQ{\mbox{\Large $\mathfrak{q}$}}

\def\FrU{\mbox{\boldmath$\mathfrak{U}$}}

\def\sFrM{\mbox{\boldmath\scriptsize$\mathfrak{m}$}}
\def\FrF{\mbox{$\mathfrak{F}$}}                                 
 
\def\FrM{\mbox{\Large $\mathfrak{m}$}}                         
\def\FrMgen{\mbox{\boldmath$\mathfrak{M}$}}                     
\def\FrHgen{\mbox{\boldmath$\mathfrak{H}$}}                     

\def\bFrR{\mbox{$\mathfrak{R}$}}
\def\bsFrR{\mbox{\scriptsize$\mathfrak{R}$}}

\def\FrG{\mbox{\Large $\mathfrak{g}$}}                            
 
\def\FS{\mbox{\LARGE\tt s}}                         

\def\sa{\mbox{\scriptsize a}}

\def\sd{\mbox{\scriptsize d}}
\def\se{\mbox{\scriptsize e}}
\def\sf{\mbox{\scriptsize f}}
 
\def\sh{\mbox{\scriptsize h}} 
\def\si{\mbox{\scriptsize i}}

\def\sll{\mbox{\scriptsize l}}  
\def\sm{\mbox{\scriptsize m}}
\def\sn{\mbox{\scriptsize n}} 
\def\so{\mbox{\scriptsize o}}

\def\sr{\mbox{\scriptsize r}}
\def\st{\mbox{\scriptsize t}}
\def\su{\mbox{\scriptsize u}}

\def\sw{\mbox{\scriptsize w}}

\def\sA{\mbox{\scriptsize A}} 
\def\sB{\mbox{\scriptsize B}}

\def\sD{\mbox{\scriptsize D}}
\def\sE{\mbox{\scriptsize E}}
\def\sF{\mbox{\scriptsize F}}
\def\sG{\mbox{\scriptsize G}}
\def\sH{\mbox{\scriptsize H}}

\def\sJ{\mbox{\scriptsize J}}
\def\sK{\mbox{\scriptsize K}}
\def\sL{\mbox{\scriptsize L}} 
\def\sM{\mbox{\scriptsize M}} 
\def\sN{\mbox{\scriptsize N}} 

\def\sP{\mbox{\scriptsize P}} 
 
\def\sR{\mbox{\scriptsize R}}
\def\sS{\mbox{\scriptsize S}}

\def\sW{\mbox{\scriptsize W}}
\def\sX{\mbox{\scriptsize X}}

\def\sfa{\mbox{\sffamily{\scriptsize a}}}     

\def\sfA{\mbox{\sffamily{\scriptsize A}}}      
\def\sfB{\mbox{\sffamily{\scriptsize B}}}      
\def\sfC{\mbox{\sffamily{\scriptsize C}}}      
\def\sfD{\mbox{\sffamily{\scriptsize D}}}      
\def\sfF{\mbox{\sffamily{\scriptsize F}}}      
\def\sfG{\mbox{\sffamily{\scriptsize G}}}      
\def\sfI{\mbox{\sffamily{\scriptsize I}}}      
\def\sfK{\mbox{\sffamily{\scriptsize K}}}      
\def\sfL{\mbox{\sffamily{\scriptsize L}}}      
\def\sfO{\mbox{\sffamily{\scriptsize O}}}      
\def\sfQ{\mbox{\sffamily{\scriptsize Q}}}      
\def\sfX{\mbox{\sffamily{\scriptsize X}}}      

\def\sbM{\mbox{{\bf \scriptsize M}}}

\def\td{\mbox{\tiny d}}
\def\te{\mbox{\tiny e}}
\def\tf{\mbox{\tiny f}}

\def\th{\mbox{\tiny h}}
\def\ti{\mbox{\tiny i}}

\def\tn{\mbox{\tiny n}}

\def\tr{\mbox{\tiny r}}

\def\tfA{\mbox{\sffamily{\tiny A}}}

\def\tfG{\mbox{\sffamily{\tiny G}}}

\def\tfW{\mbox{\sffamily{\tiny W}}}

\def\bigupsigma{\mbox{\Large$\sigma$}}
\def\biguptau{\mbox{\Large$\uptau$}}
\def\bigtau{\mbox{\Large$\tau$}}

\def\K{Kucha\v{r} }

\def\pa{\partial}
\def\d{\textrm{d}}

\def\Circ{\mbox{\Large$\circ$}}               
\def\5Star{\mbox{\Large$\star$}}              
\def\cr{\mbox{\scriptsize{\bf $\mbox{ } \times \mbox{ }$}}}

\def\sumi2{\sum\mbox{}_{\mbox{}_{\mbox{\scriptsize $i$=1}}}^2}
\def\sumi3{\sum\mbox{}_{\mbox{}_{\mbox{\scriptsize $i$=1}}}^3}

\def\sumj3{\sum\mbox{}_{\mbox{}_{\mbox{\scriptsize $j$=1}}}^3}
\def\sumk3{\sum\mbox{}_{\mbox{}_{\mbox{\scriptsize $k$=1}}}^3}

\begin{document}

\begin{titlepage}

\begin{center}

{\Large{\bf Problem of Time and Background Independence: the Individual Facets}}

\vspace{.15in}

{\large \bf Edward Anderson} 

\vspace{.15in}

\large {\em DAMTP, Centre for Mathematical Sciences, Wilberforce Road, Cambridge CB3 OWA.  } \normalsize

\end{center}

\begin{abstract}

I lay out the problem of time facets as arising piecemeal from a number of aspects of background independence.  
Almost all of these already have simpler classical counterparts.
This approach can be viewed as a facet by facet completion of the observation that Barbour-type relationalism is a background independent precursor to 2 of the 9 facets.
That completion proceeds in an order dictated by the additional layers of mathematical structure required to support each.
Moreover, the `nonlinear nature' of the interactions between the Problem of Time facets renders a joint study of them mandatory. 
The current article is none the less a useful prequel via gaining a conceptual understanding of each facet, 
prior to embarking on rendering some combinations of facets consistent and what further obstructions arise in attempting such joint considerations.
See \cite{ACastle, ATop, ABook} for up to date studies of this more complicated joint version.   
I also identify new facets (threading based), subfacets (of observables and of reconstructions) 
and further source of variety from how far down the levels of mathematical structure these facets extend.

\end{abstract}


\section{Introduction}

This Article covers the Problem of Time (PoT) facets and their underpinnings by aspects of Background Independence.
The main part of the text presents this from a structural rather than historical perspective.
The preliminary text is more informal and somewhat more historical.  
In this vein, see Sec  2       for some items of relational critique of Newtonian Mechanics, 
                  Sec  3       for the standard formulation of GR and some interplay between this and Background Independence.
              and Secs 4 and 5 for GR reformulated in dynamical terms.
This Article is written in terms of Wheeler's geometrodynamical form (GR as dynamics of 3-geometries), 
though at least the classical parts of it readily carry over to Ashtekar variables/Loop Quantum Gravity formulations of GR's dynamics.  
I use $8\pi G = 1$, $c = 1$, $\hbar = 1$ units throughout.

The main part of the text is then as follows.
Treating PoT facets piecemeal has its limitations due to plentiful interactions between the facets \cite{Kuchar92, Kuchar93, I93}. 
None the less, some progress can already be made from a simpler facet by facet kind of treatment.
In particular, I exposit that 

\ni 1) eight of the nine facets traditionally attributed to the PoT have classical precursors. 

\ni 2) That all nine of these arise from aspects of Background Independence demanded.

\ni 3) I provide updates on the natures, and thus also names, of these fundamental and beautiful concepts.

\ni I present these in conceptual outline, with but light use of mathematics.
See e.g. \cite{Kuchar92, Kuchar93, I93, APoT2, FileR, BI, TRiPoD, ATop, ABook} for more detail of some of the interactions between facets.

Thus starting at the classical level, Sec 6 covers Temporal and Configurational Relationalism  
(the background independent aspects that lead to the well-known Frozen Formalism and Thin Sandwich PoT facets respectively).  
Sec 7 covers Constraint Closure and Expression in terms of Beables 
(the Background Independence aspect names for a generalization of the Functional Evolution Problem and a reconceptualization of the Problem of Observables).
Sec 8 covers Spacetime Relationalism, Foliation Independence and Spacetime Reconstruction.
Sec 9 then outlines the remaining classical-level frontiers.  
Secs 10--13 are then the quantum counterparts of each of the previous four Secs.

Up to date studies of this more complicated joint version will appear in \cite{ACastle, ABook}. 
As intermediate steps currently or shortly available, see 
\cite{ARel, TRiPoD} for some indication of rendering Sec 6's material more widely compatible with other facets, 
\cite{ABeables}     as an expansion of Sec 7,
\cite{AM13, TRiFol} as regards incorporating Sec 8's material, and 
\cite{ATop}         as a substantial expansion on Sec 9.
The model arenas of 
\cite{FileR, ATriangle} (relational triangle), 
\cite{QuadI, QuadIII}   (relational quadrilateral), 
\cite{AMSS1}            (minisuperspace) and 
\cite{SIC1}             (inhomogenous perturbations about minisuperspace) all include some aspects of the joint problems; all bar the last involve quantum-level considerations as well.

There is much common ground between Secs 6 and 8 and parts of Mercati's \cite{Mercati14} tutorial on the Shape Dynamics program \cite{B11, Kos2, Proceedings}. 
This reflects that both Shape Dynamics and the current Article's distinct Background Independence program share a common ancestor: 
Barbour's program from the 1982--2001 period (see e.g. \cite{BB82, B94I, RWR}), from which the Temporal and Configurational Relationalism concepts come from.
The current program differs from Shape Dynamics by taking seriously that Barbour's program {\sl represents progress in beginning to unlock 
the Kucha\v{r}--Isham type multi-faceted conception of the PoT}, thus seeking to follow this up by systematically unlocking each of the further facets in turn. 
Because of this, the Background Independence program has further ancestors, 
both in building upon Kucha\v{r} and Isham's PoT and in making use of further partial resolutions due to Dirac \cite{Dirac}, Teitelboim \cite{T73} and Halliwell \cite{H03}, 
alongside a number of the Author's own (\cite{ACos2, Records, AKendall, AHall, FileR, A13} and the above-cited). 
In contrast to this broader perspective, Shape Dynamics takes the Barbour-type program seriously in the sense of {\sl repeating its methodology but now for 
conformal transformations of space alongside the original program's diffeomorphisms of space}.
Shape Dynamics is thus somewhat narrower in perspective, though it does have a second ancestor in York-type formulations of the GR initial value problem \cite{Yorktime1}, for 
indeed its conformal group and group action are along the same lines as York's.  
The fork between these two programs dates back to \cite{RWR, Than, ABFO, ABFKO}, and is updated and further explained in \cite{Phan, AM13, Mercati14, ACastle}; 
see Sec 8 for a brief outline.    

\end{titlepage}

\section{Absolute versus relational motion debate}\label{AORM}

When Newton was setting up his Mechanics, his bucket thought-experiment convinced him that absolute space was real.  
Observed physics was well accounted for within Newton's paradigm physics until the end of the 19th century.
Then evidence for further Physics began to accumulate and be noticed; this led to QM, SR and GR.

\mbox{ } 

\ni The first issue to consider, however, is whether Newtonian Mechanics in itself has a solid philosophical basis.    
Relationalists take issue with the immovable external character of the absolute space and time assumed in Newton's paradigm.  
An alternative relational worldview could be along the following lines.\foo{Note that precise mathematical implementation only starts in Sec \ref{Cl-PoT})}.  

\mbox{ } 

\ni Relationalism 0) {\it Physics is to be solely about the relations between tangible entities}.

\mbox{ }

\ni Note that this is a statement universal to all of Physics rather than just about Mechanics. 
Indeed, I use `tangible entities' rather than `material objects' to make it clear that this concept is open to fields and `force mediators' as well as `matter building blocks', 
as befits modern Physics.
Key properties of `tangible entities' are then along the following lines. 

\mbox{ }

\ni Relationalism 1) These tangible entities {\it act testably and are actable upon}.  

\mbox{ }

\ni Things that are {\sl not} acting testably or actable upon are held to be {\sl physical} non-entities.  
[These can still be held to be a type of thing as regards being able to {\sl philosophize} about them or {\sl mathematically represent} them.  
Absolute space is an obvious archetypal example of such a non-entity.]
The sentiment is that imperceptible objects should not be playing causal roles influencing the motions of actual bodies. 
As a first sharpening of this, in James Anderson's \cite{A67} view {\it ``the dynamical quantities depend on the absolute elements but not vice versa"}, 
and an absolute object {\it ``affects the behavior of other objects but is not affected by these objects in turn"} \cite{AG}. 
An intuition for what background fields are is fields that violate the action-reaction principle.

\mbox{ }

\ni Relationalism 2) Following Leibniz \cite{L}, any entities which are {\it indiscernible are held to be identical}. 

\mbox{ }

\ni I.e. Relationalism posits that, for physical purposes, physical indiscernibility {\sl must} trump multiplicity of mathematical representation.  
Such multiplicity can still exist mathematically, but the mathematics corresponding to the {\sl true} physics in question is the equivalence class spanning that multiplicity.
[One would then only wish to attribute physical significance to calculations of tangible entities which are independent of the choice of representative of the equivalence class. 
By this e.g. our universe and a copy in which all material objects are collectively displaced by a fixed distance surely share all observable properties and thus are one and the same.  
The archetype of this in modern Physics is Gauge Theory, 
which carries the additionally important insight that a set of part-tangible and part-non entities is often far more straightforward to represent mathematically.]

\mbox{ }

\ni One at least in part affords separate treatments of space and instantaneous configurations on the one hand, and of time on the other.  
This befits the great conceptual heterogeneity between these.
That understood, relational postulates can be stated, a coherent subset of which are sharply mathematically implementable. 

\mbox{ } 

\ni {\it Leibniz's Time Principle} is that there is no time for the universe as a whole.

\mbox{ } 

\ni This point is more usually expressed in terms of both Leibniz and Newton having a notion of time concerning temporal relations between events, 
but for Leibniz the existence of the events is independent of absolute time and the relational time notion is the {\sl only} notion.
Moreover Leibniz acknowledged time's ordering property but not its metric property \cite{L}.
Taking Leibniz's idea as far as eliminating the primary non-existence of time -- so that one is to start with a formulation of Physics in which time is absent -- 
is much clearer in Barbour's two approaches \cite{B94I, B94II} than in Leibniz's own works.
[See also discussion of the Frozen Formalism Problem in Secs \ref{TR-Intro} and \ref{Facets-1}.] 

\mbox{ }

\ni {\it Leibniz's Perfect Clock} is the distinct suggestion that the whole universe is the only perfect clock.
(See \cite{Bfqxi} and around p 41 of \cite{Whitrow} for details.) 
This is to be contrasted with Newton's position that the universe {\sl contains} clocks, which are regarded as substantially localized objects such as a pendulum clock.
Four objections are as follows.
1) it would be operationally impractical to use the whole universe as a clock. 
It would take a considerable effort to monitor the whole universe and one only has very limited knowledge of many of its constituent parts.  
Including scantly known information from remote parts of the universe would {\sl lower} the accuracy of one's timestandard.
2) Were one to go so far as to include the entirety of the universe's contents in one's quest to `perfect' one's clock, 
one would be treating the entirety of a closed system, at which precise point apparent frozennesses materializes as per Leibniz's Time Principle.
3) Adopting this principle would additionally open Pandora's box as regards how to reconcile Leibniz's meaning of `universe' with that of modern GR cosmology.
This is because this principle's intent is constructive, so it would require active use of the meaning of the word `universe' in its statement. 
[Contrast with Leibniz's Time Principle {\sl not} entailing detailed consideration of how one conceives of `universe'.]
4) In any case perfect clocks are not possible in QM \cite{SW58,UW89}. 
Thus we do not adopt this further suggestion of Leibniz.
We do, however (Sec \ref{Ephemeris}, \cite{ABook}), consider clocks that are considerably more extensive than a pendulum clock, pocket watch or atomic clock, 
such as those which are based on the solar system, at least for calibration purposes (Sec 2.2). 
This can be considered to carry some vestige of Leibniz's perfect clock idea, now realistically balanced with how Physics is about {\sl precision} rather than about perfection.  

\mbox{ } 

\ni Mach then pointed out some flaws in Newton's bucket argument. 
Firstly, the rotation is with respect to the `fixed stars', which Mach envisages as a need to include the effects of distant matter. 
Secondly, Mach envisaged that allowing for the bucket to be materially significant -- {\it ``several leagues thick"} \cite{M} -- 
is outside of the situation overruled by observation and practically possible future experiments.  
This has led to to a variety of strengths of statement concerning the hypothetical origin of inertia, 
along the lines of `the distribution of masses in the universe determines inertia at each point'. 
Although such a {\it `Mach's Principle for the Origin of Inertia'} is Mach's best-known insight in the foundations of Mechanics, 
it plays a limited role in the current Article beyond the above refutation.
Mach's foundational suggestions are somewhat disjoint; the ones that the current Article does largely pick up instead are, rather, as follows.  

\mbox{ } 

\ni {\it Mach's Space Principle} is that {\it ``No one is competent to predicate things about absolute space and absolute motion. 
These are pure things of thought, pure mental constructs that cannot be produced in experience. 
All our principles of mechanics are, as we have shown in detail, experimental knowledge concerning the relative positions of motions and bodies."} \cite{M}

\mbox{ }  

\ni {\it Mach's Time Principle}, on the other hand, is based upon {\it ``It is utterly beyond our power to measure the changes of things by time. 
Quite the contrary, time is an abstraction at which we arrive through the changes of things."} \cite{M}.
I.e. `time is to be abstracted from change'.  

\mbox{ } 

Indeed, it is change that we directly experience, and time notions are merely an abstraction from that, albeit a very practically useful abstraction if chosen with due care. 
A further linking  idea then is that Mach's Time Principle {\sl resolves} the timelessness of Leibniz's Time Principle.
I also caution that the well-known philosopher Charles Broad \cite{Broad} declared that time and change are ``{\it the hardest knot in the whole of philosophy}".

\subsection{Relational formulations of Mechanics}

Historically, however, there was a lack in viable relational theories/formulations of Mechanics.  
The comparatively recent {\it relational particle mechanics (RPM)} theories of Barbour--Bertotti (1982) \cite{BB82} and Barbour (2003) \cite{B03} have made up for this deficiency.
See Sec \ref{Cl-PoT} for a brief account and \cite{Buckets, FileR} for further commentary.

\subsection{Ephemeris time as a realization of Mach's Time Principle}\label{Ephemeris}

Around the turn of the 20th century, departures from predicted positions of celestial bodies were noted, especially for the Moon. 
These were most succinctly accounted for not by modifying lunar theory but by considering the rotation of the Earth to inaccurately read off the dynamical time. 
(This, to the accuracy of that epoque's Celestial Mechanics observations, remains conceivable of as the time of Newtonian Mechanics.) 
de Sitter \cite{DS27} explained this as follows.  
``{\it The `astronomical time', given by the Earth's rotation, and used in all practical astronomical computations, differs from the `uniform' or `Newtonian' time, 
which is defined as the independent variable of the equations of celestial mechanics.}"

This was then addressed by using the Earth--Moon--Sun system as a superior timestandard provider. 
Here, Clemence's eventual proposal in 1952 \cite{Clemence} involved a particular way
of iteratively solving for the Earth--Moon--Sun system for an increasingly-accurate timestandard that came to be known as the {\it ephemeris time}.
Whereas such an ephemeris time has long been in use, that is an example of Machian time has only relatively recently been remarked upon \cite{B94I, ARel2}.  

\mbox{ }

\ni As regards conceptualizing about timestandards, ephemeris time is a notable exception to basing clocks upon periodic motions.   
This is via its factoring in irregularities. 
It is also an example of a high-accuracy but inconvenient process, as opposed to consulting a convenient `reading hand'.  
The change away from a sidereal time based {\sl time-unit} followed suit in the late 1950's.
This was chosen to relate to the definition of the tropical year 1900, which definition was in use until 1967.
Subsequently a further redefinition of time-unit occurred to bring it in line with the atomic clock paradigm of timekeeping.

\subsection{Universality of relational thinking}

Furthermore, the arguments of Leibniz and Mach are philosophically compelling enough that they should apply to not just Mechanics but to Physics as a whole.
I.e. a {\it universal} position over the set of laws of Physics, which, as subsequent chapters shall reveal, was an important aspect of Einstein's thinking in developing SR and GR.
See Secs \ref{distinction} and \ref{Cl-PoT} for the extent to which SR and GR succeed in addressing and resolving the absolute versus relational motion debate.

\subsection{SR and the absolute versus relational motion debate}\label{distinction}

Since the aether rest-frame was a candidate for Newton's absolute space (e.g. p3 of \cite{Rindler}), SR being well-known for removing the aether concept is of relevance here.
SR, furthermore changes the notion of simultaneity and in doing so brings the light cone/causal structure to the forefront. 
It is also often said that in SR space and time can/must be regarded as fused into SR spacetime; 
Minkowski himself argued that the individual notions of time and space were ``{\it doomed to fade away}" \cite{Mink}.  
However, Broad \cite{Broad} retorted that SR breaks only {\sl isolation} of space and time, not their {\sl distinction}. 
E.g. signature continues to distinguish timelike and spacelike, and time retains many of its specific properties that space does not possess.  
It will be relevant below that SR's spacetime introduces a dichotomy, even if the spacetime horn might not be adhered to in subsequent physics.

Moreover, the above SR changes in worldview are but a relatively minor change of paradigm as compared to the advent of GR.
In particular, time and space in SR are {\sl also} external and absolute in the sense of SR having its own presupposed set of privileged inertial frames. 
`Acts but cannot be acted upon' objections to absolute space continue to afflict SR by applying just as well to the {\sl class of inertial frames}) \cite{Rindler}. 
The new privileged structures are underlied by SR's Minkowski spacetime's possessing suitable Killing vectors.
A fortiori, Minkowski spacetime possesses the maximal number of Killing vectors (10 in 4-$d$).    
These correspond to the (time and space) translation, rotation and boost generators of the Poincar\'{e} group.  
That passing to GR is a major paradigm in this sense is argued for in Secs \ref{T-GR}, \ref{Cl-PoT} and \ref{Cl-PoT-3}.

\section{The standard formulation of GR in spacetime terms} \label{T-GR}

This is outlined below so as to argue for its indirectness as regards addressing a number of relational issues.
Further geometrodynamical and then fully relational formulations are then presented in Secs \ref{Gdyn} and \ref{Cl-PoT}.  

\mbox{ }

\ni Einstein's historical approach to GR started from his hopes for universality of SR being thwarted by gravitation. 
He then recognized the importance of the Equivalence Principle (see \cite{Will} for further experimental evidence found since).
He approached this via his elevator thought-experiment, in which an observer in a small enclosed laboratory is not able to discern 
whether gravitational fall or rocket acceleration is being experienced.
This is tied to universality of free fall \cite{EG13} (i.e. independently of the material composition of falling test bodies).  
Einstein took this universality to point to a single underlying geometry being experienced by all the test particles: a curved generalization of SR's  4-$d$ spacetime notion.
That gravitation in one sense can be transformed away at any particular point is then represented by the mathematics of the spacetime connection.
By this feature, freely falling frames are but local concepts, thus deserving the name `local frames'.
The combination of Newton's second law and Newton's law of gravitation can be reformulated as a geodesic equation with a spacetime connection whose only nonzero components are 
${\Gamma^{^{(4)}i}}_{00} = \pa_i\phi$ for $\phi$ the Newtonian gravitational potential.

Local agreement with SR is also required.
A natural hypothesis here is Einstein's that SR inertial frames are global idealizations of GR's local inertial frames that are attached to freely falling particles.
Furthermore, in parallel with the development of SR, Einstein retained a notion of metric\footnote{$\mg$ is then the spacetime metric's determinant,  
and $\nabla_{\mu}$ is the spacetime covariant derivative.}
$\mg_{\mu\nu}$ on spacetime to account for observers in spacetime having the ability to measure lengths and times if equipped with standard rods and clocks.
I.e. the inner product character of length and angle carries over from SR to GR and, moreover, manifests the same nontrivial signature as in SR to encode the distinction between 
time and space.
Thus one is dealing with an in general curved semi-Riemannian metric.  
%
%
In another sense, this represents the gravitational field, replacing  the single Newtonian scalar field by a geometrical decuplet of fields.
It turned out to suffice the metric connection associated with this sufficed in aforementioned role for a connection in the theory.  
As $\mg_{\mu\nu}$ reduces locally to SR's $\eta_{\mu\nu}$ everywhere locally the other laws of Physics take their SR form.  
Indeed, notions of timelike, spacelike and null carry over to here, 
as does using the first and third of these to interpret massive and massless particle based matter respectively.
The straight timelike lines followed by free particles in SR's Minkowski spacetime are bent into the curves followed by relatively-accelerated freely-falling particles.  
The straight null lines constituting the lightcones of Minkowski spacetime are likewise bent.

Two remaining issues are firstly the natural question of how is one to interpret the spacetime curvature associated with the affine connection and the metric.  
Secondly, what are suitable field equations 
(in parallel with Maxwell's equations for Electromagnetism, and so as to recover the Poisson form of Newton's Law of Gravitation in a suitable limit).
These are to be tensorial: the general covariance assumption.

Spacetime curvature models gravitation in a third sense; in particular, curvature (unlike a connection) is a tensor, 
so it cannot be made to go away at the point of interest by coordinate transformation. 
Thus the notion of gravitation that this models is the one which one cannot free one's local picture of physics from.
Note that this use of `local' requires a neighbourhood rather than a point, 
e.g. the curvature tensor manifests itself though {\sl finite-region} vector transport or geodesic deviation involving {\sl finitely separated} geodesics.

As regards setting up field equations, a hypothesis that turns out to be useful and used two SR paradigm steps is viewing the source term in Poisson's law in terms of energy, 
and then extending this to sourcing by the energy--momentum--stress tensor $\mT_{\mu\nu}$.
Thus Einstein conjectured that energy--momentum--stress sources some notion of spacetime curvature and hence of gravitation.
He eventually realized that he had to find a curvature tensor that matches the properties of the energy--momentum--stress tensor $\mT_{\mu\nu}$: symmetry and divergencelessness.
This is the Einstein tensor $\mG_{\mu\nu} := \mR_{\mu\nu} - \mg_{\mu\nu} \mR/2$ \cite{Ein15},\footnote{Here $\mR_{\mu\nu}$ is the Ricci curvature tensor 
and $\mR$ the Ricci scalar curvature.}
which is then equated with $\mT_{\mu\nu}$ up to proportionality (set by the Poisson equation) 
\beq
\mG_{\mu\nu} =  \mT_{\mu\nu} \mbox{ } . 
\label{EFE}
\eeq
Various comments are in order at this point.

\mbox{ } 

\ni 1) A cosmological constant term $\Lambda \mg_{\mu\nu}$ can also be incorporated since this also fits the symmetry and conservation criteria.

\ni 2) One also has the good fortune that the number of these Einstein field equations matches the number of independent components of $g_{\mu\nu}$ so the system is, at least, well-determined.

\ni 3) N.B. that the above is not only a paradigm shift from flat Minkowski geometry to in general curved geometry, but also from {\it pre-determined} Minkowski geometry to a geometry 
that is curved by a property of the matter distribution, and so {\it determined by the contents of the universe}.   
Thus one has moved from `actors on a rigid stage' to `material blobs moving around on a rubber sheet that they deform' \cite{MTW}.  
In this loose analogy, neither the blobs nor the rubber sheet have an underlying privilege, whereas the actors can presuppose the existence of a rigid stage upon which to perform.  
Thus (\ref{EFE}) is a `geometry = matter' equation: geometry (in the form of curvature) and matter (in the form of energy--momentum--stress) influence each other in this theory. 

\ni 4) Also note that GR explains the limited extent in practise of SR's inertial frames in terms of the sources of gravitation, 
by which inertial frames cease to be structures that cannot be acted upon \cite{Rindler}.

\ni 5) I leave the well-known and yet disputed role of {\it Mach's principle} \cite{M} in the inception of GR to Secs \ref{GR-as-BI} and \ref{Cl-PoT}.  

\ni 6) The Einstein tensor includes the same amount of information as the Ricci tensor, but that this is less than the amount of information in the Riemann curvature tensor.
The Weyl tensor precisely picks out the remainder, and admits interpretation as gravitational waves.

\mbox{ } 

\ni 7) For later comparison, this Sec is a 'discover connections, discover curvature' account, 
in that connections arise first and are suggestive of consideration of the associated notions of curvature.  

\ni 8) A useful further formalization is that GR spacetime is a pair $({\FrM}, \mg_{\mu\nu})$, for $\FrM$ the underlying topological manifold; 
one additionally assumes that $\FrM$ carries differentiable manifold structure.  

\ni 9) As regards casting GR in terms of a variational principle, the Einstein--Hilbert action for pure GR is\foo{I use round brackets for function dependence, 
square ones for functional dependence, and ( ; ] for mixed function dependence (before the semicolon) and functional dependence (after it).}  
\beq
\FS^{\sG\sR}_{\sE\sH} = \mbox{$\frac{1}{2}$} \int_{\sFrM} \d^4x\sqrt{|\mg|} \, \mR(\vec{X}; \mg] \mbox{ }  .  
\label{S-EH}  
\eeq
One is to introduce here also a matter action, combined additively with (\ref{S-EH}).
This takes an $\eta_{\mu\nu}$ to $\mg_{\mu\nu}$ of its SR form, which amounts to a minimal coupling postulate (a type of local Lorentz invariance). 
Much as one can cast all the observationally-established non-gravitational classical laws of Physics in SR form, 
one can cast them (now free from this non-gravitational caveat) in GR form \cite{MTW}.  
Then varying this additive combination gives Einstein's field equations for GR, (\ref{EFE}).
The energy--momentum--stress tensor arises here as 
\beq 
\mT^{\mu\nu} := 2|\mg|^{-1/2}{\delta \FS_{\sm\sa\st\st\se\sr}}/{\delta \mg_{\mu\nu}} \mbox{ } .
\eeq 
Finally, to include the cosmological constant, pass from $\mR$ to $\mR - 2 \Lambda$ in the action, 
by which the left hand side of the Einstein field equations picks up $+ \Lambda\mg_{\mu\nu}$.

\subsection{Spacetime diffeomorphisms}

\ni 1-to-1 maps $\phi: \FrM \rightarrow \FrM$ that are differentiable and possess differentiable inverses are {\it spacetime diffeomorphisms}, Diff($\FrM$).

\mbox{ } 

\ni Note that in the case of GR spacetime, the distinction between passive and active diffeomorphisms given there acquires further significance 
Passive diffeomorphisms are coordinate transformations, tied to the well-known notion of Jacobian matrix.
On the other hand, active diffeomorphisms are the moving around of points of a manifold, tied to the notion of {\it Lie derivative}, 
which indeed provides a means of moving -- Lie-dragging -- points around.  
It is active diffeomorphisms that are the main concern in the study of GR, for Background Independence reasons that we postpone until Sec \ref{Cl-PoT-2}.
Also note the step-up from Electromagnetism, whose transformations occur at a fixed spacetime point, whereas in GR the diffeomorphism group moves points around \cite{I93}. 

\mbox{ } 

\ni Given that GR spacetime is also equipped with a metric, a subsequently useful notion are the {\it isometries}: metric-preserving 1-to-1 maps. 
Isometries are furthermore related to both Lie derivatives and Killing vectors as follows:
\beq
\pounds_{\vec{\sX}}\mg_{\mu\nu} = 2\nabla_{(\mu}\mX_{\nu)} := (\mathbb{K}\mX)_{\mu\nu} \mbox{ } .
\label{crux}
\eeq
Here the first equality is computational, and illustrates a common trend: 
that Lie derivatives can be re-expressed as covariant derivatives in the presence of sufficient structure to define the latter.
Moreover, the Lie derivative notion is more minimalistic: it pertains to just differential geometry whereas the covariant derivative also requires an affine connection.
The second equality ties the Lie derivative to the also widely useful notion of Killing vectors, $\mathbb{K}$ denoting the Killing form.  
When this last expression is equal to zero, it is Killing's equation. 
Then the vectors (in spacetime $\vec{\mX}$) which solve this are the Killing vectors. 
Killing vectors are then a crucial part of the extension of the notions of symmetries and conserved quantities to more general settings than in the flat spaces of Newtonian 
and Minkowskian Physics.

\subsection{Further comments on the background independent side of GR}\label{GR-as-BI}

\ni Moreover, GR may be viewed as not only a relativistic theory of gravity but also as a freeing from absolute or background structures. 
Indeed along such lines, GR is often interpreted as providing a physically meaningful explanation of the privileged frames of SR.  
I.e. that SR inertial frames are really idealized arbitrarily large versions of GR's local inertial frames, 
that are, furthermore, determined in turn by matter content as per Sec \ref{T-GR}.

Einstein was influenced by Mach in developing GR along these lines \cite{Einstein-1-2}, albeit not in a straightforward manner \cite{WheelerGRT, DOD}.
Initially, he misinterpreted Mach's Origin of Inertia Principle  (due to confusion between `inertia' in the usages `inertial mass' and `inertial frames').  
He then eventually abandoned his `Machian' approach for a more indirect approach -- Sec \ref{T-GR}'s -- involving spacetime frames rather than spatial frames).  
The resulting theory of GR can be investigated as regards whether various Machian criteria apply to it.  
Some do, e.g. frame dragging. 
And some do not, such as some GR solutions are in some senses un-Machian, 
e.g. in GR universes with overall rotation are physically distinguishable from universes which are not rotating \cite{HE}).

Note here Sec \ref{AORM}'s argument that `Machian' is a somewhat disjoint set of attributes that a theory might have, rather than a connected package that a theory need possess all of. 
Only some parts of Mach's insights have durability to the more advanced physical theories of the subsequent century, starting with GR. 
Those parts which are crucial in this work (and Barbour's) are Mach's Time Principle and Mach's Space Principle (Sec \ref{AORM} and expanded upon below).
These are dynamical tenets; Mach's work pre-dated spacetime largely and the correct Einstein's field equations of GR entirely.
Thus there is a sense in which a spacetime approach such as Einstein's at most indirectly addresses Mach's criteria.

To set up a background independent theory along dynamical lines, it helps to have seen the standard and traditional dynamical/canonical formulation of GR of the next Sec.

\subsection{Many routes to GR and dynamically primary physics}\label{Many-Routes}

Moreover, according to Wheeler's many-routes perspective \cite{Battelle, MTW}, Einstein's derivation of GR (Sec \ref{T-GR}) is but the first of a number of routes to GR.  

\mbox{ }

\ni Isenberg and Wheeler furthermore argued that from Galileo until Einstein's conception of SR, Physics was held to be {\it ``dynamics in the great tradition"} \cite{IW79}.   
In dynamics, time and space are treated in a distinct manner. 
In particular, dynamics concerns configurations and momenta evolving with respect to some notion of time, 
and treats derivatives with respect to time differently from those with respect to space.

However Minkowski then introduced spacetime \cite{Mink}, and the paradigm shifted from the primality of dynamics to that of spacetime (SR, QFT, GR).
The dynamical perspective does not directly fit in with the SR and GR spacetime perspective, 
in the sense that spacetime itself neither evolves in time nor plays the role of timeless configuration.
That Broad limited Minkowski's argument for this paradigm is of particular relevance here; 
both Minkowskian and Einsteinian spacetimes are co-geometrizations of space and time, rather than an end to the actual distinction between the two concepts.  
As regards further such arguments, see e.g. Whitrow \cite{Whitrow};  
Dirac \cite{Dirac} also questioned spacetime's acquisition of primary status.  
{\it ``One cannot, however, pick out the six important components from the complete set of 10 in any way that does not destroy the four-dimensional symmetry.
Thus if one insists on preserving four-dimensional symmetry in the equations, one cannot adapt the theory of gravitation to a discussion of measurements in the way Quantum Theory 
requires without being forced to a more complicated description than is needed by the physical situation.
This result has led me to doubt how fundamental the four-dimensional requirement in physics is."}
Barbour provided further arguments for spacetime non-primality in e.g. \cite{EOT, B11}.  
Wheeler additionally supplied misgivings about the status of GR spacetime at the quantum level (see Sec \ref{PoT-SRP}), 
and addressed these by conceiving of GR as Geometrodynamics so as to take a step back from GR spacetime back into the `great tradition'. 
[I.e. from GR spacetime via the ADM split into the dynamical formulation of GR that is Geometrodynamics.]

By such arguments one does not {\sl have to give up} dynamics, but rather one inspects how dynamics manifests itself in theories that also admit spacetime formulations.

Then in particular, GR spacetime is found to {\sl contain} notions of both spatial configuration and of time.
We shall see in the next two Secs how these can be extracted by splitting the spacetime metric up, 
and this additionally induces a split of GR's Einstein field equations along dynamical lines \cite{ADM, Battelle, MTW}. 
Moreover, one may consider dynamics to be primary, and thus ask from first principles what GR is a dynamics of, i.e. what its configurations are.
Either way, GR's configurations are spatial (i.e. positive-definite Riemannian) 3-metrics, $\mh_{ij}$; in the spacetime-first picture, these are the `spatial slice' part of spacetime.  
In fact, as we shall see in Sec \ref{Gdyn}, this is a redundant presentation; 
less redundantly, GR's configurations are spatial {\it 3-geometries}: 3-metrics `minus coordinate information'.  
Thus, as well as a spacetime formulation, GR also admits a dynamical formulation in terms of evolving spatial 3-metrics or 3-geometries; 
Wheeler termed the latter {\it geometrodynamics} \cite{WheelerGRT, Battelle, MTW}.

\mbox{ } 

\ni As further developments, 1) Not only then does dynamics furnish another route, 
but also the two-way passage between spacetime and dynamical formulations of GR count as a pair of routes (\cite{Battelle, MTW} and Sec \ref{Cl-PoT-3}).  

\ni 2) The relational approach has furthermore developed (\cite{RWR, BI, AM13} Secs \ref{Cl-PoT}, \ref{Cl-PoT-3}) 
so as to be able to arrive at the geometrodynamical formulation of GR without ever passing though spacetime, i.e. never departing from the `great tradition'.    
[In this picture spacetime is still a useful reformulation, just not the primary ontology.]

\section{Kinematics of split space-time}\label{GR-Kin}

\subsection{Topological manifold level structure}\label{Top-Sigma}

\ni 0) In dynamical approaches to GR, one has to preliminarily choose a residual notion of space in the sense of a 3-surface that is a fixed topological manifold $\bupSigma$. 
 
\ni 1) In the current Article, $\bupSigma$ is compact without boundary for simplicity.  

\ni 2) The current Article's specific examples, use $\bupSigma = \mathbb{S}^3$: the 3-sphere, or $\mathbb{T}^3$: the 3-torus, which are two of the simplest cases. 

\mbox{ } 

\ni By 0), a fixed $\bupSigma$ is to be shared by {\sl all} the spatial configurations in a given geometrodynamics.
I.e. dynamical GR approaches such as geometrodynamics are built subject to the restriction of not allowing for topology change.
This means that geometrodynamics covers a more restricted range of spacetimes than the spacetime formulation of GR does: those of spacetime topology $\bupSigma \times \bigtau$. 
Thus geometrodynamics is just a `manifold topolostatics' rather than being a `manifold topolodynamics' as well, an issue to which we return in Sec \ref{7-Hells} and \cite{ATop}.
This has some superficial resemblance with Newtonian space-time, e.g. as a stringing together by time variable labels.
However, the spatial slices are now in general different from each other at the metric level. 
Additionally GR spacetime carries further connotations of the following.

\ni A) Unified co-geometrization: an overall 4-metric rather than separate spatial and temporal metrics in the Newtonian case.

\ni B) Causality structure encoded by the indefiniteness of the 4-metric rather than Mechanics' slices being privileged surfaces of absolute simultaneity. 

\ni Also GR's time variable is nonunique; different choices of this in general correspond to different foliations, 
each of which is valid and with the Physics involved turning out to be foliation independent (see Sec \ref{Cl-PoT-3} for more).

Further restrictions are placed on $\bupSigma$ in Sec \ref{Ev-Eq}. 
For now at least, one accepts confinement to a subset of GR's solution space so as to be able to study its dynamics.
This is in the sense that if one does so, currently-standard mathematical structures and tools apply.

Finally, I use the notation $\bigupsigma$ in place of $\bupSigma$ if a 3-space is treated in isolation rather than as a slice within spacetime (in a sense made precise in Sec \ref{Single-Hyp}). 
For many purposes, one can also take a finite piece of space $\mS \subset \bupSigma$, rather than a whole space $\bupSigma$.\footnote{For a subset of these purposes, 
one does not need to concern oneself with such piece having boundaries.
These involve `local' considerations in a sense to be made precise at the metric level in the next Sec.}  

\subsection{Differential-geometric level structure, including spatial diffeomorphisms}\label{DLS}

One additionally assumes that $\bigupsigma$ (or whichever of the preceding Sec's variants) carries differentiable manifold structure.   
The maps preserving this level of structure are {\it diffeomorphisms}, Diff($\bupSigma$); 
many properties of these parallel those of Diff($\FrM$), because at this level of structure there is not yet a metric present whose signature distinguishes between spacetime and space.  
E.g. Diff($\bupSigma$) are again actively interpreted (nothing in the given active--passive argument is {\sl signature} dependent). 
Also, 
\beq
\pounds_{\upxi}\mh_{ab} = 2\mD_{(a}\upxi_{b)} := (\mathbb{K}\upxi)_{ab} \mbox{ } .
\label{crux2}
\eeq
is the counterpart of (\ref{crux}) and with the same ties to Killing vectors. 
Sec \ref{Cl-PoT-2} outlines further similarities and some differences.

\subsection{Metric-level structure}\label{MLS}

($\bupSigma, \mh_{ij})$ inherits spatiality from how it sits within $(\FrM, \mg_{\mu\nu})$. 
On the other hand, to ensure that $\bigupsigma$ is indeed cast in a spatial role, this is directly equipped with a specifically Riemannian (positive-definite) 3-metric, $\bh$ with 
components $\mh_{ij}(x^k)$.\footnote{As further useful notation, $\mh_{ij}$ has determinant $\mh$, inverse $\mh^{ij}$ and 3-metric-compatible covariant derivative $\mD_i$.}  
It is natural to consider 3-metrics if one is coming either from Newtonian Mechanics or SR, and to continue to model distance and angle. 
Such formulations were developed by Dirac, Arnowitt--Deser--Misner (ADM) and Wheeler.
Here, GR is reformulated as a dynamics of evolving spatial 3-geometries which stack together to form the standard formulation of GR's notion of spacetime.

\subsection{Single-hypersurface concepts}\label{Single-Hyp}
%
{            \begin{figure}[ht]
\centering
\includegraphics[width=0.35\textwidth]{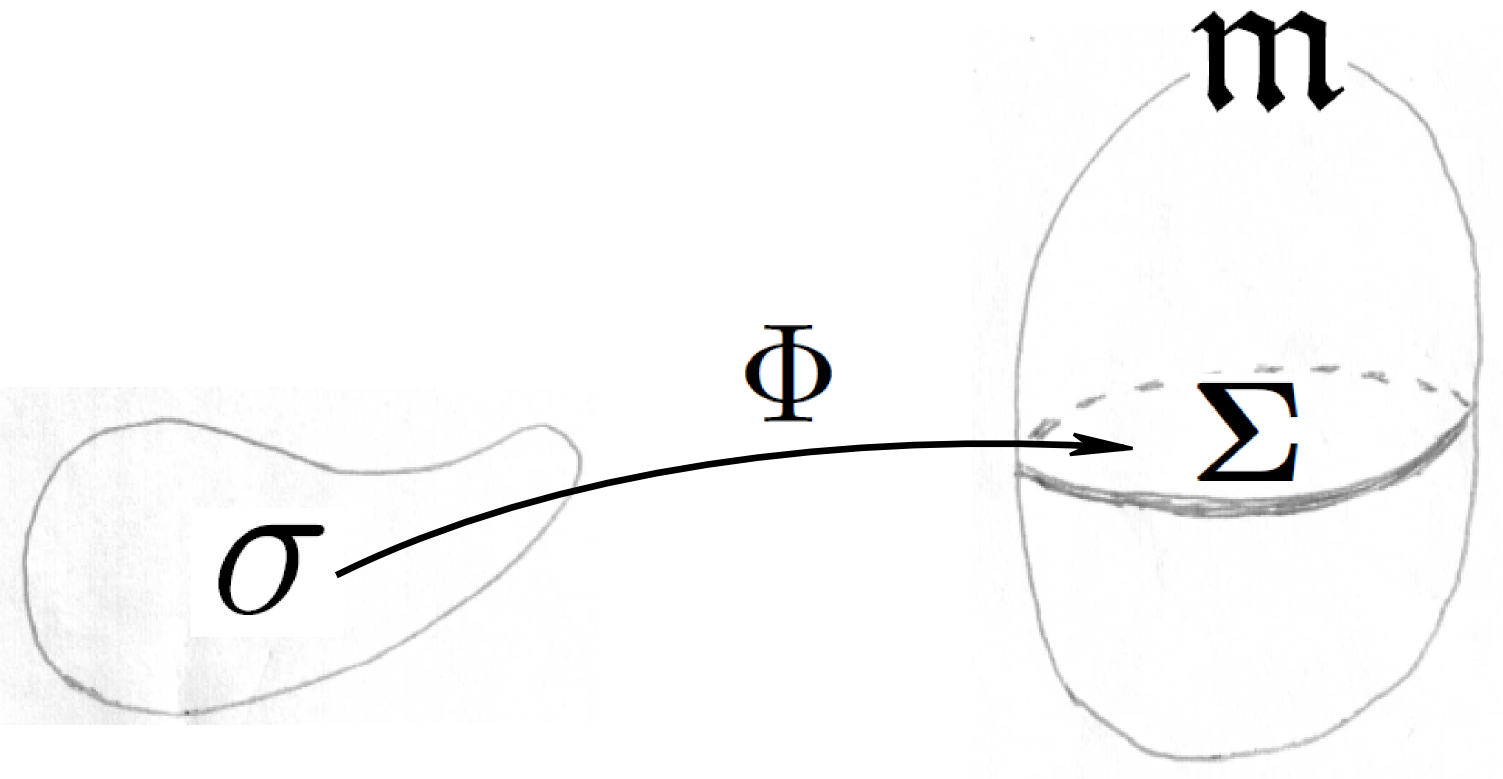}
\caption[Text der im Bilderverzeichnis auftaucht]{        \footnotesize{Embedding $\Phi$ from 3-space $\bigupsigma$ to hypersurface $\bupSigma$ within spacetime $\FrM$.         } }
\label{Space-to-Hypersurface}\end{figure}            }

\ni Let us now consider passing \cite{Gour} from a 3-space $\bigupsigma$ to a spatial hypersurface $\bupSigma$ embedded in a spacetime $\FrM$ -- some `surrounding' space it `sits' in.   
More formally, a {\it hypersurface} $\bupSigma$ within $\FrM$ -- Fig \ref{Space-to-Hypersurface}.a) 
-- is the image of a plain spatial 3-manifold $\bigupsigma$ under a particular kind of map -- an {\it embedding}, $\bupPhi$.    
This construction can also be applied locally \cite{Wald}: embedding a piece  $\mbox{\Large $s$}$ of spatial 3-surface as a piece $\mS$ of hypersurface.  
The notion of hypersurfaces within $\mathbb{R}^3$ is intuitively clear and well-known, e.g. a sheet of paper bent in whichever way, or the surface of a globe. 
Hypersurfaces are more generally characterized as surfaces $\FrHgen$ within an arbitrary-$d$ manifold $\FrMgen$ that are of {\it codimension 1}: dim($\FrMgen$) -- dim($\FrHgen$) = 1.

We next define the {\it normal} $\mn^{\mu}$ to the hypersurface $\bupSigma$ and 
               the {\it projector} $\mP^{\mu}\mbox{}_{\nu} := \delta^{\mu}\mbox{}_{\nu} + \mn^{\mu}\mn_{\nu}$  onto $\bupSigma$.

The spacetime metric is furthermore said to {\it induce} the spatial metric on the hypersurface.
This induced metric is both an intrinsic metric tensor on space, $\mh_{ij}$ and a spacetime tensor $\mh_{\mu\nu}$. 
It attains such a  `dual nationality' by being a {\it hypersurface tensor}.  
I.e. a tensor such that for each `independent index' $0 = \mT_{\mu\nu...\omega}n^{\mu} =:\mT_{\perp\nu...\omega}$. 
%
%
Since $\mh_{\mu\nu}$ is symmetric, $\mh_{\mu\nu}\mn^{\mu} = 0$ is a sufficient condition for this. 
Finally note that once metric geometry become involved, it is isometric embeddings that one is dealing with.

\mbox{ } 

\ni The {\it extrinsic curvature} of a hypersurface is its bending relative to its ambient space. 
For instance, a sheet of paper retains its flat intrinsic $\mathbb{R}^2$ geometry when it is rolled up into a cylinder. 
And yet it has nontrivial curvature relative to the surrounding $\mathbb{R}^3$. 
It is useful to further formalize this concept of extrinsic curvature as the rate of change of the normal $\mn^{\mu}$ along a hypersurface,     
\beq
\mK_{\mu\nu} := \mh_{\mu}\mbox{}^{\rho}\nabla_{\rho}\mn_{\nu} \mbox{ } .  
\label{K-def}
\eeq
Extrinsic curvature can furthermore be cast in the form of a Lie derivative, 
\beq
\mK_{\mu\nu} = \pounds_{\underline{\sn}}\mh_{\mu\nu}/2 \mbox{ } .
\label{K-Lie}
\eeq
\ni 
Extrinsic curvature is, additionally, symmetric: manifestly from (\ref{K-Lie}), and a hypersurface tensor: $\mK_{\mu\nu}\mn^{\nu} = 0$ follows from (\ref{K-def}) and suffices by symmetry.  
The trace tr$\,\mK := \mK$ and the determinant det$\,\mK$ are then useful invariants built from the tensor $\mK_{ab}$.

Additionally, despite being defined in very different ways, it turns out that the intrinsic and extrinsic notions of curvature of a surface are related.  
For a 2-surface embedded in $\mathbb{R}^3$, this is by the well-known Gauss' outstanding Theorem.  
Moreover, this result substantially generalizes, e.g. to the case in which the embedding space itself is curved and higher-dimensional (maintaining codimension 1).
The generalized result thus obtained can also be viewed in terms of projections of the Riemann tensor,\footnote{From here on, spacetime objects have (4) subscripts added where distinction 
is necessary between them and their spatial counterparts.}
\be
\mbox{(Gauss equation) }     \mbox{ }   \mR^{(4)}_{abcd} = \mR_{abcd} + 2 \mK_{a[c}\mK_{d]b} \mbox{ } , 
\label{Gauss-as-proj} 
\ee
\be
\mbox{(Codazzi equation) }   \mbox{ }    \mR^{(4)}_{\perp abc} =  2 \mD_{[c} \mK_{b]a} \mbox{ } .
\label{Cod-as-proj}  
\ee
I.e. the left hand side is here viewed as a projection which is then computed out to form the right hand side.

\subsection{Two-hypersurface and foliation concepts}\label{Fol-Intro}

Some kind of thin one-sided infinitesimal neighbourhood of $\bupSigma$ (Fig \ref{Infinitesimal-Fol}) is required for a number of further notions \cite{Gour}.    
On the other hand, the notion of foliation (see Fig \ref{Infinitesimal-Fol}.c) applies to an extended piece of spacetime, 
i.e. usually involving more than just two infinitesimally-close slices.  

{            \begin{figure}[ht]
\centering
\includegraphics[width=0.8\textwidth]{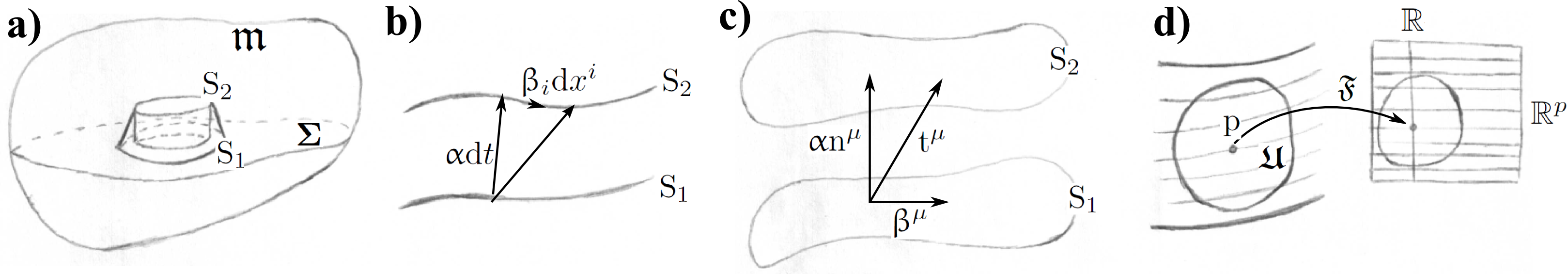}
\caption[Text der im Bilderverzeichnis auftaucht]{        \footnotesize{a) The set-up for $\mS_1$, $\mS_2$ local-in-space pieces of spatial slice; 
$\mS_2$ has to lie within the domain of dependence $\mD^+(\mS_1)$, so one does not have a `direct product' worldtube.  
b) Arnowitt--Deser--Misner 3 + 1 split of a region of spacetime, with lapse $\upalpha$ and shift $\upbeta^i$.
c) Local presentation of the $\mt^{\mu}$, $\mn^{\mu}$, $\upbeta^{\mu}$ split. 
d) {\it Foliation} $\FrF$ is a rigged/decorated version of the standard differential-geometric definition of chart from some neighbourhood $\FrU$ of point $\mp$.}  }
\label{Infinitesimal-Fol}\end{figure}          }

\ni Each foliation by spacelike hypersurfaces is to be interpreted in terms of a choice of time $\mt$ and associated `timeflow' vector field $\mt^{\mu}$.  
There are an infinity of choices of such a $\mt$ (termed a `global time function' in e.g. \cite{Wald}).  
The spatial hypersurfaces correspond to constant values of that $\mt$.  

\ni For Minkowski spacetime, $\mt$ and $\mt^{\mu}$ already exist as fully general entities, but they are usually chosen via a global inertial coordinate system \cite{Wald},  
and of course this ceases to exist in the general GR case.  

\ni For dynamical formulations of GR, one usually demands the spacetime to be time-orientable so that it is always possible to consistently allocate notions of past and future. 

\ni $\mt^{\mu}$ is restricted by $\mt^{\mu}\nabla_{\mu} \mt =  1$ and  $\ms^{\mu}\nabla_{\mu} \mt =  0$ for any tangential $\ms^{\mu}$.
Then if these hold, it is consistent to
\beq
\mbox{identify } \mt^{\mu}\nabla_{\mu}  \mbox{ with } \pa/\pa \mt
\eeq
and then
\beq
\mbox{identify }  \pa/\pa \mt \mbox{ with }  \pounds_{\underline{\st}} 
\label{pa-as-Lie}
\eeq
[in the sense of (string of projectors) $\times \, \pounds_{\underline{\st}} \mT$].  

\mbox{ } 

\ni ADM \cite{ADM} split the spacetime metric into induced metric $\mh_{ij}$, shift $\upbeta^i$ and lapse $\upalpha$ pieces (Fig \ref{Infinitesimal-Fol}):
\beq
\mg_{\mu\nu} =
\left(
\stackrel{    \mbox{$ \upbeta_{k}\upbeta^{k} - \upalpha^2$}    }{ \mbox{ }  \mbox{ }  \upbeta_{j}    } \stackrel{    \mbox{$\upbeta_{i}$}    }{  \mbox{ } \mbox{ }  \mh_{ij}    }
\right)
\mbox{ } . \label{ADM-split}
\eeq
This is often presented for a foliation, though 2 infinitesimally close hypersurfaces suffices (or even less for some parts and some weakenings of some parts.  
The corresponding split of the inverse metric is  
\beq
\mg_{\mu\nu} =
\left(
\stackrel{    \mbox{$ - {1}/{\upalpha^2}$}    }{ \mbox{ }  \mbox{ }   {\upbeta^j}/{\upalpha^2}    }
\stackrel{    \mbox{$   {\upbeta^i}/{\upalpha^2}$}    }{  \mbox{ } \mbox{ }  \mh^{ij}  - {\upbeta^i\upbeta^j}/{\upalpha^2}  }
\right) \mbox{ } ,
\eeq
and that of the square-root of the determinant is $\sqrt{|\mg|} = \upalpha\sqrt{\mh}$.  
\ni In the ADM formulation, $\mt^{\mu}$ has the tangential--normal split: $\mt^{\mu} = \upbeta^{\mu} + \upalpha \mn^{\mu}$.
This serves to define $\upbeta^{\mu} := \mh^{\mu\nu}\mt_{\nu}$ as the {\it shift} (displacement in identification of the spatial coordinates between 2 adjacent slices, 
which is an example of {\it point identification map} \cite{Stewart}). 
Additionally, $\upalpha := - \mn_{\gamma}\mt^{\gamma}$ is the {\it lapse} (`time elapsed'), which may be interpreted as the change in proper time $\d\uptau = \upalpha(t, x^i)\d \mt$.

In ADM terms, if $- \upalpha^2 + \mg_{\mu\nu}\upbeta^{\mu}\upbeta^{\nu} < 0$, the hypersurface within spacetime is spacelike and the normal direction is timelike.
In particular $\upalpha$ cannot vanish anywhere, and one is to take $\upalpha > 0$ everywhere for a future-directed normal.  
Also now the normal in ADM terms is $\mn^{\mu} = \upalpha^{-1}[1, - \uupbeta]$,   
and the extrinsic curvature in ADM terms is 
\beq
\mK_{ij} = \{\mh_{ij}^{\prime} - \pounds_{\underline{\upbeta}}\mh_{ij}\}/{2\upalpha} = \{\dot{\mh}_{ij} - 2\mD_{(i}\upbeta_{j)}\}/{2\upalpha} 
                                                                                     = \delta_{\vec{\upbeta}}{\mh}_{ij}/{2\upalpha}  \mbox{ } , 
\label{Kij-1}
\eeq
where $\mbox{}^{\prime} := \pa/\pa \mt$ for $\mt$ the coordinate time.  
A final useful construct at this level of structure is the {\it hypersurface derivative} \cite{Kuchar76I, Kuchar76Other} given by 
$\delta_{\vec{\upbeta}} := \pa/\pa\st - \pounds_{\underline{\upbeta}}$. 
Note that the correction to $\pa\mh_{ab}/\pa\mt$ is (\ref{crux2}) under the substitution $\upbeta$ for $\upxi$. 

\mbox{ }

\ni The ADM prescription for a split of spacetime is, moreover, far from unique. 
The {\it Kaluza--Klein split} parallels the {\sl inverse} ADM split in form but with new names and interpretations in place of the lapse and shift pieces.    
There is also an alternative {\it threading split} \cite{HE} in which the 1-$d$ threads are primary rather than the 3-$d$ hypersurfaces, so it is termed a 1 + 3 split to ADM's 3 + 1 one.
The ADM split's distinction is in its well-adaptedness to dynamical calculations (as laid out in more detail in the next Sec).
As well as it being built around the dynamical objects of GR (the spatial hypersurfaces), 
it picks out four multiplier coordinates (the lapse and shift), which simplifies the dynamical equations and cleanly split them into a constrained system and an evolution system.
On the other hand, the threading split is well-adapted to observational concepts such as past light cones and fluxes of gravitational waves.

\subsection{Interpretation of foliations in terms of families of possible observers}
%
{            \begin{figure}[ht]
\centering
\includegraphics[width=0.7\textwidth]{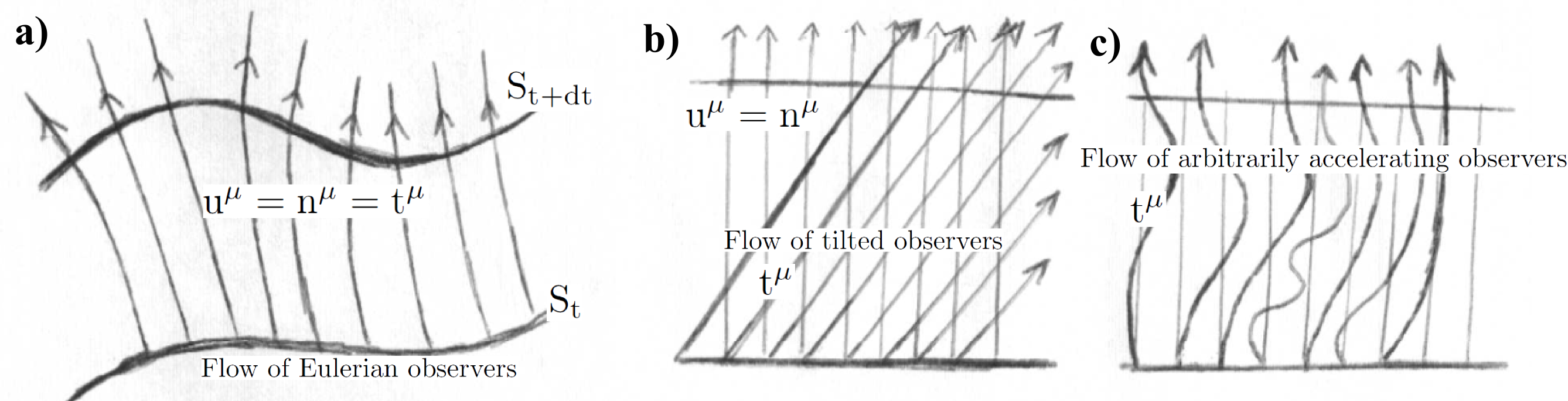}
\caption[Text der im Bilderverzeichnis auftaucht]{        \footnotesize{$\mt^{\mu}$ flow examples.
\ni a) Euler observers.
\ni b) Observers tilted away from fluid flow.
\ni c) Observers moving along an arbitrary congruence (combing spacetime with a fleet of rockets, each of which can accelerate according to the wishes of its observer).}  }
\label{Observer-congruences}\end{figure}          }

\ni Each normalized $\mt^{\mu}$ [$\mt^{\mu}/||\mt|| = \{\upalpha \, \mn^{\mu} + \vec{\upbeta}^{\mu}\}/          \sqrt{\upalpha^2 - \upbeta^2} = 
                                                            \{\upalpha \, \mn^{\mu} + \vec{\upbeta}^{\mu}/\upalpha \}/\sqrt{   1 - \{\upbeta/\upalpha\}^2   } :=
                                                            \{\upalpha \, \mn^{\mu} + \vec{\mv}^{\mu} /   \upalpha\}/\sqrt{1 - \mv^2} =
															\upgamma [1, \mv]$ for $\mv = \upbeta$]  
represents a distinct possible motion of a cloud of observers.  
These are held to be combing out spacetime rather than travelling on mutually-intersecting worldlines.
Elsewise, they have freedom of motion (`rocket engines' permitting each to accelerate independently of the others).

One needs to be careful at this point because both $\mt^{\mu}$ and the foliation itself can differ.
Let us then start by considering the simplified situation for the {\sl Eulerian observers} \cite{Gour, BojoBook} that correspond to each foliation.
Here $\muu^{\mu} = \mn^{\mu} =$ \{a particular normalized $\mt^{\mu}$ orthogonal to the foliation\}.
Thus here there is one motion of observers per foliation and in a sense that is meaningfully dual to that foliation. 
This is in parallel to the standard `ray to wavefront duality' in geometrical optics in space or its configuration space analogue in the Hamilton--Jacobi theory of mechanics \cite{Lanczos}.

Next, consider the general case, for which $\mt^{\mu}$ is unaligned with $\mn^{\mu}$. 
Nor is it necessary for the observers to `go with the flow' $\muu^{\mu}$, since 
1) the predominant matter flow in the universe is indeed not made out of observers. 
2) Observers can instead be regarded as being in independently-moving rockets.
3) These rockets can then to good approximation be idealized as a type of test particle (negligible back-reaction on the spacetime geometry from the rockets).  

\ni Foliations can be thought of as the as level surfaces of the scalar field notion of time, $\mt$.  
$\mt$ is here taken to be smooth, with a gradient that is nonzero everywhere, ensuring that these level surfaces do not intersect anywhere.

\ni Note finally that GR's well-known notion of {\it many-fingered time}, and the ray-to-wavefront dual concept of {\it bubble time}, are conceived of in these terms.

\subsection{Completion, and further interpretation, of the curvature projection equations}

The {\it Ricci equation} 
\be
\mR^{(4)}_{\perp a\perp b} =  \{\delta_{\vec{\upbeta}} \mK_{ab} + \mD_b\mD_a\upalpha\}/\upalpha + {\mK_a}^c\mK_{cb} \mbox{ } .
\label{Ricci-as-proj}
\ee
is the final projection of the Riemann tensor.
Note how this one is specifically an at least infinitesimal foliation concept: whilst $\mK_{ab}$ belongs to a surface, `$\d\mK_{ab}$' requires an infinitesimal foliation.

Three possible conceptually-distinct interpretations of the Gauss, Codazzi and Ricci equations are then as follows.

\mbox{ }

\ni 1) Top-down. 
Given a higher-$d$ manifold containing a hypersurface, how do its curvature components project onto that hypersurface 
(as a combination of intrinsic and extrinsic curvatures of that hypersurface)?

\ni 2) Down-up. 
Given a 3-surface's intrinsic geometry and how it is bent within its ambient 4-manifold, what can be said about the ambient manifold's intrinsic curvature?  

\ni 3) Intrinsic to extrinsic.  
Given the intrinsic geometry of both a 3-surface and of a 4-manifold, is there a bending by which this 3-surface fits within the 4-manifold as a hypersurface?

\mbox{ } 

\ni I.e. 1) constructs the geometry of a hypersurface within a given manifold. 
2) constructs the manifold locally surrounding the hypersurface. 
3) determines whether a given surface can be realized as a hypersurface within an also-given manifold.

\subsection{A further type of diffeomorphism: Diff($\FrM$, Fol)} 

These apply to foliated spacetimes, and are taken to involve {\sl all possible} foliations Fol for a given $\FrM$.
It is important from Sec \ref{Cl-PoT} onward to be aware that these does {\sl not} share some of the simpler mathematical similarities common to 
Diff($\FrM$) and Diff($\bupSigma$) [see  Sec \ref{CC-Intro} and those Secs].

\section{Dynamical formulation of GR}\label{Gdyn}

\subsection{Space-time split of the GR action}

Under the ADM split, the Einstein--Hilbert action takes the form 
\beq
\FS^{\sG\sR}_{\sA\sD\sM} = \int \d\mt \int_{\Sigma}\d^3x \, \mL^{\sG\sR}_{\sA\sD\sM} = \int \d\mt \int_{\Sigma}\d^3x\sqrt{\mh} \, \upalpha \{\mK_{ab}\mK^{ab} - \mK^2   +  \mR\} \mbox{ } .
\label{S-ADM}  
\eeq
This is obtained by {\sl decomposing} $\mR^{(4)}$ using a combination of contractions of the Gauss and Ricci equations 
and discarding a total divergence since $\bupSigma$ is without boundary.

\ni The result of varying with respect to this action can be recognized as 3 projections of the spacetime Einstein tensor.
[I.e. various combinations of contractions of the above Gauss, Codazzi and Ricci equations viewed as projection equations.]
These can also be obtained by projecting the Einstein field equations themselves.

However many applications require adherence to the Principles of Dynamics in laying these out.    
Begin by considering the manifestly-Lagrangian form of the action (i.e. in terms of configurations and velocities, which are here $\mh_{ij}$ and $\dot{\mh}_{ij}$).

\subsection{Configuration space as a starting point for dynamics}\label{Q-Primary}

In such dynamical approaches, one starts with the notion of configuration space $\FrQ$ of the configurations $Q^{\sfA}$.  
I.e. configurations are the generalization of the `position' half of `positions and momenta' in the case of basic point-particle mechanics. 
One is then to build composite objects out of the $Q^{\sfA}$.
These include the velocities $\dot{Q}^{\sfA}$ or the {\it changes} of configuration $\md Q^{\sfA}$, actions, and the momenta $P_{\sfA}$ corresponding to the $Q^{\sfA}$.

`Configurational minimalism alternatives' \cite{FileR, JBook} then entertain the further possibility of the configurations being the primary entities 
(above and beyond the momenta or combinations of what are physically coordinates and momenta).   
I.e. here the configuration space $\FrQ$ has a primary role in the conceptualization of Physics.  
There are various strengths of configurational minimalism.
One can consider just $Q^{\sfA}$ but also consider these alongside a notion of velocity $\dot{Q}^{\sfA}$ or of change $\d Q^{\sfA}$, or extended paths in configuration space.
We shall see later that upgrading the inclusion of change to extended paths (Fig \ref{Web-Cl}) or histories (Sec \ref{PH-Intro}), or the commonplace primary-level 
inclusion of momenta $P_{\sfA}$, furnish 3 further alternatives with more structure that one can adopt at this stage.

Some simple examples of configuration spaces are as follows.  
These also introduce some of the current Article's model arenas.

\mbox{ } 

\ni As a first example, for {\it scaled relational particle mechanics (RPM)} \cite{BB82, FileR, QuadI} is a mechanics in which only relative times, relative angles and relative separations are 
meaningful. 
It is also known as Euclidean RPM or Barbour--Bertotti (1982) theory \cite{BB82}; see also the reviews \cite{Buckets, FileR}.
Then the configuration space is $\FrQ(N, d) = \mathbb{R}^{Nd}$ of $N$ particles (Fig \ref{Spazenda}.a)).
This has an obvious Euclidean metric on configuration space (i.e. the $\mathbb{R}^{Nd}$ one rather than the spatial $\mathbb{R}^d$ one).
%
{            \begin{figure}[ht]
\centering
\includegraphics[width=0.9\textwidth]{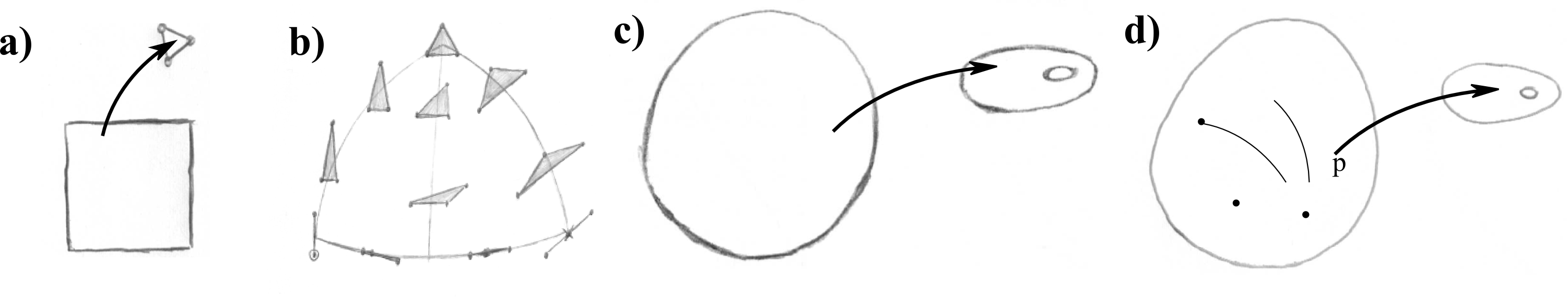}
\caption[Text der im Bilderverzeichnis auftaucht]{\footnotesize{a) The redundant $\mathbb{R}^6$ or $\mathbb{R}^4$ 
(translations trivially quotiented out) presentation of the space of triangles in the plane.
b) Kendall's $\mathbb{S}^2/\mathbb{Z}_2 \times \mathbb{Z}_3$ `spherical blackboard' presentation \cite{Kendall} for the shape space of triangles formed by 3 unlabelled points. 
The corresponding scaled RPM configuration space is the `cone' of this (product with $\mathbb{R}_+$ alongside a special point at the origin).  
c) Riem($\mathbb{T}^3$), each point $\mp$ of which represents a smooth spatial metric $\mh_{ij}$ on the 3-torus topological manifold, $\mathbb{T}^3$. 
d) A much more heuristic sketch of Superspace($\mathbb{T}^3$), 
each point $\mp$ of which represents a spatial geometry (equivalence class of spatial metrics under spatial diffeomorphisms) on $\mathbb{T}^3$.  
The bold points and lines here schematically indicate stratification, by which different parts of Superspace can be of different dimensionality \cite{Fischer70}.
N.B. that 3-$d$ RPM configurations and gauge theory configuration spaces (orbit spaces) also exhibit stratification, so this is a commonly encountered complication.} }
\label{Spazenda} \end{figure}          }

\subsection{The GR action endows Riem with a metric geometry}\label{DeWitt-supermetric}

As a second example of configuration space, I return to action (\ref{S-ADM}) and recast it in terms of the configuration space geometry for GR.
Its kinetic term contains $\sqrt{\mh}\{\mh^{ac}\mh^{bd} - \mh^{ab}\mh^{cd}\}$ contracted into $\mK_{ab}\mK_{cd}$ and thus into $\dot{\mh}_{ab}\dot{\mh}_{cd}$.  
This is a metric $\mM^{abcd}$ on the configuration space Riem($\bupSigma$) of all $\bh$'s on that particular fixed topological manifold $\bupSigma$.\foo{If one does not 
wish to presuppose spacetime, I use instead the notation Riem($\bigupsigma$).  
The latter occurs in investigation of geometrodynamical theories in general,
rather than in treatment of specifically the geometrodynamics that is obtained by splitting GR spacetime and GR's Einstein field equations.}
%
Fig \ref{Spazenda}.c) portrays this as an example of a {\it space of spaces}.

Since $\mM^{abcd}$ possesses four indices, it is termed a {\sl supermetric}.
Moreover DeWitt's \cite{DeWitt67} 2-index to 1-index map $\mh_{ij} \mapsto \mh^A$ casts it in the standard form of two downstairs indices.
Then $\mM^{ijkl} \mapsto \mM_{AB}$ and then $\mT_{\sG\sR}$ takes the form $\mM_{AB}\delta_{\vec{\upbeta}}{\mh}^A\delta_{\vec{\upbeta}}{\mh}^B/4$.  
[The capital Latin indices in this context run from 1 to 6.]
Pointwise, this is a --+++++ metric,
and so, overall it is an infinite-dimensional version of a semi-Riemannian metric.  
I.e. the GR configuration space metric alias inverse DeWitt supermetric.  
This indefiniteness is associated with the expansion of the universe giving a negative contribution to the GR kinetic energy.  
This is entirely unrelated to the indefiniteness of SR and GR spacetimes themselves.  
Its inverse is 
\be
\mN^{AB} = \mN_{abcd} = \{\mh_{ac}\mh_{bd} - \mh_{ab}\mh_{cd}/2\}/\sqrt{\mh}  \mbox{ } , 
\ee
which is the DeWitt supermetric itself.
DeWitt additionally studied the more detailed nature of this geometry in \cite{DeWitt67}.

Thus, all in all, using an explicitly indefinite inner product notation, the geometrical DeWitt presentation of the manifestly Lagrangian form of the ADM action works out to be
\beq
\FS^{\sG\sR}_{\sA\sD\sM-\sL\sD} = \int\d\mt\int_{\sbSigma}\d^3x\sqrt{\mh}\,\upalpha 
\left\{
\mT^{\sG\sR}_{\sA\sD\sM-\sL\sD}/\upalpha^2 +  \mR(\ux; \bh]   - 2\Lambda
\right\} \mbox{ } , \mbox{ } \mbox{ for }
\mT^{\sG\sR}_{\sA\sD\sM-\sL\sD} = ||\dot{\bh}     - \pounds_{\suupbeta}\bh||^2_{\mbox{\scriptsize ${\bM}$}}/4 \mbox{ } .  
\label{ADM-L}
\eeq

\subsection{Minisuperspace}

As a third example, a simpler subcase of Example 2) is {\it minisuperspace} ($\bupSigma$) \cite{Magic}: the space of homogeneous positive-definite 3-metrics on $\bupSigma$. 
These are notions of space in which every point is the same.  
Here full GR's ${\mM}^{ijkl}(\mh_{mn}(x^i))$ has collapsed to an ordinary $6 \times 6$ matrix, ${M}_{\sfA\sfB}(h_{ab})$: 
an overall (rather than per space point) curved (--+++++) `minisupermetric'.  
Nested simpler subcases within this are diagonal minisuperspace with $3 \times 3$ matrix $M_{\sfA\sfB}$ \cite{Magic}, which is a flat (--++) minisupermetric, 
and isotropic minisuperspace [flat (--) minisupermetric], for instance for $\bupSigma = \mathbb{S}^3$ with standard hyperspherical metric.  
[Thus it is a closed cosmological model, and anisotropy is being ignored.]

The specific minisuperspace model used in the current Article's detailed examples is a spatially closed model on Machian grounds 
(to avoid undue influence of boundary or asymptotic physics -- a criterion also argued for by Einstein \cite{Ein21}).
Then the choice of $\bupSigma = \mathbb{S}^3$ is then simplest, and the most conventional for closed-universe cosmologies. 
This is here to contain fundamental rather than phenomenological matter, due to having quantization in mind \cite{BI75, HH83}.
One needs at least 2 degrees of freedom, and cosmology conventionally makes use of scalar fields.  
The simplest case is 1 minimally-coupled scalar field.  
It is not hard to extend the current program's treatment to extend this to n scalar fields (a source of further cosmological modelling versatility).  
A cosmological constant term is needed to support the spatially-$\mathbb{S}^3$ FLRW cosmology with scalar field matter in the case with matter effects are presumed small \cite{Rindler}.
The configuration space metric's specific form for this Minisuperspace is, up to a conformal factor of $a^3$, just 2-$d$ Minkowski space $\mathbb{M}^2$ equipped with its standard 
indefinite flat metric.

\subsection{The geometrodynamical momenta}

Now as well as extrinsic curvature being an important characterizer of hypersurfaces, it is relevant due to its bearing close relation to the GR momenta, 
\beq
\mp^{ij} := {\delta \mL^{\sG\sR}_{\sA\sD\sM}}/{\delta \dot{\mh}_{ij}} = - \sqrt{\mh}\{ \mK^{ij} - \mK \mh^{ij} \} = \mM^{ijkl}\delta_{\vec{\upbeta}}\mh_{ij}/2\upalpha
\label{Gdyn-momenta}
\eeq
I.e. the gravitational momenta are a densitized version of $\mK_{ab}$ with a particular trace term subtracted off.  
Also note that, taking the trace, 
\beq
\mp/\sqrt{h} = - 2\mK \mbox{ } . 
\label{p-K}
\eeq

\subsection{The GR constraints}\label{GR-cons-and-sp-diff}

The ADM-Lagrangian action then encodes the GR Hamiltonian constraint 
\beq
\scH := \mN_{ijkl}\mp^{ij}\mp^{kl} - \sqrt{\mh}\{  \mR(\ux; \bh] - 2\Lambda\} = 0
\label{Hamm}
\eeq
from variation with respect to the lapse $\upalpha$.
From variation with respect to $\upbeta^{i}$, it also encodes the momentum constraint 
\beq
\scM_{i} := - 2\mD_{j}{\mp^{j}}_{i} = 0  \mbox{ } .  
\label{Momm}
\eeq
\ni The GR momentum constraint can be straightforwardly interpreted as physicality residing in the choice of coordinatization or point-identification but rather solely in terms of 
the remaining information of the 3-geometry of space that is also contained in the 3-metric.  
This is how GR is, more closely, a dynamics of 3-geometries in this sense (a geometrodynamics \cite{Battelle, DeWitt67} on the quotient configuration space (Fig \ref{Spazenda}.d)
\beq
\mbox{Superspace($\bupSigma$) := Riem($\bupSigma$)/Diff($\bupSigma$)} \mbox{ } . 
\eeq
\mbox{ } \mbox{ } However, interpreting the GR Hamiltonian constraint is tougher. 
It has a `purely-quadratic in the momenta' form, meaning it is a quadratic form plus a zero-order piece but with no linear piece.
We shall see in Sec \ref{Facets-1} that this property leads to the Frozen Formalism Facet of the PoT.

Note that in terms of $\mK_{\mu\nu}$ (and setting $\Lambda = 0$), the constraints are 
\beq
\mbox{Hamiltonian constraint, } \scH := 2{\mG}^{(4)}_{\perp\perp} = \mK^2 - \mK_{ij} \mK^{ij}  + \mR(\ux; \bh]   = 0  \mbox{ } ,   
\eeq
\be
\mbox{momentum constraint, } \scM_{i} :=  2{\mG}^{(4)}_{i\perp} = 2\{\mD_{j}{\mK^{j}}_{i} - \mD_{i}\mK\} = 0 \mbox{ } .
\label{Mom}
\eeq
These forms of GR equations were already known to Darmois in the 1920's \cite{Darmois}. 
As indicated, the $\mK_{ij}$ forms of these constraints serve to identify \cite{Wald} these as 
contractions of the Gauss--Codazzi equations for the embedding of spatial 3-slice into spacetime ({\it Constraint--Embedding Theorem of GR}).

Geometrodynamics is usually described by evolving spatial 3-metrics which, unlike the description in terms of 3-geometries, 
have Diff($\bupSigma$) redundancy but also the benefit of explicit computibility.

Finally note that the degrees of freedom count works out as $6 \times 2 - 3 \times 2 - 1 \times 2 = 2 \times 2$.

\subsection{GR Evolution equations}\label{Ev-Eq}

Sec \ref{T-GR} already laid down some topological restrictions, e.g. orientability 
and Sec \ref{Top-Sigma} considered simple-product spacetimes.
One may require further restrictions on the spacetime to ensure good causal behaviour.  
We assume $\bupSigma \times \biguptau$ preventing consideration of topology change. 
The local in space version builds a surface within the domain of dependence of the initial $\mS$.  
That is still a direct product at the level of topological manifolds, at least in the present Article's context. 

\ni The hypersurfaces are held to be everywhere spacelike. 
Going into such a split entails orientability and no closed timelike curves -- the conditions for Cauchy surfaces to exist. 

\mbox{ } 

\ni In terms of gravitational momenta, the ($\Lambda = 0$) evolution equations are
\be
\{\delta_{\vec{\upbeta}}\mp^{ij} = \sqrt{\mh} \{ \mR \, \mh^{ij}/2  -  \, \mR^{ij} + \mD^j\mD^i - \mh^{ij} \mD^2 \}\upalpha - 2 \, 
                                   \upalpha   \{ \mp^{ic}{\mp_c}^j - \mp \, \mp^{ij}/2 \}/\sqrt{\mh} + \upalpha \mh^{ij}\{\mp_{ij}\mp^{ij} - \mp^2/2\}2\sqrt{\mh}   \mbox{ } .  
\label{BFO-Evol}
\ee
In terms of the extrinsic curvature, 
\be
\left\{
- \{\delta_{\vec{\upbeta}} \mK_{ab} - \mh_{ab}\delta_{\vec{\upbeta}}\mK\} - \mD_b\mD_a\upalpha + \mh_{ab}\mD^2\upalpha
\right\}/{\upalpha} - 
\left\{
2{\mK_a}^c\mK_{bc} - \mK\mK_{ab} + \{\mK_{ij} \mK^{ij} + \mK^2\}\mh_{ab}/2
\right\}
+ \mG_{ab} = \mG^{(4)}_{ab} = 0 \mbox{ } . 
\label{Gdyn-Evol}
\ee
which form completes the constraint equations as regards forming the remaining projection of ${\mG}^{(4)}_{\mu\nu}$,
The three of them can also be interpreted in terms of contractions of the Gauss--Codazzi--Ricci embedding equations 
(thus extending the Constraint--Embedding Theorem to the {\it Constraint--Evolution--Embedding Theorem of GR}).  

\mbox{ }

\ni Also note the success in deriving these equations as regards removing all Riemann tensor (and thus Weyl tensor) projections from the system of projection equations.  
However, some other formulations -- e.g. the threading formulation -- use other linear combinations which do cause some such terms to be kept.  

\vspace{10in}

\section{Classical-level Background Independence and the Problem of Time. I. Time and Configuration}\label{Cl-PoT}

\cite{BI} and the current Article then argue furthermore that GR succeeds in meeting criteria of Background Independence as well as of a relativistic theory of gravitation. 
See also \cite{Dirac, Kuchar92, I93, PrimaFacie, I95, Rov97, BI00b, Carlip01, Thie03, Smolin0304, AL04, KieferBook, BI}.
In this way, it is a {\sl gestalt theory}.
Furthermore, adopting the gestalt position by which GR has relational foundations has consequences as regards subsequent conceptualization of `Quantum Gravity'. 
Note that the wording `Quantum Gravity' itself is a misnomer from the perspective of GR being a gestalt entity, since it refers only to the first aspect of GR.
In some approaches, these words do indeed reflect what is attempted. 
However, a number of other approaches (in particular Loop Quantum Gravity and M-Theory) do in fact consider GR as a gestalt entity. 
I choose to make this clear by terming background-independent programs, as not just `Quantum Gravity', but rather `Quantum Gestalt'; my acronym QG below refers to the latter. 
Finally, this highlights the complementary possibility (see below) of studying quantum Background Independence in the absense of 
any theory of gravitation that is compatible with relativity.\footnote{As to other names, `Quantum GR' will not do in this role due to implying the specific Einstein field equations.
On the other hand, the Quantum Gestalt position remains open-minded as to {\sl which} relativistic theory of gravitation is involved.  
Quantum Gestalt is also in contradistinction to `Background Independent Quantum Gravity' since the latter may carry connotations of Background 
Independence and gravitational inputs being {\sl separate} rather than part of a coherent whole whose classical counterpart, GR, is already also a coherent whole.} 
%
In this approach, adopting Background Independence entails the notorious PoT as a direct consequence to be faced.  
This is as opposed to simply beginning from a position of denying (parts of) Background Independence/Relationalism so as to avoid (parts of) the PoT from occurring in one's scheme.
This second kind of approach tends to stay within more standard conceptualizations (usually SR's and QM's), where more standard and tractable types of calculations can be considered, 
as opposed to more even-handedly combining nontrivial subsets of QM concepts and of GR-as-gestalt concepts.

\subsection{A first dichotomy: constraints are all versus constraint providers}\label{Prov}

Having noted that geometrodynamics is a constrained formulation, one can ask for underlying reasons why the theory of nature might have such constraints. 
This leads to the notion of {\it constraint providers}; 
a well-known example of such is that if one takes a Lagrangian with particular symmetries, then Gauge Theory -- with its gauge constraints -- ensues. 
Sec \ref{CR-Intro} is a variant of this.
However, we first consider another kind of constraint provider (Sec \ref{TR-Intro}).

The alternative attitude to constraints is that one is prescribed a set of such ab initio. 
Indeed, Applied Mathematicians have a theory of constrained systems in general (see \cite{Souriau, Goldstein, Arnol'd, Sni} and Sec \ref{Cl-PoT-2} 
which does not require asking where the constraints came from.
The questions of where the constraints came from and what they represent physically are, however, natural to question-asking, foundational theoretical physicists such as Wheeler. 
Indeed, Wheeler \cite{Battelle} additionally asked the following question, which readily translates to asking for first-principles reasons for the form of the 
crucially important GR Hamiltonian constraint, $\scH$.
\beq
\stackrel{\mbox{\it \normalsize ``If one did not know the Einstein--Hamilton--Jacobi equation, how might one hope to derive it straight off from}} 
         {\mbox{\it \normalsize plausible first principles without ever going through the formulation of the Einstein field equations themselves?"}} 
\label{Wheeler-Q}
\eeq
Here, one is no longer thinking just of GR's specific geometrodynamics, but rather of a multiplicity of geometrodynamical theories.  
One is then to look for a {\sl selection principle} that picks out the GR case of geometrodynamics.

Finally, the reverse process to `provide' is `encode'.  
I.e. to subsequently build into one's action auxiliary variables variation with respect to which encodes the constraints; further reasons for encoding become apparent in 
Sec \ref{Cl-PoT-2}.

\subsection{Temporal Relationalism}\label{TR-Intro}

The {\it Temporal Relationalism Postulate} \cite{BB82, FileR} is Sec \ref{AORM}, `Leibniz's Time Principle'.
It is useful via admitting a mathematically sharp implementation (and with the below useful consequences).  
At the classical level, this implementation the following selection principles for the Principles of Dynamics action.

\mbox{ } 

\ni i) This action is not to include any extraneous times (such as $t^{\sN\se\sw\st\so\sn}$) or extraneous time-like variables (such as the ADM lapse of GR, $\upalpha$).

\ni ii) Time is not to be smuggled into the action in the guise of a label either.

\mbox{ } 

\ni One formulation of ii) is for a label to be present but physically meaningless because it can be changed for any other (monotonically related) label 
without changing the physical content of the theory.   
I.e. the action in question is to be {\it manifestly reparametrization invariant}. 
This is the formulation we stick to in the current Article.

For later reference, it is useful to envisage ${\d}/{\d\lambda}$ as the Lie derivative $\pounds_{{\d}/{\d\lambda}}$ in a particular frame \cite{Stewart} -- c.f. (\ref{pa-as-Lie}).

\mbox{ } 

\ni I first give the example of temporally-relational but spatially-absolute mechanics.     
A reparametrization-invariant action for this is\footnote{Here 
indices $I, J$ run over $1$ to $N$ (particle number), indices $i$ run over 1 to $d$ (spatial dimension) and $m_I$ are the particles' masses.}
\beq
\FS_{\sJ} := \int \d\lambda \, L_{\sJ} := 2 \int \d\lambda \sqrt{T W} \mbox{ } .  
\label{S-Jacobi}
\eeq
I.e. {\it Jacobi's action principle} \cite{Lanczos} (`J' stands for Jacobi); 
see e.g. \cite{FileR} for how this is indeed physically equivalent to the more commonly used Euler--Lagrange action principle.
Therein, the kinetic energy $T := ||\dot{\mbox{\boldmath{$q$}}}||_{\mbox{\scriptsize\boldmath{$m$}}}\mbox{}^2/2$ where the configuration space metric $\mbox{\boldmath{$m$}}$ is just the 
`mass matrix' with components $m_{I}\delta_{IJ}\delta_{ij}$, and with {\it potential factor} $W := E - V(\mbox{\boldmath{$q$}})$ for $V(\mbox{\boldmath{$q$}})$ the potential energy and 
$E$ is the total energy of the model universe.

The conjugate momenta are then $\mbox{\boldmath{$p$}} := \pa L_{\sJ}/\pa \dot{\mbox{\boldmath{$q$}}} =  \sqrt{W/T} \, \dot{\mbox{\boldmath{$q$}}}$. 
The main way in which actions implementing ii) work is that they necessarily imply {\it primary constraints} \cite{Dirac},  
i.e. relations between the momenta that are obtained without use of the equations of motion.
This is because, as Dirac argued \cite{Dirac}, an action that is reparametrization-invariant is homogeneous of degree 1 in the velocities.  
Thus the $k = Nd$ conjugate momenta are (by the above definition) homogeneous of degree 0 in the velocities. 
Thus they are functions of at most $k - 1$ ratios of the velocities. 
So there must be at least one relation between the momenta themselves (i.e. without any use made of the equations of motion).  
But this is the definition of a primary constraint.   
Hence Temporal Relationalism indeed acts as a constraint provider.

The next concern is what sort of constraints this provides. 
The purely quadratic form of the above mechanics action \cite{B94I} causes the constraint in question to be purely\footnote{This `purely' 
means in particular that there is no accompanying linear dependence on the momenta.  
At the level of conic sections and higher-dimensional quadratic forms, this means a `centred' choice of coordinates, so that the ellipse or whatever else is centred about the origin; 
see also Sec \ref{Strat-Intro}.} 
quadratic: 
\beq
\scE := ||\mbox{\boldmath{$p$}}||_{\mbox{\scriptsize\boldmath{$n$}}} + V(\mbox{\boldmath{$q$}}) = E \mbox{ } .  
\label{E-Constraint}
\eeq
Here $\mbox{\boldmath{$n$}} = \mbox{\boldmath{$m$}}^{-1}$, with components $\delta_{IJ}\delta_{ij}/m_I$.  
\ref{E-Constraint}) usually occurs in Physics under the name and guise of an {\it energy constraint}. 
However, in the relational formulation, its function is as an {\it equation of time}, as per the argument three paragraphs down. 
Because of this, I denote such constraints by $\scC\scH\scR\scO\scN\scO\scS$.  
Finally, this approach's equations of motion are of the form $\sqrt{W/T} \, \dot{\mbox{\boldmath{$p$}}} = -\pa V / \pa \mbox{\boldmath{$q$}}$.

Is there a paradox between the Temporal Relationalism Postulate and our appearing to `experience time'?  
To begin to form an answer, `time' is a useful concept for everyday experience. 
However its nature is less clear, and also everyday experience is of subsystems rather than concerning the whole universe.

To answer further, recollect Mach's Time Principle (Sec \ref{AORM}): that `time is to be abstracted from change'. 
Thus timelessness for the universe as a whole at the primary level is {\sl resolved} by time emerging from change at the secondary level from Mach's Time Principle. 
This Machian position is particularly aligned with the second and third formulations of Temporal Relationalism ii), which are in terms of change rather than velocity.
More specifically, \cite{ARel2, ABook} argues furthermore that one is best served by adopting a conception of time along the lines of the astronomers' ephemeris time, 
from including a sufficient totality of locally relevant change.

A specific implementation of a Machian emergent time is then as follows. 
It is a time that is distinguished by its simplification of the model's momentum--velocity relations and equations of motion using 
$\pa/\pa t^{\se\sm(\sJ)} = \sqrt{W/T}\pa/\pa\lambda$.  
This can be integrated up to give
\beq
t^{\se\sm(\sJ)} = \int \d\lambda \sqrt{T/W}  \mbox{ } . 
\label{t-em-J}
\eeq
This equation is indeed the manifest `equation of time' rearrangement of $\scE$. 
Moreover, recasting this in Manifestly Parametrization Irrelevant form, this is an equation for obtaining time from change, i.e. in compliance with Mach's Time Principle.
I.e. one has arrived at a recovery of Newton's time on a temporally-relational footing.

Furthermore, in the GR counterpart of this working, it is the crucial Hamiltonian constraint ${\scH}$ that arises at this stage as a primary constraint.
We shall see that this case amounts to a recovery of GR proper time.  
To emphasize that Temporal Relationalism provides a centrally important constraint whose interpretation is as an equation of time, 
we name the general case of the constraint provided in this manner as $\scC\scH\scR\scO\scN\scO\scS$.
$\scE$ and $\scH$, as arrived at within the Relational Approach to Physics, are then both subcases of this.

\subsection{Configurational Relationalism}\label{CR-Intro}

This term covers both a) {\it Spatial Relationalism} \cite{BB82}: no absolute space properties.
b) {\it Internal Relationalism} is the post-Machian addition of not ascribing any absolute properties to any additional internal space that is associated with the matter fields.
Configurational Relationalism is addressed as follows.

\mbox{ } 

\ni i) One is to include no extraneous configurational structures either (spatial or internal-spatial metric geometry variables that are fixed-background rather than dynamical).

\ni[Since time-parametrization is really a 1-$d$ metric of time, the i) condition of both Temporal and Configurational Relationalism reflect a single underlying relational 
conception of Physics: no fixed-background metric-level geometry.] 
 
\ni ii) Physics in general involves not only a $\FrQ$ but also a $\FrG$ of transformations acting upon $\FrQ$ that are taken to be physically redundant.

\mbox{ } 

\ni This is a matter of practical convenience: often $\FrQ$ with redundancies is simpler to envisage and calculate with.
For ii) the Internal Relationalism case is another formulation of Gauge Theory from that presented in books on QFT. 
The spatial case is similar, in that it can also be thought of as a type of Gauge Theory for space itself.
This includes modelling translations and rotations relative to absolute space as redundant in Mechanics, for which $\FrQ = \mathbb{R}^{Nd}$,   
                                                         or Diff($\bupSigma$) as redundant in GR,        for which $\FrQ = \mbox{Riem}(\bupSigma)$.
In accord with Sec \ref{DLS}, these are actively interpreted.
See \cite{ABook} for restrictions on $\FrQ$, $\FrG$ pairings.  

\mbox{ }

\ni{\bf Best Matching implementation of Configurational Relationalism} 

\mbox{ } 

\ni Implementing Configurational Relationalism at the level of Lagrangian variables ($\mbox{\boldmath $Q$}$, $\dot{\mbox{\boldmath $Q$}}$) is known as {\it Best Matching} \cite{BB82}.
This involves pairing $\FrQ$ with a $\FrG$ such that $\FrQ$ is a space of redundantly-modelled configurations.
Here $\FrG$ acts on $\FrQ$ as a shuffling group: one considers pairs of configurations, keeping one fixed and shuffling the other 
(i.e. an active transformation) until the two are brought into maximum congruence.

In more detail, one proceeds via constructing a $\FrG$-corrected action.
For the examples considered here, this involves replacing each occurrence of $\dot{\mbox{\boldmath $Q$}}$ with 
$\dot{\mbox{\boldmath $Q$}} - \stackrel{\rightarrow}{\FrG_g}\mbox{\boldmath $Q$}$, where $\stackrel{\rightarrow}{\FrG_g}$ indicates group action.
Note that whereas these correction terms can be interpreted as fibre bundle connections \cite{Mercati14}, they can also be interpreted as Lie derivatives (see e.g. \cite{Stewart, FileR}).
The latter is more minimalistic since it is at the level of differential geometry without having to assume connection or bundle structures.

Then varying with respect to the $\FrG$ auxiliary variables $g^{\sfG}$ produces the {\it shuffle constraints} $\scS\scH\scU\scF\scF\scL\scE_{\sfG}$. 
These arise as {\sl secondary constraints}: via use of equations of motion. 
Being linear in the momenta, their form is denoted by $\scL\scI\scN_{\sfG}$ (a subcase of such linear constraints since not all $\scL\scI\scN_{\sfL}$ arise from shuffling).

The {\sl intent} of Best Matching is that $\FrG$ acts as a gauge group.
(`Gauge' is here meant in the sense of Dirac: at the level of constant-time data, see \cite{ABeables} for contrast with other notions of gauge.)
However, {\it demonstrating} the realization of this intent requires having obtained the brackets between the constraints [see Aspect 3) below].
Thus we need a {\sl candidate} name for the constraints provided by the Best Matching shuffle, and it is $\scS\scH\scU\scF\scF\scL\scE_{\sfG}$ that I deploy in this role.

The initial introduction of $\FrG$ corrections appears to be a step in the wrong direction, due to its extending the already redundant space $\FrQ$ of the $\mbox{\boldmath $Q$}$ 
to some joint space of $\mbox{\boldmath $Q$}$ and $g^{\sfG}$.  
However, if the above attempt is successful upon checking Aspect 3), the end-product of the next step $\scS\scH\scU\scF\scF\scL\scE_{\sfG}$ are of the form $\scG\scA\scU\scG\scE_{\sfG}$, 
which are a type of constraint that, as the next Sec explains, uses up {\sl two} degrees of freedom per $\FrG$ degree of freedom.  
In this case, each degree of freedom appended wipes out not only itself but also one of $\FrQ$'s redundancies. 
By this, one indeed ends up on the configuration space that is free of these redundancies: 
the quotient space $\FrQ/\FrG$, as is required to successfully implement Configurational Relationalism.

In the Best Matching procedure, one furthermore takes $\scS\scH\scU\scF\scF\scL\scE_{\sfG}$ as equations in the Lagrangian variables to solve for the $g^{\sfG}$ auxiliaries themselves.
One then substitutes the extremizing solution back into the original action to obtain a reduced action on the $\FrQ/\FrG$ configuration space.  

\ni Let us further clarify this with the example of scaled RPM.  
These are taken to be fundamental rather than effective mechanics problems for which potentials of the form 
$V(\mbox{\boldmath{$q$}}) = V(\underline{q}_I \cdot \underline{q}_J  \mbox{ alone})$ apply.
This form then guarantees that auxiliary translation and rotation corrections applied to this part of the action straightforwardly cancel each other out within.  
[Because of this, there is no need in this example to correct the $\mbox{\boldmath{$q$}}$ themselves.]  
The situation with the kinetic term is more complicated because $\d/\d\lambda$ is not a tensorial operation under $\lambda$-dependent translations and rotations. 
This leads to the translation and rotation corrected kinetic term 
\ni
\be
T^{\sR\sP\sM} = ||\Circ_{\underline{A}, \underline{B}}\mbox{\boldmath$q$}||_{\mbox{\scriptsize\boldmath$m$}}\mbox{}^2/2 
\mbox{ } \mbox{ for } \mbox{ } \Circ_{\underline{A}, \underline{B}}\mbox{\boldmath$q$} := \dot{\mbox{\boldmath$q$}} - \underline{A} - {\underline{B}} \cr \mbox{\boldmath$q$} \mbox{ } .    
\label{T}
\ee
The action is then
\be
\FS^{\sR\sP\sM} = 2\int \d\lambda \sqrt{W \, T^{\sR\sP\sM}}  \mbox{ } .
\label{Jac} 
\ee

\ni Then the momenta conjugate to the $\mbox{\boldmath{$q$}}$ are 
$
\mbox{\boldmath{$p$}} = \sqrt{W/T^{\sR\sP\sM}} \, \{\dot{\mbox{\boldmath{$q$}}} - \underline{B} \cr \mbox{\boldmath{$q$}}\}.    
$
Then by virtue of the Manifest Reparametrization Invariance and the particular square-root form of the Lagrangian, the momenta obey a primary constraint quadratic in the momenta,  
\be
\scE := ||\mbox{\boldmath $p$}||_{\mbox{\scriptsize\boldmath $n$}} + V(\mbox{\boldmath $q$}) = E \mbox{ } . 
\label{calE}
\ee
Next, variation with respect to $\underline{A}$ and $\underline{B}$ give, respectively, the secondary   
\be
\underline{\scP} := \sum\mbox{}_{\mbox{}_{\mbox{\scriptsize I = 1}}}^{N} \underline{p}_{I} = 0 \mbox{ } \mbox{ } \mbox{ (zero total momentum constraint) }
\label{ZM} \mbox{ } , \mbox{ } \mbox{ and }
\ee
\be
\underline{\scL} := \sum\mbox{}_{\mbox{}_{\mbox{\scriptsize I = 1}}}^{N} \underline{q}^{I} \cr \underline{p}_{I} = 0 \mbox{ } \mbox{ } \mbox{ (zero total angular momentum constraint) }
\label{ZAM} \mbox{ } .
\ee
These constraints are linear in the momenta. 
The first can furthermore be interpreted as the centre of mass motion for the dynamics of the whole universe being irrelevant rather than physical.  
All the tangible physics is in the remaining relative vectors between particles.     
See Fig \ref{Spazenda}.b) an example of corresponding configuration spaces.
The evolution equations here take the form $\sqrt{W/T^{\sR\sP\sM}} \, \mbox{\boldmath{$p$}} = - \pa V/\pa \mbox{\boldmath{$q$}}$.

Returning to the Best Matching procedure, the constraints (\ref{ZM}, \ref{ZAM}), rewritten in `Lagrangian' configuration--velocity variables 
($\mbox{\boldmath{$q$}}, \dot{\mbox{\boldmath{$q$}}}$), are to be solved for the auxiliary variables $\underline{A}$, $\underline{B}$ themselves.
This solution is then to be substituted back into the action, so as to produce a final Tr and Rot-independent expression that {\sl directly} implements Configurational Relationalism.  
This action is to be elevated this new action to be one's primary starting point.
It turns out \cite{FileR, Kendall} that one can do this explicitly for many RPM's \cite{FileR}. 
%
{            \begin{figure}[ht]
\centering
\includegraphics[width=0.75\textwidth]{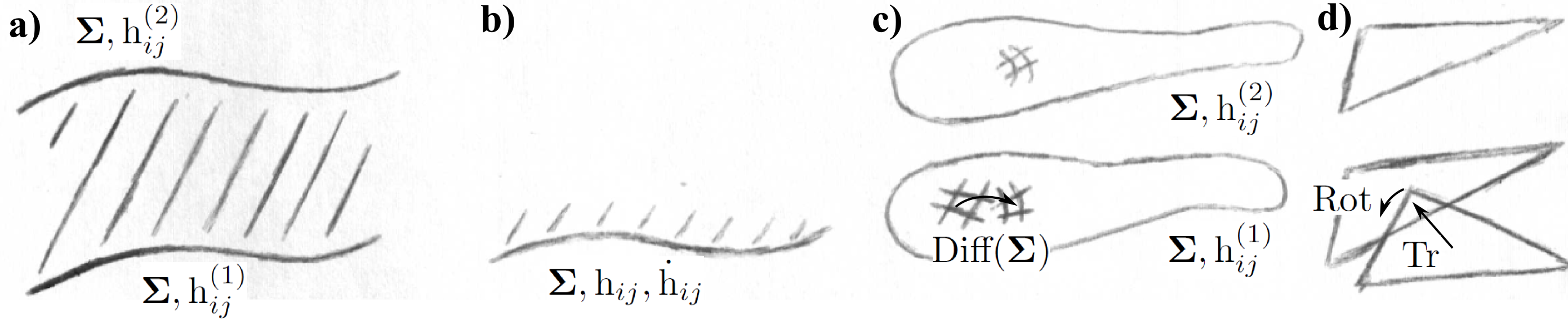}
\caption[Text der im Bilderverzeichnis auftaucht]{        \footnotesize{a) Thick sandwich bounding bread-slice data to solve for the spacetime `filling'. 
This failed to be well-posed.
b) It was succeeded by Wheeler's idea of Thin Sandwich data to solve for a local coating of spacetime \cite{WheelerGRT}.
This is the `thin' limit of taking the bounding `slices of bread': the hypersurfaces $\mh_{ij}^{(1)}$ and $\mh_{ij}^{(2)}$ as knowns.
This solved for the spacetime `filling' in between, in analogy with the QM set-up of transition amplitudes between states at two different times \cite{WheelerGRT}.
c) The Thin Sandwich can then be re-interpreted in terms of Best Matching Riem($\bupSigma$) with respect to Diff($\bupSigma$), which amounts to the shuffling as depicted.
I.e. keep one fixed  and shuffle the other to seek out how to minimize the incongruence between the two.  
d) The analogous triangleland Best Matching for relational triangles with respect to the rotations Rot and translations Tr.
This is included to indicate that Best Matching can be applied to a wide range of theories rather than just to geometrodynamics or the corresponding spatial diffeomorphisms.} }
\label{Facet-Intro-4} \end{figure}          }

\ni Consider next the geometrodynamical subcase of Best Matching. 
This is the so-called {\it Thin Sandwich}: Fig \ref{Facet-Intro-4}.b) and \cite{WheelerGRT, TSC1, TSC2, Kuchar92, I93}. 
In parallel with the above set-up for RPM, the Baierlein--Sharp--Wheeler (BSW) \cite{BSW} action is 
\be
\FS^{\sG\sR}_{\sB\sS\sW} = \int\d\lambda\int_{\Sigma}\sqrt{\overline{\mT}^{\sG\sR}_{\sB\sS\sW} \sqrt{h}\{\mR - 2\Lambda \}} \mbox{ } , 
\bar{\mT}^{\sG\sR}_{\textrm{BSW}} := \left|\left|\delta_{\stackrel{\rightarrow}{\upbeta}}\mbox{\boldmath$h$}\right|\right|_{\sbM}^{\mbox{ }\mbox{ } 2} \mbox{ } . 
\label{S-BSW}
\ee
[The overline denotes densitization, i.e. inclusion of a factor of $\sqrt{\mh}$.]
The primary constraint in this case is indeed the usual $\scH$.
For equivalence of the ADM and BSW actions, and more about the BSW action, see \cite{FileR}.  
This example indeed also gives rise to an emergent Machian time, though this Article only has scope for the Minisuperspace case to which we next turn.  
This is because of Temporal and Configurational Relationalism interfering with each other; see \cite{FEPI, FileR, AM13, TRiPoD} for resolution of this.

\subsection{Minisuperspace Model Arena example} \label{MSS-Intro}

Here Temporal Relationalism occurs but Configurational Relationalism does not (spatial homogeneity renders $\scM_i$ trivial due to its dependence on spatial derivatives).
Thus we can below provide and comment upon the formula for GR's Machian emergent time in the case of minisuperspace. 
On the other hand, RPM also exhibiting Configurational Relationalism renders it more useful as a model arena.
Further reasons for this are that RPM's also possess notions of structure and hence of structure formation, whereas minisuperspace does not.  
[In RPM's, linear constraints and inhomogeneities are logically-independent features. 
This is in contrast with how in minisuperspace both are simultaneously trivialized by homogeneity rendering the spatial derivative operator meaningless.]
On the other hand, minisuperspace, by being a restriction of GR, inherits some features that RPM's do not possess, 
e.g. kinetic indefiniteness and imposition of more specific restrictions on the form of the potential.

In treating minisuperspace, there are two different sets of modelling assumptions to consider. 
1) The matter physics is light and fast ($l$) as compared to the gravitational physics being heavy and slow ($h$). 
2) Both are $h$ and only further entities -- anisotropy or inhomogeneity -- are $l$.

The action (of Misner type \cite{Magic}) for the current Article's choice of model is (bars denote conformally transformed quantities)
\beq
\FS_{\sr\se\sll\sa\st\si\so\sn\sa\sll}^{\mbox{\scriptsize isotropic-MSS}} = \mbox{$\frac{1}{2}$}\int \d \lambda \, \sqrt{\overline{T} \, \overline{W}}    \mbox{ } , \mbox{ }  \mbox{ }  
        \overline{T} :=  \mbox{exp}(3\Omega)\left\{- \left\{\frac{\d\Omega}{\d \lambda}\right\}^2 + \left\{\frac{\d\phi}{\d \lambda}\right\}^2\right\}                                                                          \mbox{ } , \mbox{ }  \mbox{ } 
\overline{W} : = \mbox{exp}(3\Omega)\{\mbox{exp}(-2\Omega) - V(\phi) - 2\Lambda\}                                                                       \mbox{ } . 
\label{MSS-Action}
\eeq
Here, the {\it Misner variable} $\Omega := \mbox{ln} \, a$, for $a$ the usual cosmological scale factor. 
The corresponding Hamiltonian constraint is then 
\beq
\scH := \mbox{$\frac{\mbox{\scriptsize exp}(-3\Omega)}{2}$}\big\{-\pi_{\Omega}^2 + \pi_{\phi}^2 + \mbox{exp}(6\Omega)\{V(\phi) + 2\Lambda - \mbox{exp}(-2\Omega)\}\big\} = 0 \mbox{ } . 
\label{MSS-H}
\eeq 
Finally, the Machian classical emergent time for this model takes the form 
\beq
t^{\se\sm(\sJ)}_{\mbox{\scriptsize isotropic-MSS}} = \int  \sqrt{- \d\Omega^2 + \d\phi^2}\left/\sqrt{\mbox{exp}(-2\Omega) - V(\phi) - 2\Lambda} \right. \mbox{ } . 
\label{plain-tem}
\eeq

\vspace{10in}

\subsection{Temporal Relationalism leads to the Frozen Formalism Problem}\label{Facets-1}

The {\it Schr\"{o}dinger picture Quantum Frozen Formalism Problem} comes from elevating an equation -- of the form 
$\scQ\scU\scA\scD := N^{\sfA\sfB}(\mbox{\boldmath{$Q$}})P_{\sfA}P_{\sfB}/2 + V(\mbox{\boldmath{$Q$}}) = 0$ that encompasses GR's $\scH$ and RPM's $\scE$ -- to a quantum equation
\beq
\widehat{\scQ\scU\scA\scD}\Psi = 0 \mbox{ } . 
\label{E=0-TISE} 
\eeq
Here, $\Psi$ is the wavefunction of the universe. 
It is commonly viewed as the $E = 0$ case of a time-independent Schr\"{o}dinger equation 
\be
\widehat{H}\psi = E\psi \mbox{ } ,
\label{TISE}
\ee
-- i.e. a stationary alias timeless alias frozen wave equation -- occurring where one would expect a time-dependent equation such as a time-dependent Schr\"{o}dinger equation   
\be
\widehat{H}\psi = i\pa\psi/\pa t \mbox{ } .
\label{TDSE} 
\ee
A finite form for $\widehat{H}$ is $-\triangle + V$.
For GR, (\ref{E=0-TISE}) is the {\it Wheeler--DeWitt equation} \cite{Battelle, DeWitt67} $\widehat{\scH}\Psi = 0$, 
or in more detailed form\foo{Here `\mbox{ }' implies in general various well-definedness and operator-ordering issues.} 
\beq
\widehat{\scH}\Psi := - \frac{1}{\sqrt{\mM}}  \frac{\delta}{\delta \mh^{\mu\nu}}
\left\{
\sqrt{\mM}\mN^{\mu\nu\rho\sigma}  \frac{\delta\Psi}{\delta \mh^{\rho\sigma}}
\right\} 
\Psi + \sqrt{\mh}\{2\Lambda - \mR(\ux; \bh]\}\Psi + \hat\scH^{\mbox{\scriptsize matter}}\Psi = 0   \mbox{ } . 
\label{WDE} 
\eeq
Note that this (and for field theories more generally) contains in place of a partial derivative $\pa/\pa Q^{\sfA}$ a functional derivative $\delta/\delta h_{ij}(x^k)$ (Sec \ref{FFP-Intro}). 
$\scQ\scU\scA\scD$ itself can arise the ADM way from presupposing spacetime, or as an equation of time $\scC\scH\scR\scO\scN\scO\scS$ resulting from implementing Temporal Relationalism.
From the latter perspective, the Frozen Formalism Problem is already present at the classical level for the whole universe.

Also, in \cite{BI, AM13, ABook} the implementation of Temporal Relationalism is upgraded along the following lines.
It is a further conceptual advance to formulate one's action and subsequent equations without use of any meaningless label at all.  
I.e. a {\it manifestly parametrization irrelevant} formulation in terms of {\sl change}          $\d Q^{\sfA}$             rather than 
     a      manifestly reparametrization invariant          one in terms of a label-time velocity $\d Q^{\sfA}/\d \lambda$.  
But it is better still to formulate this without even mentioning any meaningless label/parameter, by use of how the preceding is dual to a configuration space geometry formulation. 
These upgrades lead to Temporal Relationalism being tractable at each level of structure rather than just at the level of Lagrangian (or Jacobian) formulations \cite{TRiPoD}.
In particular then, Temporal Relationalism can then be treated in tandem with other PoT facets \cite{TRiFol, ABook}.

\subsection{More general consideration of Configurational Relationalism}\label{Facets-2}

The Lagrangian variables form of the GR momentum constraint $\scM_i$ is the {\it thin sandwich equation} \cite{TSC1}.
Solving this is the {\it Thin Sandwich Problem}, which is another of Isham and Kucha\v{r}'s listed PoT facets.  
This particular facet always was presented as a manifestly classical problem.
That it is indeed a problem concerning time via, firstly, its connection with the local construction of a spacetime foliation.
Secondly, it is a pre-requisite for various PoT strategies, including the emergent time one above and the internal time one below.
This prerequisiteness stems from how the momentum constraint interferes with resolutions of the Frozen Formalism Problem, which, as an interference between facets, 
lies outside of the scope of this Paper.  
That it is indeed a major mathematical problem is clear from e.g. \cite{TSC2}.

Here one can consider $\dot{h}_{ij} - \pounds_{\beta}h_{ij}$ free from its hypersurface derivative conceptual packaging,  
and obtain an action (\ref{S-BSW}) viewed along such lines, e.g. by eliminating the lapse Lagrange multiplier from the ADM action \cite{BSW}.  
Moreover, one can pass yet further to $\pa{h}_{ij} - \pounds_{\pa\sF}h_{ij}$: 
in terms of a `frame variable' $\mF^i$ replacement for the shift $\upbeta^i$ that implements manifest parametrization irrelevance, 
and thus to an action in terms of a kinetic arc element $\pa s$ in place of $\d\lambda \sqrt{T}$. 
This is then the relational action for GR, 
and can be conceived of from Temporal and Configurational relational first principles rather than ever passing through ADM's formulation \cite{RWR, FileR}.

\mbox{ } 

\ni Moreover, the Thin Sandwich can be generalized by, sequentially,  

\ni a) Best Matching, since this applies to a wider range of theories than just geometrodynamics, for which foliation and sandwich names cease to be appropriate).

\ni b) Configurational Relationalism then applies to resolutions at levels other than that of the Lagrangian variables. 
E.g. this can also apply at the Hamiltonian level or at the level of solving the quantum equations. 
This most general implementation involves $\FrG$ acting on the object in question and then an operation over all of $\FrG$ cancelling this out.
I emphasize that not only is Best Matching a case of this whose latter operation is an extremization, 
but also that other cases of this more general procedure are known elsewhere in the literature, from Kendall's shape comparer \cite{Kendall} to 
                                                                                                 the Gromov--Hausdorff distance between metric spaces and 
																		    the even more well-known group averaging technique from Group Theory and Representation Theory. 
The latter operations for the last three are extremize, inf and average.
It is via this observation that Configurational Relationalism can be considered in the context of whatever other structures one's theorizing requires, 
in particular leading to its incorporation in the treatment of the other PoT facets \cite{FileR, BI, ACastle, ABook}.  

\vspace{10in}

\subsection{Other timefunction-based PoT strategies}\label{Strat-Intro}

This Subsection is included for comparison with other approaches.
Consider a time-dependent Schr\"{o}dinger equation (\ref{TDSE}) as being preceded by a {\sl parabolic} equation in the momenta \cite{Kuchar81b},
\beq
p_t = p_{\sfA} p^{\sfA} + C \mbox{ } .
\label{parab}
\eeq
Here $p_t$ is the momentum conjugate to whatever $t$ plays the role of time in this model (Newtonian time in the conventional case).
On the other hand, a time-independent Schr\"{o}dinger equation (\ref{TISE}) is preceded by an {\it elliptic} equation
\beq
p_{\sfA} p^{\sfA} + C = 0
\label{ellip}
\eeq
One problem is that the GR Hamiltonian constraint $\scH$ looks more like (\ref{ellip}) than (\ref{parab}). 
[The classical form
\beq
p_t = f(Q^{\sfA}, P_{\sfA}) \mbox{ } ,
\label{general-p}
\eeq
would produce an equation in which the role of time is similar to that arising from (\ref{parab}), with $f$ playing the role of Hamiltonian for the system.  
However, this can lead to major quantum-level difficulties -- ambiguities, implausibilities -- due to operator ordering and well-definedness issues.]   

\mbox{ }

\ni One might next argue that the specific form of the quantum $\scH$ (\ref{WDE}) looks even more like the Klein--Gordon equation's classical precursor \cite{Kuchar81} -- 
the {\it hyperbolic} equation
\beq
p_t\mbox{}^2 = p_{\sfa} p^{\sfa} + C \mbox{ } 
\label{hyperb}
\eeq
for $\fa$ taking one value less than $\fA$.
I.e. the indefiniteness of Riem($\bupSigma$) might furnish a time along the lines of Minkowski spacetime's indefiniteness doing so.
However, in the case of GR, this approach goes awry at the quantum level as per Sec \ref{Tfn-Strat}.

\ni Canonical transformations can map into and out of the general form (\ref{general-p}) \cite{Kuchar81}.
In this case, specialize the notation to
\beq
p_{t^{\th\ti\td\td\te\tn}} + H_{\st\sr\su\se}(Q^{\sfO}, P_{\sfO}, t^{\sh\si\sd\sd\se\sn}) = 0 \mbox{ } .
\label{hidden-true}
\eeq 
for $Q^{\sfO}$ the theory's other variables. 
(\ref{hidden-true}) corresponds to \cite{Kuchar92} `finding a hidden time $t^{\sh\si\sd\sd\se\sn}$ -- the conjugate of $p_{t^{\th\ti\td\td\te\tn}}$ -- 
alongside a `true Hamiltonian' $H_{\st\sr\su\se}$.

\mbox{ }

\ni Again more simply, one might solve one's original equation for a particular $p$ that is designated to be a $p_t$. 
However this approach has an added problem in justifying this choice. 
Common choices made are the scale variable or the matter field.  
The scale variable has the virtue of being singled out among all the other variables -- there is precisely one such variable. 
However, it is clearly not monotonic for the important case of recollapsing universes.
On the other hand, `the matter variable' may be unique in simple models, but there are obviously multiple matter variables in more general models, 
and it is then in general unclear how to pick `the time' amongst these.
Such ambiguities are problematic since, as we shall see in Sec \ref{PoT-MCP}, different choices of time are capable of leading to inequivalent quantum theories.  

\mbox{ } 

\ni Interestingly, the dilational momentum (conjugate to scale): the {\it York time} \cite{Yorktime1} $\propto \mp/\sqrt{\mh} \propto \mK$ by \ref{p-K}, 
so it is a  constant mean curvature (CMC) slice -- {\sl is} monotonic in sizeable regimes; this is as an example of (\ref{hidden-true}).
Another suggested way forward is introducing a matter field that specifically results in time -- so called {\it reference matter approaches}. 
Then (\ref{general-p}) is realized as
\beq
p_{t^{\tr\te\tf}} + H_{\st\sr\su\se}(Q^{\sfO}, P_{\sfO}, t^{\sr\se\sf}) = 0  
\label{appended-reference}
\eeq
for $Q^{\sfO}$ now the original theory's other variables {\sl and} the other appended variables.  
This differs from (\ref{hidden-true}) in that time is to be found among fields appended to one's theory rather than already hidden within.
\ni A particular case is {\it unimodular time}, which arises as the momentum conjugate to the cosmological constant upon elevating this to a dynamical variable.  

\mbox{ }

\ni I apply time and clock postulates as selection principles over the present Sec's wide range of suggestions.
By these and other criteria \cite{APoT2}, the previously introduced Machian emergent time `wins out' at the classical level.  
Moreover, not all approaches {\sl have a time}; Secs \ref{TNE-Intro} and \ref{PH-Intro} have further PoT strategies based, rather, on timelessness or on histories in place of time.  

\vspace{10in}  

\section{II. Brackets and Beables}\label{Cl-PoT-2}

Fig \ref{Input-3} considers a generalization of Sec \ref{Prov}'s dichotomy. 
%
{            \begin{figure}[ht]\centering\includegraphics[width=0.75\textwidth]{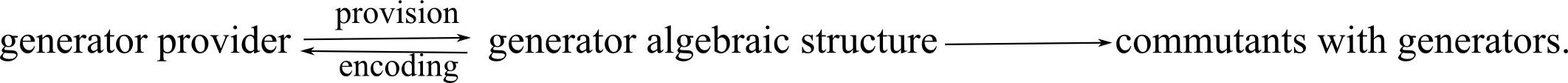}
\caption[Text der im Bilderverzeichnis auftaucht]{        \footnotesize{I phrase this in terms of generators, not constraints, to include the the next Sec's spacetime version too.  
As well as Sec \ref{Prov}'s providing and encoding, the further significant structures that the present Sec focuses 
on are the algebraic structures formed by the constraints themselves, and by those entities which commute with the constraints.}  } \label{Input-3}\end{figure}          }
 
\subsection{Aspect 3 of Background Independence: Constraint Closure}\label{CC-Intro}

Do constraints beget more constraints? 
If $\scC_{\sfA}$ vanishes on a given spatial hypersurface, what can be said about $\dot{\scC}_{\sfA}$?  
If $\dot{\cal C}_{\sfA}$ is equal to some $f({\cal C}_{\sfA})$ alone, it is said to be {\it weakly zero} in the sense of Dirac \cite{Dirac} (denoted by $\approx 0$). 
However, there is a lack of rigour in such a `Lagrangian' picture of `constraint propagation' (i.e. $\dot{\scC}_{\sfA}$ evaluated by use of Euler--Lagrange equations).

Let us next consider the joint space of $Q^{\sfA}$ and $P_{\sfA}$, 
as standardly equipped with the Poisson brackets structure $\mbox{\bf \{} \mbox{ } \mbox{\bf ,} \mbox{ } \mbox{\bf \}}$: {\it phase space}.

In this formulation, forming and handling brackets between constraints turns out to give a rigorous algorithm for handling whether constraints beget further independent constraints.
This is {\it Dirac's algorithm} \cite{Dirac, HTbook}, in which the brackets of known constraints can in general provide further constraints, specifier equations and inconsistencies.
This makes it clear that Constraint Closure is indeed a necessary check.
The end-product algebraic structure of constraints is, schematically, 
\beq
\mbox{\bf \{} {\scC}_{\sfF}\mbox{\bf ,} \,  {\scC}_{\sfF^{\prime}} \mbox{\bf \}} \approx 0 \mbox{ } .
\label{C-C}
\eeq
$\fF$ here indexes {\it first-class} constraints: those that close among themselves under Poisson brackets.
A constraint is {\it second-class} if it is not first-class; 
first-class constraints use up 2 degrees of freedom each to second-class's 1; gauge constraints are a subset of first-class constraints.
See e.g. \cite{HTbook} for means of removing second-class constraints.

The Dirac algorithm can then be applied to determine whether the constraints provided by Temporal and Configurational Relationalism 
-- $\scC\scH\scR\scO\scN\scO\scS$ and $\scS\scH\scU\scF\scF\scL\scE_{\sfI}$ respectively, which are in particular $\scH$ and $\scM_i$ in the case of GR -- are the whole picture. 
[They would not be if {\sl further} constraints or specifier equations arise upon forming the brackets among these provided constraints.]  
In particular, suppose the $\scS\scH\scU\scF\scF\scL\scE_{\sfI}$ close among themselves in the form 
\beq
\mbox{\bf \{} \scS\scH\scU\scF\scF\scL\scE_{\sfI} \mbox{\bf ,} \, \scS\scH\scU\scF\scF\scL\scE_{\sfI^{\prime}} \mbox{\bf \}} = 
                                       f_{\sfI\sfI^{\prime}}\mbox{}^{\sfI^{\prime\prime}} \scS\scH\scU\scF\scF\scL\scE_{\sfI^{\prime\prime}}                  \mbox{ } , 
\eeq
where $f_{\sfI\sfI^{\prime}}\mbox{}^{\sfI^{\prime\prime}}$ are constants, so this is a Lie algebra.
Then the $\scS\scH\scU\scF\scF\scL\scE_{\sfI}$ arising from the attempted rendering irrelevant of $\FrG$ is vindicated, 
insofar as it is playing the role of a gauge algebra $\scG\scA\scU\scG\scE_{\sfG}$ that realizes the physical irrelevance of $\FrG$.
If then $\mbox{\bf \{} \scG\scA\scU\scG\scE_{\sfG} \mbox{\bf ,} \, \scC\scH\scR\scO\scN\scO\scS \mbox{\bf \}}$ also closes, 
then $\scC\scH\scR\scO\scN\scO\scS$ is established as a good $\FrG$ object.
The final consideration is whether $\mbox{\bf \{} \scC\scH\scR\scO\scN\scO\scS \mbox{\bf ,} \, \scC\scH\scR\scO\scN\scO\scS \mbox{\bf \}}$ closes; 
e.g. if this were to produce a new linear constraint, it would be asserting that an enlarged $\FrG$ is a necessary consideration.
If the constraints $\scC\scH\scR\scO\scN\scO\scS$ and $\scS\scH\scU\scF\scF\scL\scE_{\sfI}$ do beget unexpected others, 
then one's attempted formulation has a Constraint Closure Problem: Facet 3) of the Problem of Time [see Comment 3) below].  
Moreover, at the classical level for the current Article's various toy models and full GR, this has the status of a {\sl solved} problem.

As a first example, consider Electromagnetism.  
Note that field theory constraint algebras are most usefully presented in terms of smearing functions.\footnote{In general, I use 
$({\scC}_{\tfW}| \sA^{\tfW}) := \int d^3z \, {\scC}_{\tfW}(z^i; \upphi_{\tfA}] \sA^{\tfW}(z^i)$ (an `inner product' notation) 
for the {\it smearing} of a $\fW$-tensor density valued constraint ${\cal C}_{\tfW}$ by an opposite-rank $\fW$-tensor smearing with no density weighting: $A^{\tfW}$. 
$\upphi_{\tfA}$ is here a full repertoire of the fields involved.
The undefined letters in this Sec's presentations of Electromagnetism, Yang--Mills Theory and GR are just such smearings.}  
%
Then Electromagnetism has the Abelian algebra of constraints $\mbox{\bf \{}  (\scG|\upxi)   \mbox{\bf ,} \,  (\scG|\upzeta)   \mbox{\bf \}} = 0$. 
Its Yang--Mills generalization has the Lie algebra of constraints $\mbox{\bf \{}  (\scG_I|\upxi^I) \mbox{\bf ,} \,  (\scG_J|\upxi^J \mbox{\bf \}} = f_{IJ}\mbox{}^K(\scG_K|\upxi^I\upzeta^J)$.  
As a first example with a $\scC\scH\scR\scO\scN\scO\scS$ as well, scaled RPM's constraint algebra's nonzero brackets are 
$\mbox{\bf \{}  \scL_i \mbox{\bf ,} \,  \scL_j \mbox{\bf \}}  = \epsilon_{ij}\mbox{}^k \scL_k$, 
$\mbox{\bf \{}  \scP_i \mbox{\bf ,} \,  \scL_j \mbox{\bf \}}  = \epsilon_{ij}\mbox{}^k \scP_k$. 
By the first of these the $\scL_i$ close as a Lie algebra, which is a subalgebra of the full constraint algebra (itself a larger Lie algebra in this case).
The second means that $\scP_i$ is a good object (a vector) under the rotations generated by the $\scL_i$.
$\scE$ then closes with these gauge constraints, establishing it as a scalar under the corresponding transformations.

For full GR, the algebraic structure formed by the constraints is 
[using $X \overleftrightarrow{\pa}^i Y := \{ \pa^i Y \} X - Y \pa^i X$]   
\be
\mbox{\bf \{} (    \scM_i  |    \upxi^i    ) \mbox{\bf ,} \, (    \scM_j    |    \d \upchi^j    ) \mbox{\bf \}} =  (    \scM_i    | \, [ \upxi, \upchi ]^i )  \mbox{ } ,
\label{Mom,Mom}
\ee
\be
\mbox{\bf \{} (    \scH    |    \upzeta    ) \mbox{\bf ,} \, (    \scM_i    |    \upxi^i    ) \mbox{\bf \}} = (    \pounds_{\underline{\upxi}} \scH    |    \upzeta    )  \mbox{ } , 
\label{Ham,Mom}
\ee
\be 
\mbox{\bf \{} (    \scH    |    \upzeta    ) \mbox{\bf ,} \,(  \scH  |    \upomega  )\mbox{\bf \}}  = (  \scM_i \mh^{ij}   |   \upzeta \, \overleftrightarrow{\pa}_j \upomega )  \mbox{ } . 
\label{Ham,Ham}
\ee
Note that this does close in the sense that there are no further constraints or other conditions arising in the right-

\ni hand-side expressions. 
The first bracket means that Diff($\bupSigma$) on a given spatial hypersurface themselves close as an (infinite-dimensional) Lie algebra.
The second bracket means that $\scH$ is a good object (a scalar density) under Diff($\bupSigma$).  
Both of the above are kinematical rather than dynamical results.    
The third bracket, however, is more complicated in both form and meaning \cite{T73}.
The presence of $\mh^{ij}(\mh_{kl}(\underline{x}))$ in its right hand side expression causes 

\ni 1) the transformation itself to depend upon the object acted upon -- contrast with the familiar case of the rotations!  

\ni 2) The GR constraints to form a more general algebraic entity than a Lie algebra: a {\it Lie algebroid}.   
More specifically, (\ref{Mom,Mom}--\ref{Ham,Ham}) form the {\it Dirac algebroid} \cite{Dirac}.     
(See \cite{BojoBook} if interested in a bit more on algebroids in general and the Dirac algebroid in particular.)  

\ni 3) If one tried to do GR without Diff($\bupSigma$) irrelevance, (\ref{Ham,Ham}) would in any case enforce this.

\ni 4) Finally, by not forming a Lie algebra, clearly the constraints -- and Diff($\FrM$, Fol) -- form something other than Diff($\FrM$).  
Indeed, the vast difference in size corresponds to the variety of possible foliations.  

\ni As a last example, for minisuperspace one just has the Abelian constraint algebra, $\mbox{\bf \{}     \scH     \mbox{\bf ,} \, \scH    \mbox{\bf \}} = 0$.  
This turns out to be much simpler than (\ref{Mom,Mom}--\ref{Ham,Ham}) due to $\mD_i$ here annihilating everything by homogeneity.

\subsection{Background Independence Aspect 4: Expression in terms of Beables}\label{Beables-Intro}

Having found constraints and introduced a classical brackets structure, one can then ask which objects have zero classical brackets with `the constraints'.
These objects, termed {\it observables} or {\it beables}, are more physically useful than just any $Q^{\sfA}$ and $P_{\sfA}$ due to containing physical information only.  
The Jacobi identity applied to two constraints and one observable/beable determines that the input notion of `the constraints' is a closed algebraic structure of constraints.
Applied instead to one constraint and two observables/beables determines that the observables/beables themselves form a closed algebraic structure.  
In this sense, the observables/beables form an algebraic structure that is associated with that formed by the constraints themselves.

Also note the contextual distinction between observables, which `are observed', and beables, which just `are'. 
Bell \cite{Bell} then argued that the latter are more appropriate for whole-universe cosmology, 
via carrying no connotations of external observing; there is further quantum-level motivation for beables; see also \cite{ABeables}.

In particular, {\it Dirac beables} \cite{DiracObs} are quantities that (for now classical) brackets-commute with all of a given theory's first-class constraints, 
\beq
\mbox{\bf \{}    \scC_{\sfC}    \mbox{\bf ,} \, \iD_{\sfD}   \mbox{\bf \}} \approx 0 \mbox{ } .
\eeq 
Examples include commutation with $\scG$ for Electromagnetism, with $\scE$, $\scL_i$, $\scP_i$ for RPM, or with $\scH$, $\scM_i$ for GR.

On the other hand, {\it \K beables} \cite{Kuchar93, Kuchar99} are quantities that form zero brackets with all of a given theory's first-class linear constraints, 
\beq
\mbox{\bf \{}    \scF\scL\scI\scN_{\sfI}    \mbox{\bf ,}  \, \iK_{\sfK}    \mbox{\bf \}} \approx 0 \mbox{ } .
\eeq
In common examples, $\scF\scL\scI\scN_{\sfG} = \scG\scA\scU\scG\scE_{\sfG}$ (see \cite{HTbook} for a counter-example), 
in which case \K beables coincide with {\it gauge-invariant quantities}. 
The gauge-invariant quantities in question correspond to the successful realization of Configurational Relationalism's $\FrG$ 
as a means of passage to a less redundant quotient configuration space $\FrQ/\FrG$.  
As examples, \K beables are trivial for minisuperspace and spatially-absolute Mechanics, RPM examples include pure shapes and scales.
There is no Dirac--Kucha\v{r} beables distinction for Electromagnetism, and \K beables for GR include, formally, the 3-geometries themselves.

Rovelli's {\it partial observables} \cite{RovelliBook} do not require commutation with any constraints.
These contain unphysical information but one is to consider correlations between pairs of them that are physical.  
One often imagines each as being measured by a localized observer, hence this concept is usually expressed as observables rather than as beables. 
Moreover, via these correlations, this scheme also comes to involve `total observables' similar to Dirac's notion of observables.

Finally, the {\it Problem of Beables} -- more usually termed `Problem of Observables': Facet 4) of the PoT -- 
is that it is hard to construct a set of beables, in particular for gravitational theory.  
This is problematic because \K and especially Dirac beables are hard to find \cite{ABeables}, and even more so at the quantum level.   
Dirac beables are sufficiently hard to find for full GR that \K \cite{Kuchar93} likened postulating having obtained a full set of these to be comparable to catching a Unicorn.
It concerns time (and in some approaches the name `constants of the motion') via theories of this kind having Hamiltonians of form
$H = \int_{\Sigma}\d \Sigma \, \Lambda^{\sfX}\scC_{\sfX}$ for Lagrange multiplier coordinates $\Lambda^{\sfX}$, so that 
\beq
\d \iD_{\sfD}/\d t = \mbox{\bf \{}    \iD_{\sfD}    \mbox{\bf ,} \, H    \mbox{\bf \}} = 
                     \mbox{\bf \{}    \iD_{\sfD}    \mbox{\bf ,} \, \mbox{$\int_{\Sigma}$}\d \Sigma \, \Lambda_{\sfA}{\cal C}^{\sfF}    \mbox{\bf \}} = 
\mbox{$\int_{\Sigma}$}\d \Sigma \, \Lambda_{\sfF}    \mbox{\bf \{}    \iD_{\sfD}    \mbox{\bf ,} \, {\cal C}^{\sfF}    \mbox{\bf \}} \approx 0 \mbox{ } . 
\label{Falla} 
\eeq
This looks to lead to a frozen picture of the world (beables being unable to change value), though \cite{PSS10} reveals this to be a fallacy.

\subsection{Timeless strategies}\label{TNE-Intro}

A number of approaches take Temporal Relationalism's primary timelessness at face value (especially at the quantum level).
These involve considering only questions about the universe `being', rather than `becoming'. 
This can cause at least some practical limitations, but nevertheless can address at least {\sl some} questions of interest.  
These are approaches for which quantum-level motivation is stronger (Sec \ref{QM-Nihil-Intro}). 

\vspace{10in}

\section{III. Spacetime and the two-way passage between it and Space}\label{Cl-PoT-3}

\subsection{Spacetime versus space dichotomy as one reason for variety of PoT facets}\label{Sp-vs-Sp}

GR has more background independent features than theories of Mechanics do.
This is due to GR having a spacetime notion, which has more geometrical content than Mechanics' space-time notion does.
The latter is far more of an amalgamation of separate space and time notions 
-- multiple copies of a spatial geometry strung together by being labelled with a time variable -- whereas the latter is a co-geometrization of space and time. 
In the GR case, spacetime possesses its own versions of generator-providing Relationalism, closure of these generators and beables as associated commutants with these generators.

See Fig \ref{Crux-2} for an outline of some of the consequences. 
%
{            \begin{figure}[ht]
\centering
\includegraphics[width=0.7\textwidth]{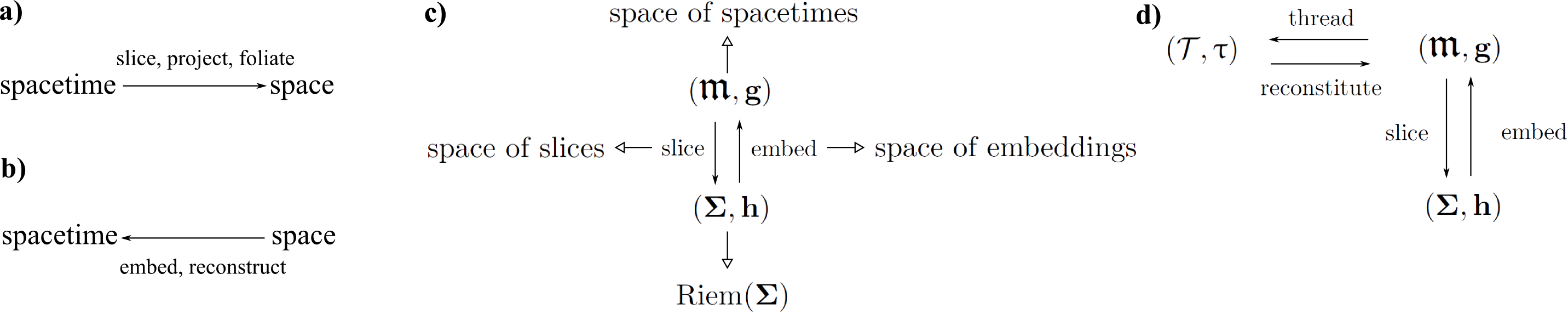}
\caption[Text der im Bilderverzeichnis auftaucht]{        \footnotesize{
\ni a) One passes from spacetime to space by considering a slice and projecting spacetime entities onto it, or by foliating the spacetime with a collection of spaces.
\ni b) Moving in the opposite direction involves embedding rather than projecting, and is a reconstruction (harder due to assumption of less structure).  
Compare this two-way passage with two of Wheeler's routes and with Facets 6) and 7).   
c) Moreover, one is now to interpret spacetime versus space as a dichotomy of primality; the case for space primality is a subcase of that for the primality of dynamics.
Including the spaces of each of the four preceding entities (linked by white arrows), 
one arrives at the eightfold that is crucial for understanding many of the facets of the Problem of Time.  
\ni d) We now include {\it threading}, which I take to mean a single thread, in the same manner as a single spatial slice in the 3 + 1 split.
${\cal T}$ is the thread's topological manifold and $\tau$ is a time metric along each thread. 
One can also consider threading spacetime with a congruence of timelike curves; I term this {\it filling} because it is a spacetime-filling congruence; 
this is analogous to foliation in the 3 + 1 split. 
Finally, I use {\it reconstituting} spacetime for the process of devolping the flow along such a congruence; this is analogous to spacetime reconstruction in the 3 + 1 split.
A parallel of d) is also realized in canonical-and-covariant histories approaches such as Kouletsis' \cite{Kouletsis08} as `time maps' alongside `space maps'.} }
\label{Crux-2}\end{figure}            }

\subsection{Background Independence Aspect 5: Spacetime Relationalism}\label{SR-Intro}

The first position adopted in this Sec is to start afresh with primality now ascribed to spacetime rather than to space.  
Spacetime's own relationalism is then as follows. 

\mbox{ } 

\ni i) The are to be no extraneous spacetime structures, in particular no indefinite background spacetime metrics. 
Fixed background spacetime metrics are also more well-known than fixed background space metrics. 

\ni ii) Now as well as considering a spacetime manifold $\FrM$, consider also a $\FrG_{\sS}$ of transformations acting upon $\FrM$ that are taken to be physically redundant.

\mbox{ }

\ni For GR,  $\FrG_{\sS}$ = Diff($\FrM$).
\ni Also note that $\FrM$ can be equipped with matter fields in addition to the metric. 
Then i) can be extended to include no extraneous internal structures, 
whereas ii)'s $\FrG_{\sS}$ can have a part acting internally on a subset of the fields, corresponding to a spacetime/path format Gauge Theory.
Then the internal part of ii) is closer to QFT's more usual spacetime presentation of Gauge Theory than the internal part of Configurational Relationalism is.
On the other hand, Configurational Relationalism is more closely tied to Dirac observables/beables, out of these all being configuration-based notions.

\subsection{Background Independence Aspect 5.b: Closure of Diff($\FrM$)}

Diff($\FrM$) indeed straightforwardly forms a Lie algebra, in parallel to how Diff($\bupSigma$) does:   
\be
\mbox{\bf |[} (    \mD_{\mu}    |    X^{\mu})   \mbox{\bf ,} \, (    \mD_{\nu}|Y^{\nu}    ) \mbox{\bf ]|} = (    \mD_{\gamma}    | \, [X, Y]^{\gamma}    ) \mbox{ } . 
\label{Lie-2} 
\ee
Diff($\FrM$) also shares further specific features with Diff($\bupSigma$), such as its right hand side being of Lie derivative form.
Thus all three types of Relationalism considered so far are implemented by Lie derivatives.

However, whereas Diff($\bupSigma$)'s generators are conventionally associated with dynamical constraints, Diff($\FrM$)'s are not. 
Additionally, Diff($\bupSigma$)'s but not Diff($\FrM$)'s classical realization of the Lie bracket is conventionally taken to be a Poisson bracket.
This furthermore implies that there is conventionally no complete spacetime analogue of the previous Sec's notion of beables/observables.
These differences are rooted in time being ascribed some further distinction in dynamical and then canonical formulations than in spacetime formulations.   
(\ref{Lie-2}) is to be additionally contrasted with the Dirac algebroid (\ref{Mom,Mom}--\ref{Ham,Ham}).  
Clearly there are two very different algebraic structures that can be associated with GR spacetime: 
the first with unsplit spacetime and the second with split space-time including keeping track of how it is split.

\subsection{Further detail of background independent concepts and terminology}

We next consider the nomenclature `absolute', `relational' and `background-dependent'. 
Giulini \cite{Giu06}, building upon James Anderson's precedent \cite{A64, A67}, defines `absolute' and `background-dependent' to be exactly the same notion.  
As given, this applies to what I term spacetime non-relationalism, though it can be extended to spatial and temporal non-relationalism as well. 
On the other hand, I identify classical Background Independence as the multi-aspect precursor of the multi-faceted classical PoT. 
Relationalism -- viewed as the triple of Temporal, Configurational and Spacetime Relationalisms\footnote{\cite{ARel} considers other notions of Relationalism.} -- 
is a portion of the preceding.
Because of this, I take on board Giulini's conceptualization, but re-name his `background-dependence' as `non-relationalism' (thus obtaining an `expected synonym' absolute = non-relational), 
whilst I continue to define Background (In)dependence in the more general way exposited in Secs \ref{Cl-PoT}--\ref{Cl-PoT-3}.

Furthermore, I caution that Sec \ref{Cl-PoT}--\ref{Cl-PoT-3}'s 8-aspect classical Background Independence
refers to `differentiable structure through to metric-level Background Independence'.\footnote{This is usually referred to as just metric-level Background Independence, 
though this does not to justice to how much of its content rests upon notions of diffeomorphism, which themselves are meaningful down to the level of differentiable structure.} 
%
Moreover, Giulini's definition is not straightforward, nor even claimed to be a completed item, much less one that is adhered to in other parts of the literature, 
which can make yet other distinctions between uses of `absolute' and `background dependent'.  
Finally, note that I am laying even less claims to completeness of this item by considering absolute, relational, and background (in)dependent to remain valid concepts 
at each of Fig 12.b)'s hitherto largely unexplored levels of structure.

The preceding Sec's Diff($\FrM$) is to be interpreted actively -- point-shuffling transformations -- as opposed to passively (coordinate changes). 
In setting up GR, Einstein originally valued general covariance, which has passive connotations. 
However, Kretschmann pointed out that any theory could be cast in generally covariant form.  
On the other hand, active diffeomorphisms continue to play a foundational role in the {\sl physics} of GR (below). 
This can cause confusion because there is a sense in which active and passive diffeomorphisms are {\sl mathematically} equivalent \cite{Wald}.
The claim, however, is that there is physical difference between conceiving in terms of passive and active diffeomorphisms, 
with the latter being a more insightful position \cite{I93, RovelliBook}.  
Understanding this requires a more careful look at the active diffeomorphisms in the GR context.

Start by considering a topological manifold--metric pair ($\FrM, \mg$). 
Then, following Isham, {\it ``Invariance under such an active group of transformations robs the individual points in $\FrM$ of any fundamental ontological significance"} \cite{I93}. 
To further understand what this means in the specific context of GR spacetime in the presence of matter fields, Isham continues with  
{\it ``for example, if $\varsigma$ is a scalar field on $\FrM$ the value $\varsigma(X)$ at a particular point $X\in\FrM$ has no invariant meaning"}. 
See also in this regard the `hole argument' in the Philosophy of Physics literature \cite{EN87, Sta89}, 
though it should be cautioned that this argument has many other parts that this paragraph does not refer to.\footnote{E.g. a hole argument was used early on 
(by Kretschmann against Einstein) to overrule attributing physical significance to what are now known as passive diffeomorphisms. 
Indeed, this argument started before a sharp active--passive distinction was made, then served as one reason to make such a distinction, 
and only then became an argument focusing upon the role of the active diffeomorphisms.}

Next, note that the Einstein field equations are invariant under the group of spacetime diffeomorphisms Diff(${\FrM}$) of ${\FrM}$. 
Preliminarily contrast this with SR, where the Poincar\'{e} invariance group is much smaller.
The main issue, however, is that if $p$, $q$ are two points in $\FrM$ and $\mg_{\mu\nu}(x)$, $\widetilde{\mg}_{\mu\nu}$ are two metrics which solve Einstein's field equations, 
they are related by a diffeomorphism (locally $\phi: X^{\mu} \rightarrow \phi^{\mu}(\vec{X})$)
\beq
\widetilde{\mg}_{\mu\nu}(\vec{X}) = \frac{\pa \phi^{\rho}}{\pa X^{\mu}}\frac{\pa \phi^{\sigma}}{\pa X^{\nu}}\mg_{\rho\sigma}(\phi(\vec{X})).
\eeq
Active diffeomorphism invariance of the theory then amounts to diffeomorphisms $\phi$ being guaranteed to map {\sl solutions to solutions}. 
This property clearly remains the case even if the theory if formulated in a coordinate-independent manner, and so has nothing to do with spacetime coordinate transformations. 
So whereas any theory can be recast in a form invariant under passive spacetime diffeomorphisms, active spacetime diffeomorphism invariance is a property of theories themselves. 
It is possessed by metric-level background independent theories such as GR, but not by dynamical field theories that live upon metric-level fixed backgrounds \cite{RovGaul00}.

Diffeomorphisms and their interplay with equations of motion now having been introduced, the following definitions can be given \cite{A64, A67, Giu06}.    

\ni Definition 1.  An equation of motion on $\FrM$ is termed {\it diffeomorphism invariant} if and only if Diff($\FrM$) is a permitted invariance group for it.  

\ni Definition 2.  Any field which is either non-dynamical or whose solutions are all locally diffeomorphism equivalent is termed an {\it absolute structure}.

\ni Definition 3.  Finally, a theory is termed {\it background independent} if and only if its equations are Diff($\FrM$)-invariant as per Definition 1,  
and its fields do not include absolute structures as per Definition 2.

\ni I next back-track to consider corresponding statements in terms of Diff($\bupSigma$) for geometrodynamics.  
At this level, the role of the Einstein field equations as equations to be solved is replaced by the momentum constraint of GR.
Moreover, solutions are now {\it pairs} ($\bh, \bK$) or ($\bh, \bp$).
Thus the next supposition concerns $\phi \in$ Diff($\bupSigma$) mapping solutions of the form e.g. ($\bh_1, \bK_1$) to solutions ($\bh_2, \bK_2$).
Statements about this remain the case even if the theory if formulated in a coordinate-independent manner, and so has nothing to do with spatial coordinate transformations. 
So whereas any theory can be recast in a form invariant under passive spatial diffeomorphisms, active spatial diffeomorphism invariance is a property of theories themselves. 
It is possessed by metric-level background independent theories such as GR-as-geometrodynamics, and not by any dynamical field theories on metric-level fixed backgrounds.

\mbox{ } 

I finally note that with the algebra (\ref{Lie-2}) and GR known to respect Diff($\FrM$)-invariance, Spacetime Relationalism is a resolved problem at the classical level.  
Upon solving the Einstein field equations, the resulting Lorentzian metric on $\FrM$ provides meaning to each of timelike-, null- and spacelike-separated, 
and to causality \cite{I93}, even if these notions themselves are not preserved by Diff($\FrM$) itself. 
Involving Diff($\bupSigma$) at the classical level is even more straightforward.  
However, diffeomorphism-invariance at the quantum level is not at all straightforward \cite{I84}; indeed this is a major unresolved part of the PoT.

\subsection{Background Independence Aspect 5.c: Spacetime Observables}

Diff($\FrM$) is closely related to {\it spacetime observables} in GR.
Such objects would be manifestly Diff($\FrM$)-invariant. 
I.e. commutants $\mS_{\sfQ}$ [e.g. using the {\sl generator} version of weakly vanishing] 
\be
\mbox{\bf |[} (    {\cal D}_{\mu}    |    X^{\mu})  \mbox{\bf ,} \, (    \mS_{\sfQ}|Y^{\sfQ}    ) \mbox{\bf ]|} \mbox{ }  `=' \mbox{ } 0 \mbox{ } \label{Sp-Obs} .
\ee

\subsection{Paths and Histories strategies}\label{PH-Intro}

Here one considers finite paths instead of instantaneous changes. 
Moreover, histories carry further connotations than paths: for now at classical level, they possess their own conjugate momenta and brackets structure..
These are Isham--Linden style histories \cite{IL} (the older Gell-Mann--Hartle \cite{Hartle} style histories are purely quantum).
However, histories are another approach with more quantum level than classical motivation.
The basic tenet here is as follows. 

\ni {\it Not Time but History}. Perhaps instead it is the histories that are primary ({\it Histories Theory} \cite{GMH, Hartle}).    

\ni Finally N.B. that histories have a mixture of spacetime properties and canonical properties.

\subsection{Web of classical PoT strategies}\label{Cl-Strat-Web}
%
{            \begin{figure}[ht]\centering\includegraphics[width=0.9\textwidth]{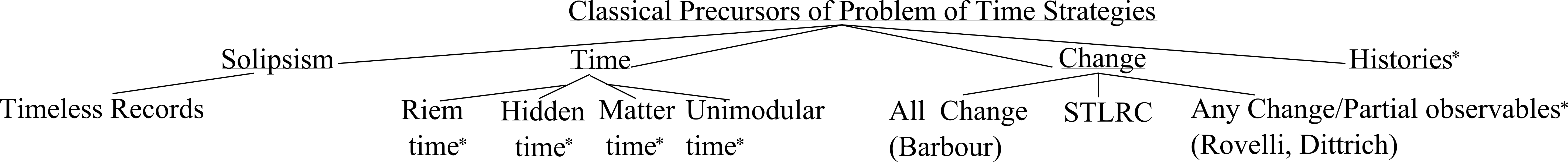}
\caption[Text der im Bilderverzeichnis auftaucht]{        \footnotesize{Web of the various types of strategy and their relations, in a diagram with `cylindrical topology'. 
Explain distinction between just $Q^{\sfA}$, $Q^{\sfA}$ and d$Q^{\sfA}$, paths in $\FrQ$ and histories.
* indicates the 6 out of the 10 strategies covered by Kucha\v{r} and Isham's reviews \cite{Kuchar92, I93} that can be taken to start at the classical level.  
C.f. Fig \ref{Web-QM-2014} for the full 10 of these and the QM-level developments since.
STLRC stands for `sufficient totality of locally relevant change' \cite{ARel2, ABook}, which is the principal type of change considered in this Article.  
}  } \label{Web-Cl}\end{figure}          }
 
\ni The branches in Fig \ref{Web-Cl} this reflect the long-standing philosophical fork between `time is fundamental' 
                                                                                           and `time should be eliminated from one's conceptualization of the world'.  
A finer classification \cite{Kuchar92, I93, APoT, FileR} of `time is fundamental' is into Time Before Quantum and Time After Quantum. 
A finer classification of elimination of time is into Time from Change, 
                                                      History in place of Time, 
												  and pure Solipsism in which questions of `becoming' are to be reduced to mere questions of `being'.

\subsection{Background Independence Aspect 6: Foliation Independence}\label{Fol-Indep-Intro}

Here one is to further develop embeddings, slices and foliations as more advanced foundations for the spacetime-assumed ADM split.
Moreover, GR spacetime admits multiple foliations.  
At least at first sight, this property is lost in the geometrodynamical formulation.

{\it Foliation Dependence} is a type of privileged coordinate dependence. 
This runs against the basic principles of what GR contributes to Physics.
{\it Foliation Independence} is then an aspect of Background Independence, and the {\it Foliation Dependence Problem} is the corresponding PoT facet. 
It is obviously a time problem since each foliation by spacelike hypersurfaces being orthogonal to a GR timefunction.
I.e. each slice corresponds to an instant of time for a cloud of observers distributed over the slice.
Each foliation corresponds to the cloud of observers moving in a particular way.

{\it Refoliation Invariance} is then that evolving via Fig \ref{Refol-4-SRP's}.e)'s dashed or dotted hypersurfaces gives the same physical answer \cite{T73}. 
This happens to be a property possessed by split GR space-time \cite{Bubble, T73}.
This is by bracket (\ref{Ham,Ham})'s pictorial form telling one how to close Fig \ref{Refol-4-SRP's}.a) in the form of Fig \ref{Refol-4-SRP's}.b).  
Thus the Dirac algebroid not only guarantees constraint closure but also resolves the Foliation Dependence Problem at the classical level (Refoliation Invariance Theorem of GR).  
Encoding Diff($\FrM$, Fol) for arbitrary (rather than fixed) foliation Fol can now be seen as the reason why the much larger Dirac algebroid has replaced unsplit spacetime's Diff($\FrM$) 
[the Dirac Algebroid {\sl is} Diff($\FrM$, Fol)].
%
{            \begin{figure}[ht]
\centering
\includegraphics[width=0.85\textwidth]{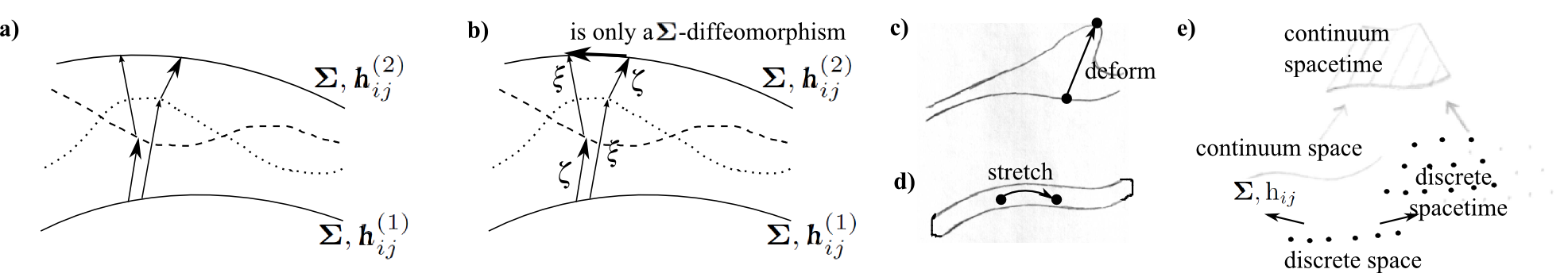}
\caption[Text der im Bilderverzeichnis auftaucht]{\footnotesize{a) The Foliation Dependence Problem is whether evolution via each of the dashed and the dotted spatial hypersurfaces 
give the same physical answers.  
b) Teitelboim's \cite{T73} classical `Refoliation Invariance' resolution of this via the pictorial form of the Dirac algebroid's bracket (\ref{Ham,Ham}). 
c) The pure deformation and d) the pure stretch (just the spatial diffeomorphism again).
e) Pictorial form of 3 types of Spacetime Reconstruction: spacetime from space, from discrete spacetime and from discrete space.} }
\label{Refol-4-SRP's} \end{figure}          }

By the Refoliation Invariance property, GR spacetime is not just a strutting together of spaces like Newtonian space-time. 
Rather, it manages to be many such struttings at once that are physically mutually consistent.
Furthermore, this strutting is tied to the freedom of families of observers to move as they please.
Even the relational form of Mechanics fails to manifest this property; minisuperspace is often just considered in terms of the foliation privileged by spatially homogeneous surfaces. 
Both these limitations are unsurprising since Refoliation Invariance is a diffeomorphism-specific issue.

Finally, Hojman-Kucha\v{r}-Teitelboim \cite{HKT} obtained as a  first answer to Wheeler's question (\ref{Wheeler-Q}). 
Their first principles are the {\it deformation algebroid} of the two operations in Fig \ref{Refol-4-SRP's}c)-d) for a hypersurface, which take the same form as the Dirac algebroid.  
This set-up does however still presuppose spacetime (more specifically, embeddability into spacetime).

Note that most usually one makes a choice to work with split or unsplit spacetime.
However, a few approaches do involve both at once  \cite{Savvidou04a, Kouletsis}. 
Thus all of Temporal, Spatial and Spacetime Relationalism can be manifested at once.

\subsection{Background Independence Aspect 7: Spacetime Reconstruction}\label{SRP-Intro}

Next consider assuming less structure than spacetime.  
In general, if classical spacetime is not assumed, one needs to recover it in a suitable limit. 
This can be hard; in particular, the less structure is assumed, the harder a venture it is.
See Sec \ref{PoT-SRP} for Wheeler's \cite{Battelle} further quantum-level motivation for considering this. 

\mbox{ } 

\ni Already at the classical level, Spacetime Reconstruction can be considered along two logically independent lines: 

\ni a) from space (as an `embed rather than project' `inverse problem' to the previous Sec's, which is harder since now only the structure of space is being assumed).

\ni b) From making less assumptions about continua, giving a total of four reconstruction procedures (Fig \ref{Refol-4-SRP's}.e); 
we expand on this in the sense of `less layers of mathematical structure assumed' in Fig \ref{Bigger-Set}.

\ni The `spacetime from space' case a) that the current Article concentrates upon additionally provides \cite{RWR, AM13} a second answer to Wheeler's question (\ref{Wheeler-Q}).
This clearly goes beyond Hojman--Kucha\v{r}--Teitelboim's first answer's assumption embeddability into spacetime, 
being based rather on 3-spaces $\bigupsigma$ in place of hypersurfaces $\bupSigma$ and proceeds from Temporal and Configurational Relationalism first principles.
Then the consistency of the ensuing constraints' algebraic structure along the lines of the Dirac algorithm returns, from a more general $\scH_{\st\sr\si\sa\sll}$. 
GR's particular $\scH$ alongside local Lorentzian relativity and embeddability into GR spacetime here emerges one of very few consistent possibilities. 
Thus Relationalism is not just a demonstration of the existence of a formulation in which GR is relational, but, a fortiori, is its own route to GR 
(in Wheeler's sense, as per Sec \ref{Many-Routes}).
The few alternatives to this arising in this working differ substantially in causal structure and as regards whether they admit Refoliation Invariance.   
Indeed, as \cite{AM13} details, these few alternatives are foundationally interesting through having, in turn, 
local Galilean-type relativity, local Carrollian relativity, and the CMC condition.  
These are now found to arise {\sl from the Dirac procedure} as the choice of factors that need to vanish in order to avoid the constraint algebroid picking up an obstruction.
[Contrast the form of Einstein's historical dichotomy between universal local Galilean or Lorentzian relativity.] 

\mbox{ } 

\ni It is at this point that Shape Dynamics arises, out of choosing to focus on the option in which the CMC condition emerges, 
and then often further exploiting its connection to conformal mathematics and the York time (Sec \ref{Strat-Intro}).
I however choose to focus instead on the recovery of GR and its spacetime structure at this point, which enables the preceding Subsecs on Spacetime Relationalism and Refoliation Invariance 
to be recast upon temporally relational foundations \cite{ACastle, TRiFol}. 
This completes the incorporation of background independent features into the relational program.
On the other hand, the status of spacetime, refoliation invariance and its link to multiple families of observers 
(as outlined for GR in Sec \ref{Fol-Indep-Intro}) remain unclear on the Shape Dynamics side of this fork.

\subsection{Model arenas, diffeomorphism specifics and slightly inhomogeneous cosmology}\label{Diff-HH-Intro}

Whereas minisuperspace exhibits all but one of the aspects of Background Independence, I have explained how 
homogeneity  implies {\sl very quickly resolved} Problem of Time facets for many of these. 
On the other hand, RPM's 

\ni exhibits 6 of the 9 traditional PoT aspects \cite{FileR}: see Fig \ref{Medium-Table}.  
This includes the Configurational Relationalism aspect that minisuperspace does not possess. 
Moreover, while the RPM and minisuperspace cases are simple to calculate with, they miss the subtleties specifically associated with diffeomorphisms \cite{Kuchar92, I93}.  
The diffeomorphism-specific facets are: detail of the Thin Sandwich, Spacetime Relationalism and further specifics about the Problem of Beables, 
Foliation Dependence, and Spacetime Reconstruction.  

\mbox{ } 

A first arena in which these appear nontrivially is {\it slightly inhomogenous cosmology}.  
A particular such involves inhomogeneous perturbations about the spatially-$\mathbb{S}^3$ minisuperspace with single scalar field matter model of Sec \ref{MSS-Intro}.  
At the semiclassical quantum cosmology level, this particular case becomes the {\it Halliwell--Hawking model} \cite{HallHaw, SIC1}, 
which I choose as the current Article's most complicated specific example.    
RPM's and minisuperspace complementarily support by one or the other having all {\sl other} Background Independence/Problem of Time aspects of this model.   
Here one considers the first few (usually two) orders of the perturbation of the metric.
Each of these form a simplified configuration space in place of the full Riem($\bupSigma$) that are currently under investigation.
\cite{SIC1} shows that slightly inhomogeneous cosmology already exhibits the Thin Sandwich problem, which in this case, moreover, is solvable, and also considers  
this model arena from a Histories perspective.

\section{Classical-level frontiers}\label{7-Hells}

\subsection{Facet Interference}\label{7-Gates}
%
{            \begin{figure}[ht]
\centering
\includegraphics[width=0.55\textwidth]{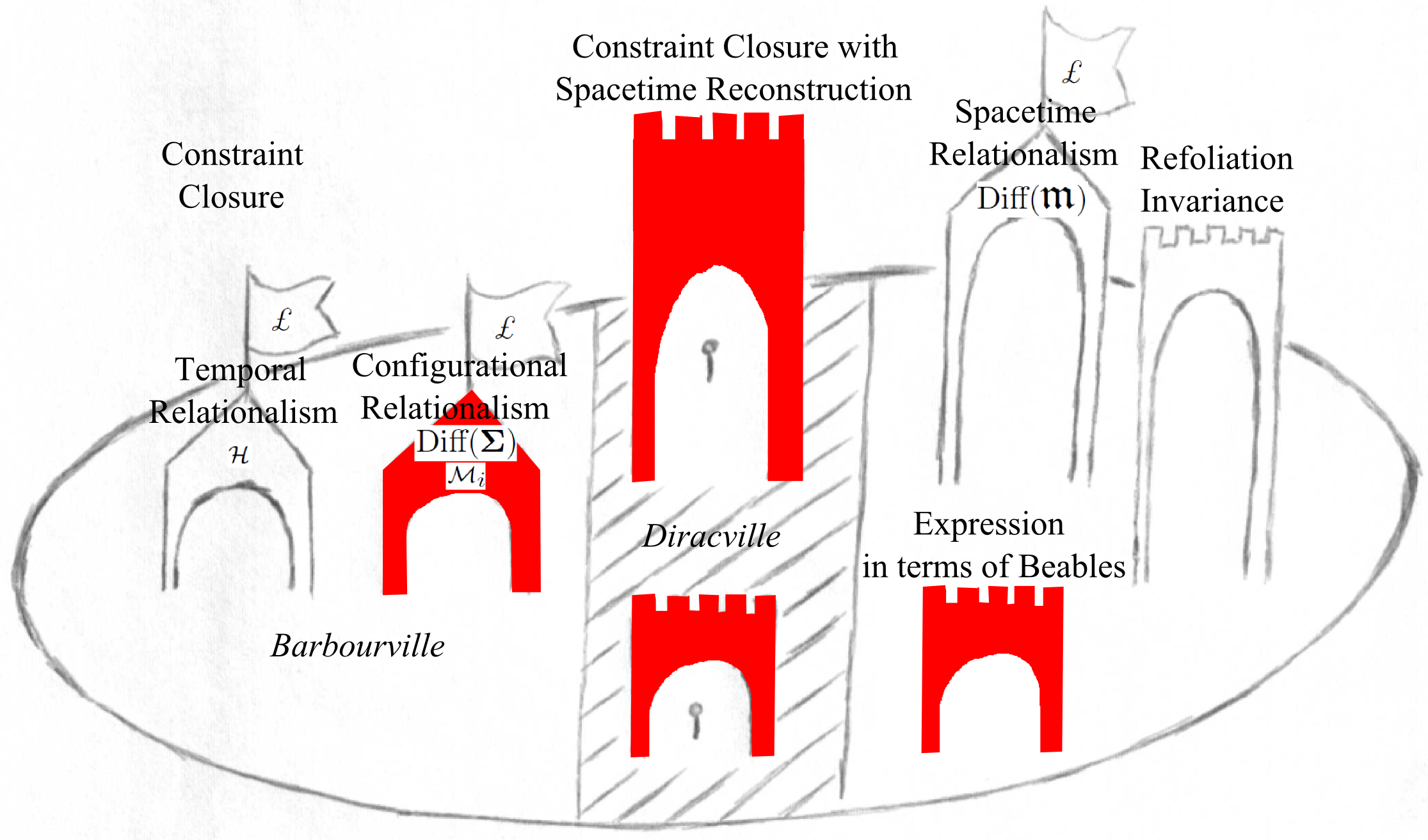}
\caption[Text der im Bilderverzeichnis auftaucht]{\footnotesize{Implementing each aspect of Background Independence causes a PoT facet to appear, so I depict these as gates.
For `a local' resolution of the PoT one needs to consistently get past seven of these (facets 1--7).
My presentation of this as a picture as well as a parable is by grouping and decorating the gates to indicate a number of significant subsets of the gates.
Groupings of gates indicated are, zonally, 
Barbourville and 
Diracville (also with keyholes indicated), 
Spacetime (tall gates) versus Space (short gates), and  
Relational (pointy, and each admitting a Lie implementation as marked on its flag). 
All gates have an algebraic element to either their definition [Diff($\bupSigma$) and Diff($\FrM$) as indicated], or to their classical resolution (listed below).  
Some means through some gates at the classical level are as follows. 
Emergent Machian time to go through Temporal Relationalism (algbraically a reparametrization or a pure deformation) and Best Matching through Configurational Relationalism.
Nontrivial termination of the Dirac Procedure to unlock Constraint Closure, and likewise with Spacetime Reconstruction for the taller double-gate version.
[This is algebraically the Dirac Algebroid = Diff($\FrM$, Fol) for the first and this singled out from a larger family of algebroids in the second; 
these are valid keys that open the keyholes].
Teitelboim's Refoliation Invariance depiction of the Dirac Algebroid = Diff($\FrM, \FrF$)'s $\{\scH, \scH\}$ bracket secures passage through Foliation Dependence. 
Note that the preceding trio are jointly resolved by the Dirac Algebroid in the case of classical GR.  
The Problem of Beables is to be resolved by finding an algebraic structure of beables associated with Dirac Algebroid = Diff($\FrM, \FrF$).
Finally, the four of these algebraic structures indicated in red involve Poisson Brackets.} } 
\label{Gates} \end{figure}          }
 
\ni Fig \ref{Gates} summarizes the present Sec's outline of each PoT facet as a gate, expanding on a quantum-level presentation of Kucha\v{r}'s \cite{Kuchar93}. 
{\sl There is a strong tendency for PoT facets to interfere with each other rather than standing as independent obstacles} \cite{Kuchar92, I93, Kuchar93}. 
The main point of the gates picture is that going through a further gate has a big tendency to leave one outside of gates that one had entered earlier.  
I.e. the facets bear rich conceptual and technical relations with each other since they arise from a joint cause: bridging the gap between background dependent and background independent 
paradigms in physics, most notably the mismatch of the notions of time in GR and Quantum Theory.
Due to this, it is likely to be advantageous to treat them as parts of a coherent package rather than disassembling them into a mere list of problems to be addressed piecemeal.
In \cite{APoT, FileR}, I portrayed that jointness via the facets being an Ice Dragon's\foo{It is specifically `ice' for the frozen formalism aspect.} 
different body parts which coordinate in defense when one confronts it.
However, addressing this problem requires many reconceptualizations, and a Kucha\v{r}-type enchanted castle gates parable is more robust to this, so this Article uses that presentation. 
Moreover, these difficulties largely lie outside the scope of the present Article.  
See \cite{ACastle, TRiPoD, ABook} for new results concerning facet interference.

\vspace{10in}

\subsection{Further multiplicity of some gates}

For instance, spacetime-related gates have histories counterparts. 
There are also multiple types of beables (Fig \ref{Bigger-Set}).
This renders furtherly clear why Background Independence and the PoT have the list of constituent parts that they do.  
%
{            \begin{figure}[ht]
\centering
\includegraphics[width=1.0\textwidth]{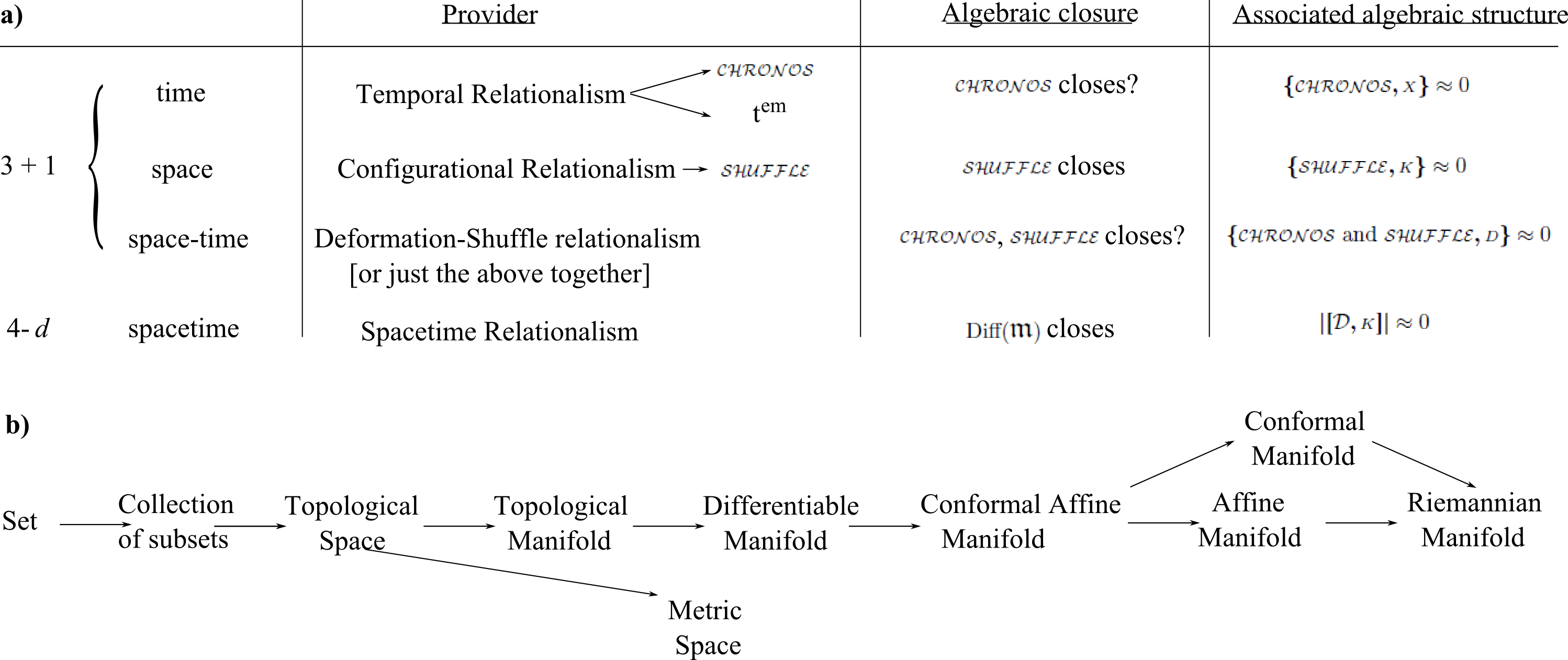}
\caption[Text der im Bilderverzeichnis auftaucht]{        \footnotesize{a) Spacetime and 3 + 1 split space-time, space and time aspects of Background Independence. 
Noting that the 1 + 3 split's pde's and brackets are distinct, a total of $\{1 + 3 + 3\} \times 3 = 21$ aspects arise in this manner, plus an extra 4 from the two-way maps 
between spacetime and each of slices and threadings respectively.  
That gives 25 aspects.
They are based on 7 spaces of spaces and 4 spaces of maps between spaces 
[This still neglects distinctions such as between individual slices and foliations. 
An additional duplicity is that threadings and histories are not expected to involve the same brackets structure.]
Note that not all of these 25 aspects are always present, due to some amounting to choices of perspective or depending upon the algebraic structure formed by the generators. 
Amongst other developments, this conceptualization re-allocates where in the theoretical scheme Dirac Beables belong, 
and points out their complement: constraints that commute with $\scC\scH\scR\scO\scN\scO\scS$ but not necessarily with $\scS\scH\scU\scF\scF\scL\scE_{\tfG}$.  
b) Levels of mathematical structure commonly assumed in Classical Physics.  
\cite{ATop} considers allowing for each of these to be dynamical in turn.  
Note that metric information = conformal structure + localized scale, and then in the indefinite spacetime metric case, conformal structure additionally amounts to causal structure.   
Note also that the above 25 aspects, 7 spaces of spaces and 4 spaces of maps apply at each level (as far down as each concept involved remains defined).  
In particular, I use `space', `time', `spacetime', `slice', `surround', `reconstruct', `thread' and `reconstitute' as level-independent concepts, though some of the properties
of the usual uses of these terms are level-dependent (e.g. spacetime signature).  
{\sl This conceptualization indeed points to many further versions of the PoT facets at each of these levels of structure.}} }
\label{Bigger-Set}\end{figure}            }

\subsection{Threading formulation counterparts}

Moreover some of the gates privilege ADM's 3 + 1 split of spacetime. 
But it is not a foregone conclusion that this is the only split to consider.
Consider first also the threading formulation.
Then comparing with the two-way passage between spacetime and space at each stage (foliating/slicing and reconstructing/embedding), 

\ni 1) if one threads a spacetime, what is the effect on its Diff($\FrM$)? 

\ni 2) What are threading observables/beables? 

\ni 3) Can classical rethreading-invariance be demonstrated?

\ni 4) Is threading-dependence a problem in passing to QM?

\ni 5) How well behaved is threading reconstitution i) conceptually and ii) as a p.d.e. problem?  

\mbox{ } 

\ni Also, what if the split is not Cauchy but rather characteristic (evolution a long a null direction), or not 3 + 1 or 1 + 3 but rather 2 + 2 (spatial 2-surfaces and 2 null directions)?

\subsection{General consideration of combined approaches}

Concurrent deployment of some combinations of the individual strategies shows capacity to remove weaknesses from each.
Also `a local' resolution of the PoT is a well-defined question and rather simpler to address in the absense of the nominally-excluded Multiple-Choice and Global facets.
Dropping facet 5) from `a local' as well, one would have a local canonical approach; instead dropping 1) and 2) it is a path, spacetime or covariant approach. 
Next, keeping both 5) and 1), 2) at once, one has a canonical-and-covariant approach. 
Kouletsis' combined approach \cite{Kouletsis08} is a particular example of this within the context of Histories Theory; so far this approach has only been presented at the classical level.
Also note that if space is primary but one's scheme succeeds at Spacetime Reconstruction \cite{AM13}, one subsequently tap into facets 5) and 6) \cite{TRiFol}.  

\vspace{10in}

\subsection{Aspect 8: Global well-definedness, leading to Facet 8: Global Problems of Time}

Kucha\v{r} \cite{Kuchar92} considered the part of this directly pertaining to timefunctions.
Two senses of globality here are that 

\ni a) timefunctions may not be globally defined in space.

\ni b) Timefunctions may not be globally defined in time itself.  

\ni [One can add `in spacetime', `in configuration space' ... to this list \cite{ATop}.

\ni Moreover, most of the other facets {\sl and} strategies can be beset by global issues! 
Thus I pass from the name `Global Problem of Time' for the timefunction issues to `Global Problems of Time', to cover this larger set.  
At the level of underlying Background Independence aspects, the statement is that (Facet 8) we would  
{\sl prefer that all our notions of Background Independence are globally well-defined}.

\ni Another classification of Global Problems of Time is into effects understandable in terms of meshing conditions of charts or of p.d.e. solutions. 
The first is basic, whereas the second is anything but. 
\ni An additional classification concerns how many layers of structure down the `global effect' occurs, as per Fig \ref{Bigger-Set}.b).

\ni Indeed, surmounting facets 1) to 7) would constitute {\it a local resolution} of the PoT, which is a well-defined sub-problem both conceptually and technically.  
The `local' avoids the Global Problems of Time and the `a' avoids the nonuniqueness of Facet 9): the Multiple Choice Problem, that, being purely quantum, 
is only introduced in Sec \ref{T-QG}.
\cite{ABook} contains a chapter on global issues in the usual conception of metric to differential geometry level Background Independence.

\subsection{Supergravity}

Supergravity is substantially different from GR already  at the classical level as regards how Background Independence manifests itself. 
I outline this in \cite{ASugra}; some points are already available in \cite{ABeables}.

\subsection{Deeper levels of structure}

Fig \ref{Bigger-Set}.b) outlines what these layers are.
Considering not only the metric level but the topological manifold level as well was also first proposed by Wheeler \cite{Wheeler64b}.
Following Isham's pioneering consideration of quantum treatments of yet deeper levels than that \cite{I89-Latt, IKR, I91}, I consider classical precursors of these in \cite{ATop, ABook}. 

\vspace{10in}

\section{Quantum Background Independence. I. Temporal and Configurational Relationalism}\label{T-QG}

Passage to QM usually proceeds from Newtonian Mechanics or SR prior to upgrading these to GR.
his amounts to a paradigm split with QM and GR lying on opposite sides.
In the early 20th Century, each of these areas moved away in a different direction conceptually and technically from Newtonian Mechanics and SR,   
without enough cross-checks to keep Physics within a single overarching paradigm. 
The paradigm split also has a practical justification whose applicability extended far beyond that epoque. 
I.e. our practical experiences concern regimes that involve at most one of QM or GR.
Indeed, one would expect a regime being both of these at once to be characterized by the decidedly otherworldly Planck units.

From the 1920's until around 1960 was a period of stagnancy for GR \cite{Will}, 
and a knock-on effect of this was that the above paradigm split remained largely unaddressed until the end of this period. 
GR was then revived by Wheeler's U.S. group, Zel'dovich's U.S.S.R group, and the U.K. groups including Bondi, Sciama, Penrose and Hawking.
It is work from the first of these groups that is relevant to the current Article (Sec \ref{Gdyn}). 
This was in part preceded by the French School's work (Sec \ref{Gdyn}), and by Dirac's 1950's work on the canonical formulation of GR (as subsequently reviewed in \cite{Dirac}).  
Wheeler in particular turned attention to the conceptualization of this approach in the 1960's, envisaging in the process some of the Problem of Time impasses \cite{WheelerGRT, Battelle}. 

\mbox{ }

\ni The greater part of the PoT \cite{Battelle, Kuchar81, Kuchar91, Kuchar92, I93, Kuchar99, KieferBook, RovelliBook, APoT, APoT2, FileR, BI} 
occurs because the `time' of GR and the `time' of Quantum Theory are mutually incompatible notions.
This causes difficulty in trying to replace these two branches of physics with a single framework in regimes in which neither Quantum Theory nor GR can be neglected. 
This is needed in parts of the study of black holes or of the very early universe.

\subsection{The Quantum Frozen Formalism Problem (Schr\"{o}dinger picture)}\label{FFP-Intro}

The Schr\"{o}dinger-picture Frozen Formalism Problem is that the stationary alias timeless alias frozen wave equations such as (\ref{TISE}) occur in a situation in which one 
would expect time-dependent wave equations such as (\ref{TDSE}) or the Klein--Gordon equation $\Box \varsigma = m^2\varsigma$.
This frozenness is well-known to be a consequence of the GR Hamiltonian constraint $\scH$ (or similar model arena objects) being quadratic and not linear in the momenta. 
It is useful in this regard to clarify that timeless equations such as the Wheeler--DeWitt equation apply {\sl to the universe as a whole}. 
On the other hand, the more ordinary laws of Physics apply to small subsystems {\sl within} the universe, which does suggest that this is an apparent, rather than actual, paradox. 
One of this Article's main points is that the origin of $\scH$ can be traced further back to Temporal Relationalism.  
The Frozen Formalism Problem then suggests that, in apparent contradiction with everyday experience, nothing at all {\sl happens} in the universe! 
Thus one is faced with having to explain the origin of the notions of time in the laws of Physics that appear to apply in the universe. 
Secs \ref{Tfn-Strat}, \ref{QM-Nihil-Intro}, \ref{PIH-Intro}, \ref{QM-Strat-Web} introduce a number of strategies for such explanations.   
It is useful to note throughout that timelessness is a problem for the universe {\sl as a whole}, rather than for the much more usually studied case of {\sl subsystems within} a universe.

\mbox{ } 

\ni Since in QM the wave equation does not suffice to obtain physical answers, which require also an inner product input: $\langle\psi_1|\widehat{O}|\psi_2\rangle$, 
we need an inner product for canonical QG too. 
Ordinary QM comes with the Schr\"{o}dinger inner product, whilst the Klein--Gordon equation has a distinct notion of inner product associated with its multi-particle interpretation. 
For canonical QG, however, there are in general further complications: the {\it Inner Product Problem} alias {\it Hilbert Space Problem}.
This is a time problem -- a subfacet of the Frozen Formalism Problem -- due to the ties between inner products, conservation of probability and unitary evolution.

\subsection{Some timefunction-based strategies to deal with frozenness}\label{Tfn-Strat} 

External time is here inappropriate because 1) It does not feature in the wave equation for GR. 

\ni 2) External time is furthermore incompatible with describing truly {\sl closed} systems, the ultimate of which is closed-universe Quantum Cosmology. 
How QM should be {\sl interpreted} for the universe as a whole is then a recurring theme in this Article.  

\ni Thus one is in need of another type of time concept.  

\mbox{ } 

\ni {\it Emergent time before quantization}? 
We have already seen that in the absense of time at the primary level in classical whole-universe physics, the classical Machian resolution by $t^{\se\sm(\sJ\sB\sB)}$ presents itself.
Unfortunately, this fails to unfreeze the quantum equation.  
This is resolved by starting the time-finding process afresh at the quantum level.

\mbox{ } 

\ni {\it Time After Quantum} Perhaps one has slow, heavy `$h$'  variables that provide an approximate 
timestandard with respect to which the other fast, light `$l$' degrees of freedom evolve \cite{HallHaw, Kuchar92, KieferBook}.  
In the Halliwell--Hawking \cite{HallHaw} scheme for semiclassical GR Quantum Cosmology, $h$ is scale (and homogeneous matter modes) and $l$ are small inhomogeneities.  
The Semiclassical Approach involves i) making the Born--Oppenheimer ansatz 
\beq
\Psi(h, l) = \psi(h)|\chi(h, l)\rangle
\label{BO}
\eeq 
and the WKB ansatz 
\beq
\psi(h) = \mbox{exp}(i\,S(h)/\hbar) \mbox{ } 
\label{WKB}
\eeq 
(each of which is accompanied by a number of associated approximations).  

\ni ii) Forming the $h$-equation
\beq
\langle\chi| \widehat{\scQ\scU\scA\scD} \, \Psi = 0 \mbox{ } .
\eeq 
The above-mentioned ans\"{a}tze and associated approximations ensure that this yields a Hamilton--Jacobi equation\foo{For simplicity,
I present the below in the case of one $h$ degree of freedom; see \cite{ACos2} for multiple such and other generalizations.}
\beq
\{\pa S/\pa h\}^2 = 2\{E - V(h)\} \mbox{ }  
\eeq
where $V(h)$ is the $h$-part of the potential. 
Then one way of solving this is for an approximate emergent semiclassical time $t^{\se\sm(\sW\sK\sB)}(h)$. 

\ni iii) One then forms the $l$-equation 
\beq
\{1 - |\chi\rangle\langle\chi|\}\widehat{\scQ\scU\scA\scD}\,\Psi = 0 \mbox{ } . 
\eeq 
This fluctuation equation can be recast (modulo some more of the approximations) into an emergent-WKB-time-dependent Schr\"{o}dinger equation for the $l$-degrees of freedom. 
The mechanics/RPM form of this is
\beq
i\hbar\pa|\chi\rangle/\pa t^{\se\sm(\sW\sK\sB)}  = \widehat\scE_{l}|\chi\rangle \mbox{ } .
\label{TDSE2}
\eeq
The emergent-time-dependent left-hand side arises from the cross-term $\pa_{h}|\chi\rangle\pa_{h}\psi$.
$\widehat\scE_{l}$ is the remaining surviving piece of $\widehat\scE$, acting as a Hamiltonian for the $l$-subsystem.

Moreover, for $\scQ\scU\scA\scD$ arising as an equation of time $\scC\scH\scR\scO\scN\scO\scS$, 
$t^{\se\sm(\sW\sK\sB)}$ can be interpreted as \cite{FileR, ACos2} a Machian emergent semiclassical time (whether for the above toy model or for GR Quantum Cosmology).
To zeroth order, this coincides with $t^{\se\sm(\sJ)}_0$, but both fail to be Machian on account of allowing neither classical nor semiclassical $l$-change to contribute.   
To first order, however, 
\beq
t^{\se\sm(\sW\sK\sB)}_1 = F[h, l, \d h, |\chi(l, h)\rangle] \mbox{ } ,
\eeq 
which is clearly distinct from the $h$, $l$ expansion of (\ref{t-em-J}), which is of form $F[h, l, \d h, \d l]$.
This pairing of the previously existing emergent semiclassical time and the Machian classical emergent time is new to this program, as is the Machian reinterpretation of the former.

N.B. moreover that the above working leading to such a time-dependent Schr\"{o}dinger equation ceases to function in the absence of making the WKB ansatz and approximation.  

Additionally, in the quantum-cosmological context, is not known to be a particularly strongly supported ansatz and approximation to make.    
This is crucial for this Article since propping this up requires considering one or two further PoT strategies from the classical level upwards, 
a point to which we return in Sec \ref{PIH-Intro} after having surveyed the rest of the individual strategies and facets. 

\mbox{ }

\ni Further options involve quantum-level continuations of Sec \ref{Strat-Intro} approaches.  
Since these involve continuing to use at the quantum level a candidate time found at the classical level, they are known as {\it Time Before Quantum} approaches.  

\ni 1) Riem time from the hyperbolic reformulation (\ref{hyperb}) of $\scH$ due to the supermetric's indefiniteness forms a Klein--Gordon-like quantum equation 
\beq
\pa_{t}^2 \Psi = - \triangle_{\st\sr\su\se} \Psi +  C[\mbox{\boldmath $h$}]\Psi \mbox{ } .  
\label{KG-type}
\eeq
Probably the most lucid summary of this approach is that 
\beq
\triangle_{\mbox{\scriptsize\boldmath$M$}} 
\mbox{ }  \mbox{ } \mbox{ is actually a } \mbox{ }   \Box_{\mbox{\scriptsize\boldmath$M$}} \mbox{ } .  
\eeq
\mbox{ } \mbox{ }  
One can then furthermore try the corresponding Klein--Gordon inner product interpretation. 
Unfortunately this fails due the GR potential not being as complicit as Klein--Gordon theory's simple mass term (see \cite{Kuchar81, Kuchar91} for details).  
This approach was traditionally billed as Time Before Quantum, through choosing to make this identification of a time after the quantization, but can just as well be set up ante quantum. 

\ni 2) If one uses instead hidden time, a parabolic reformulation for $\scH$ of type (\ref{hidden-true}) is then promoted to a hidden-time-dependent Schr\"{o}dinger equation 
\beq
i\pa \Psi/\pa {t_{\sh\si\sd\sd\se\sn}} = \hat{\fH}_{\st\sr\su\se}\Psi \mbox{ } .
\eeq
\ni 3) If one uses instead reference matter time, another parabolic reformulation (\ref{appended-reference}) 
is then promoted to a reference-matter-time-dependent Schr\"{o}dinger equation  
\beq
i\pa \Psi/\pa {t_{\sr\se\sf}} = \hat{\fH}_{{\st\sr\su\se}}\Psi \mbox{ } .
\eeq

\subsection{Which theories and model arenas exhibit other quantum PoT Facets} 
%
{            \begin{figure}[ht]\centering\includegraphics[width=1.0\textwidth]{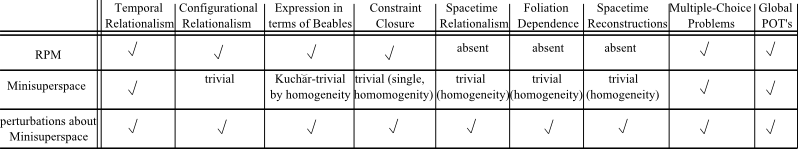}
\caption[Text der im Bilderverzeichnis auftaucht]{        \footnotesize{Which model arenas exhibit which facets.}  } \label{Medium-Table}\end{figure}          }

\subsection{Problem of Time Facet 2) Configurational Relationalism}\label{PoT-CR}

\ni 1) {\it Reduced quantization approaches}. If one succeeds in reducing out $\FrG$ at the classical level, one may not need to face a $\FrG$ again at the quantum level. 

\ni 2) {\it Dirac quantization approaches}.  
These involve, alternatively, quantizing first and only then dealing with the constraints corresponding to $\FrG$, now promoted to further quantum wave equations
\beq
\widehat{\scF\scL\scI\scN}_{\sfI} \Psi = 0 \mbox{ } . 
\eeq
\ni As regards 1), note that moreover the quantum theory requires a distinct $\FrG^{\prime}$ on some occasions (see the next Sec).
This is a second reason, in addition to practical inability to reduce at the classical level, for 2).  
Sec \ref{Facets-2}'s indirect $\FrG$-act, $\FrG$-all method continues to be applicable at the quantum level, whether as a means of formulating 2) 
or as an indirect means of expressing all subsequent objects required by one's theory in the absense of being able to solve either 1) or 2).

\section{II. Constraint Closure and Beables}

\subsection{Problem of Time Facet 3: Constraint Closure Problem}\label{PoT-CC}

Here quantum commutators play even more central a role than classical Poisson brackets did.
Moreover the QM notion of {\sl equal-time} commutation relations poses significant difficulties in the context of GR, 
due to this `equal-time' notion carrying connotations of there being a unique pre-given time, which does not meet the GR conception of time.
Then amongst other things, these impinge on the use of quantum constraint algebraic structures. 
The issue one wishes to establish at this point is whether the quantum constraints algebraically close under the commutator bracket.  


Another source of problems is that commutator algebraic structures are not in general isomorphic to their classical Poisson brackets antecedents/approximands.  
\cite{I84} traces the topological underpinnings of this discrepancy.
This gives one reason why classical constraint closure does not necessarily pass over to the quantum level:
\beq
\mbox{\bf \{}    \scC_{\Gamma}    \mbox{\bf ,} \, \scC_{\Lambda}    \mbox{\bf \}} \approx 0 
\mbox{ } \not{\vspace{-0.04in}\Rightarrow}  \mbox{ } \mbox{ } 
\mbox{\bf [}    \widehat\scC_{\Gamma}    \mbox{\bf ,} \, \widehat\scC_{\Lambda}    \mbox{\bf ]} = 0
\label{C-C-2}
\eeq
Additionally, the classical constraints are replaced by operator-valued quantum constraints. 
However, this is subject to operator-ordering ambiguities and well-definedness issues, 
and each way of handling these could in principle lead to differences in the outcome of the evaluation of the commutators between the constraints.

Breakdown of the closure of the algebraic structure of the constraints at the quantum level is termed the {\it Functional Evolution Problem} in \cite{Kuchar92, I93}. 
However, the `functional' in this name carries field-theoretic connotations -- it is the type of derivative that features in the field-theoretic form of the problem.  
To additionally include the classical case and maximally clarify the nature of this problem, I consider it best to refer to it as the {\it Constraint Closure Problem}.

Anomalies -- classical symmetries that quantum theory does not accept -- are one manifestation of non-closure.
They are additionally a means by which a classically accepted $\FrG$ may need to be replaced by a distinct $\FrG^{\prime}$ at the quantum level.
Whilst not all anomalies involve time or frame, a subset of them do, thereby forming part of the PoT.

This is an opportune point at which to note that many a quantization approach \cite{I84} is of at most limited value in GR
due to the classical GR constraints forming not a Lie algebra but rather the Dirac algebroid.

\subsection{Problem of Time Facet 4: Problem of Beables}\label{PoT-PoB}

Observables or beables at the quantum level carry the further basic connotation of being self-adjoint operators, 
so that their eigenvalues are real-valued and hence can correspond to measured or experienced physical properties.

Thinking in terms of beables rather than observables has further quantum-level benefits as regards adopting a realist interpretation (e.g. decoherence or consistent histories). 
The classical treatment also cited closed system and cosmological benefits, and, moreover these benefits are combined and further extended in Quantum Cosmology \cite{Bell}.
Quantum Cosmology being the central arena of this Article, henceforth I use `beable' unless the situation specifically requires use of the word `observable'. 

\mbox{ }  

\ni The rather general quantum definition of beables involves three changes as compared to its classical precursor. 
I.e. the quantum version involves self-adjoint operators $\widehat{\iB}$ that form zero quantum commutators with the quantum constraints, 

\beq
\mbox{Quantum Dirac beables: $\widehat{D}_{\sfD}$ such that } \mbox{ } \mbox{\bf [}\widehat{\cal C}_{\sfF} \,  \mbox{\bf ,}  \mbox{ } \widehat{D}_{\sfD} \mbox{\bf ]} = 0 \mbox{ } ,
\label{QDB}
\eeq
\beq
\mbox{Quantum \K beables: $\widehat{K}_{\sfK}$ such that } \mbox{ }   \mbox{\bf [}  \widehat{\scF\scL\scI\scN}_{\sfI}\mbox{\bf ,} \, \mbox{ } \widehat{K}_{\sfK}\mbox{\bf ]} = 0 
\mbox{ } .
\label{QKB}
\ee
Quantum partial observables are defined as a conituation of their classical definition too, 
the difference being in that their capacity to `predict numbers' now carries the inherent probabilistic connotations of Quantum Theory.

\mbox{ } 

\ni The {\it Problem of Quantum Beables} is then that it is hard to come up with a sufficient set of these for quantum-gravitational Physics.  

\mbox{ } 

\ni Finally I comment that in the Heisenberg picture of QM, the apparent manifestation of frozenness is, rather,   
\beq
\mbox{\bf [} \widehat{H} \mbox{\bf ,} B \mbox{\bf ]} = 0
\eeq
with similar connotations to the preceding classical level's Poisson brackets relation (\ref{Falla}).

\subsection{Quantum-level timeless strategies}\label{QM-Nihil-Intro}

Now instead of seeking a timefunction, instead consider setting up a new interpretation of quantum theory which makes no use at all of any concept of time.  
I.e. take the apparent timelessness of quantum QR at face value, and determine how much of Physics can be recovered therein.
The familiar forms of concepts such as temporal evolution, becoming and history would then need to arise as some kind of phenomenological {\it semblance} of dynamics or of history, 
arising from pure being.  

\mbox{ } 

\ni Example 1) The {\it Na\"{\i}ve Schr\"{o}dinger Interpretation}, which is due to Hawking and Page \cite{HP86-88}, though this name for it was coined by Unruh and Wald \cite{UW89}.  
This concerns the probabilities of being for questions about universe properties such as: what is the probability that the universe is large? 
Flat? 
Isotropic? 
Homogeneous?   
One obtains answers to these via consideration of the probability that the universe belongs to region $\bFrR$ of $\FrQ$ that corresponds to a quantification of a particular such property, 
\beq
\mbox{Prob}(R) \propto \int_{\bsFrR}|\Psi|^2\mathbb{D}\fQ \mbox{ } , 
\eeq 
for $\mathbb{D}\fQ$ the configuration space volume element.
This approach is termed `na\"{\i}ve' due to it not using any further features of the constraint equations.  

\mbox{ }

\ni Example 2) The {\it Conditional Probabilities Interpretation} due to Page and Wootters \cite{PW83} goes further by addressing {\sl conditioned} questions of being. 
The conditional probability of finding $B$ in the range $b$, given that $A$ lies in $a$, and to allot it the value
\beq
\mbox{Prob}(B\in b | A\in a; \BigupRho) = \mbox{tr}\big(\mP^B_{b}\,\mP^A_{a}\,\BigupRho\,\mP^A_a\big)/\mbox{tr}\big(\mP^A_a\,\BigupRho\big) \mbox{ } ,    
\label{Pr:BArhoProto}
\eeq
where $\BigupRho$ is a density matrix for the state of the system and the $\mP^A_a$ denote projectors.  
Examples of such questions for are `what is the probability that the universe is flat given that it is isotropic'?  

\ni \cite{PW83} gives convincing arguments in answer to Sec \ref{Tfn-Strat}'s 2) that such a system's only physical states are {\sl eigenstates} of the 
Hamiltonian operator, whose time evolution is essentially trivial.  
Note that this scheme can be extended to cover questions of being at a time, by considering correlations between the subsystem of interest and clock variables.  

\mbox{ } 

\ni Example 3) {\it Records Theory} \cite{PW83, GMH, B94II, EOT, H99, Records} involves localized subconfigurations of a single instant of time.  
It concerns issues such as whether these contain useable information, are correlated to each other, and whether a semblance of dynamics or history arises from this.  
This requires 

\ni A) notions of localization in space and in configuration space. 

\ni B) Notions of information, correlation, and of pattern more generally \cite{Records, AKendall, ATop}.  

\ni See \cite{Records, ATop} contain novel analysis of Records Theory. 

\mbox{ } 

\ni Example 4) {\it Evolving constants of the motion} (`Heisenberg' rather than `Schr\"{o}dinger' style QM) interpretation of partial observables \cite{RovelliBook}.
Note that this approach allows for change, unlike the previous 3 examples, which are all solipsist.

\section{III. Spacetime, its split and its recovery}

\subsection{Problem of Time Facet 5: Spacetime Relationalism}\label{PoT-SRel}

Is spacetime, or any of its aspects, meaningful in QG, and how do the other aspects emerge in the classical limit?  
Is (spatial or spacetime) classical geometry or any of its subaspects meaningful in QG, and how do the other aspects emerge in the classical limit. 
Is there something resembling the classical notion of {\sl causality} in QG? 
If so, which aspects of classical causality are retained as fundamental, and how do the other aspects emerge in the classical limit?
This feeds into Spacetime Relationalism versus Temporal and Configurational Relationalism. 
This in turn feeds into the Feynman path-integral versus canonical starting points, and into whether to adopt quantum-level Refoliation Invariance and Spacetime Reconstruction.

Specific quantum level issues are as follows.

\ni 1) Whether a hypersurface is spacelike depends on the spacetime metric $\mg$; however in QG this would undergo quantum fluctuations \cite{I93}. 
In this way, the notion of spacelike would depend on the quantum state, as would causal relations, including the microcausality condition 
\beq
\mbox{\bf [}\widehat{\phi}(\vec{X})\mbox{\bf ,}\widehat{\phi}(\vec{Y})\mbox{\bf ]} = 0     \mbox{ } .
\label{microcausal}
\eeq
that is crucial to standard QFT.
Then for most pairs of events $\vec{X}, \vec{Y}\in \FrM$ there is at least one Lorentzian metric by which they are {\it not} spacelike separated  \cite{FH87}.
Thus if all metrics are `virtually present' due to fluctuations and as manifested e.g. in the path integral sum) (\ref{microcausal})'s right hand side is generically not zero. 
This gives one further reason in spacetime-presupposed approaches for there being difficulties with the notion of equal-time commutation relations in QG.

\ni 2) Relativity places importance upon labelling spacetime events via physical clocks and spatial frames of reference.
Suppose one tries to do so using proper time at the quantum level \cite{I93}.  
Unfortunately, proper time intervals are built out of $\bg$, and thus are only meaningful after solving the equations of motion.
This is also rendered problematic by $\bg$'s quantum fluctuations.  
Attempting to get round this by casting time in the role of a quantum operator has to contend with that not being part of standard quantum theory for deep-seated interpretational reasons.

\ni 3) From a technical perspective, the representation theory of Diff($\FrM$) is difficult.  

\ni 4) There is further interplay if one's theory attempts to combine spacetime and canonical concepts.
Moreover, the representation theory of the Dirac algebroid is {\sl even more} difficult.

\subsection{Path Integral and Histories Theory Approaches}\label{PIH-Intro}

Whereas QM {\sl path} integrals are sometimes already called sum over {\sl histories} approaches, the present Article we adopt a more structured notion of history.

\ni The PoT facets do not take an entirely fixed form.  
The Frozen Formalism Problem occurs if one simply splits spacetime and works canonically. 
However, one has the advantages of no frozen wave equation, inner product problem or foliation issues if an unsplit spacetime formulation is used. 
For instance, path integral approaches are of this form.
However, whilst very successful in QFT, they face their own set of problems if one attempts to apply them to QG.

\mbox{ } 

\ni I) In place of an Inner Product Problem, 
there is a {\it Measure Problem}.
Moreover, this is diffeomorphism-specific and thus is itself directly a Problem of Time facet resulting from the Spacetime Relationalism Background Independence aspect.  
Diff(\FrM) invariance unfortunately also aggravates the Measure Problem.  

\ni II) The gravitational path integral is, at the very least, problematic as regards being well-defined. 
(Discrete approaches can overcome this difficulty).

\ni III) The analogue of the Wick rotation familiar from QFT (`from imaginary time to real time') in general goes awry in curved spacetimes. 
This is due to ambiguities \cite{HL90} arising as regards which contours to use for the `rotation' in the complex manifold.  

\ni Note also approaches involving both path integrals and canonical formalism. 
QFT is free to rely on the canonical approach as regards computing what its Feynman rules are. 
Thus ordinary QM `in terms of path integrals' can sometimes in fact be a combined path-integral {\sl and} canonical approaches.
Then in QG, such a combined scheme would pick up the problems of {\sl both} approaches.

\mbox{ } 

\ni Finally, one might use quantum-level Histories Theory instead, i.e. with more structures than are usually assumed in a path integral approach.  
At the quantum level, these are also adorned with projection operators and admit both discrete \cite{Hartle} and 
                                                                                      continuous (`histories projection operator' \cite{IL}) time formulations.

\vspace{10in}

\subsection{Web of quantum PoT strategies}\label{QM-Strat-Web}
%
{            \begin{figure}[ht]\centering\includegraphics[width=0.97\textwidth]{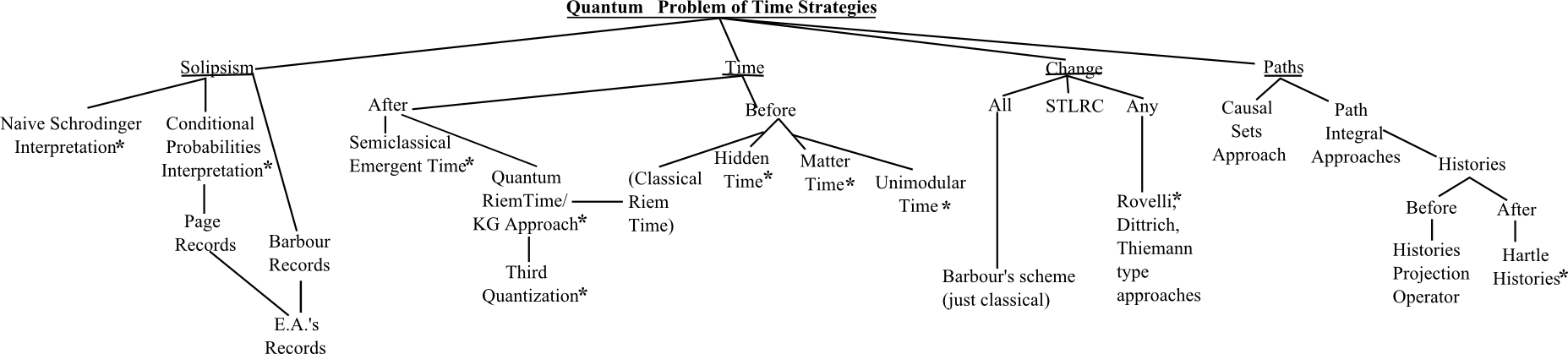}
\caption[Text der im Bilderverzeichnis auftaucht]{        \footnotesize{The quantum-level extension of Fig \ref{Web-Cl}'s web of PoT strategies. 
* indicates the 10 strategies covered by Kucha\v{r} and Isham's classic reviews \cite{Kuchar92, I93}.
Before and After are here meant relative to the quantization procedure. }  } \label{Web-QM-2014}\end{figure}          }

\subsection{A particular Combined Strategy}\label{Combo2}

Combining the semiclassical, records and histories approaches is a particularly interesting prospect. 
Firstly, the semiclassical approach can be placed on a Machian (time is to be abstracted from change) footing of the STLRC type \cite{ACos2}.
Then histories decohereing is one possible way of obtaining a semiclassical regime in the first place. 
I.e. finding an underlying reason for the WKB assumption without which the Semiclassical Approach does not work. 
The also-elusive question \cite{GMH} of which degrees of freedom decohere which others in Quantum Cosmology is to be answered by considering where the records form.
There is indeed a Records Theory within Histories Theory \cite{GMH, H99}.  
Either of semiclassicality or histories theory frees timeless records approaches from having to provide a semblance of dynamics on their own.  
The semiclassical approach also aids in the computation of answers to timeless propositions.
This mutually supporting semiclassical--histories--records perspective, minus its Machian element, was developed by Halliwell \cite{H03, H09}. 
I then showed \cite{AHall, A13} that its Machian counterpart is a natural third-generation Machian strategy following on from classical and then semiclassical emergent Machian time.

\subsection{Problem of Time Facet 6: Foliation Dependence Problem}\label{PoT-Fol}

A quantization of GR that retains the nice classical property of refoliation invariance would be widely seen as conceptually sound and appealing.   
However, at the quantum level there ceases to be an established way of guaranteeing this.    
If this is not retained, starting with the same initial state $\Psi_{\si\sn}$ 
{\it ``on the initial hypersurface and developing it to the final hypersurface along two different routesproduces inequality"},
\beq
\Psi_{\sf\si\sn-1} \neq \Psi_{\sf\si\sn-2}  \mbox{ } \mbox{ (quantum foliation-dependence criterion) }  
\label{fol-dep}
\eeq
and this {\it ``violates what one would expect of a relativistic theory."} \cite{Kuchar92}.

\subsection{Problem of Time Facet 7: Spacetime Reconstruction Problem}\label{PoT-SRP}

As Wheeler also pointed out \cite{Battelle}, at the quantum level, fluctuations of the dynamical entities are unavoidable. 
In the present case, these are fluctuations of 3-geometry, and these are then too numerous to be embedded within a single spacetime.  
The beautiful geometrical way that classical GR manages to be Refoliation Invariant breaks down at the quantum level.

\mbox{ } 

\ni Wheeler furthermore pointed out \cite{Battelle, W79} the further relevance of Heisenberg's uncertainty principle.  
Precisely-known position $\underline{q}$ and momentum $\underline{p}$ for a particle are a classical concept corresponding to a worldline.
This view of the world is entirely accepted to break down in quantum physics due to Heisenberg's Uncertainly Principle.
In QM, worldlines are replaced by the more diffuse notion of wavepackets. 
However, in GR, what the Heisenberg uncertainty principle now applies to are the quantum operator counterparts of $\mh_{ij}$ and $\mp^{ij}$. 
But by formula (\ref{Gdyn-momenta}) this means that $\mh_{ij}$ and $\mK_{ij}$ are not precisely known.   
Thus the idea of embeddability of a 3-space with metric $\mh_{ij}$ within a spacetime is itself quantum-mechanically compromised.
Thus (something like) the geometrodynamical picture (considering the set of possible 3-geometries and the dynamics of these) 
would be expected to take over from the spacetime picture at the quantum level.  
It is then not clear what becomes of notions that are strongly associated with classical GR spacetime, 
such as locality (if one believes that the quantum replacement for spacetime is `foamy' \cite{Battelle}), or causality.   
In particular, microcausality is violated in some such approaches \cite{I81, I85, I93}.
Additionally, the recovery of semiclassicality aspect of spacetime reconstruction has a long history of causing difficulties in LQG.  
Investigation of the semiclassical and quantum commutator bracket counterpart of the classical Dirac algebroid has also started \cite{Bojo12}.  
The extension of such work to a family of such algebraic structures in parallel with the current paper remains to be tackled (this Sec is part-based on \cite{AM13}).

\mbox{ } 

\ni There is also an issue of recovering continuity in suitable limits in approaches that treat space or spacetime as discrete at the most fundamental level

\section{Quantum-level frontiers}

\subsection{Facet interference and combined strategies} 

See \cite{Kuchar92, I93, Kuchar93, APoT2, ABook} for quantum level facet interference.
See Sec \ref{Combo2} for a semiclassical quantum level combined strategy.

\subsection{Problem of Time Facet 8: Global Problems of Time}\label{PoT-Glob}

To the second classification of types of Global PoT's in Sec \ref{7-Hells} one can now add e.g. `meshing together unitary evolutions', 
which is not only hard to prove existence for but also remains largely undeveloped even as regards its own conceptual meaning.
Many quantum-level facets and strategies exhibit global problems; see e.g. \cite{I93, ABook}.

\subsection{Problem of Time Facet 9:  Multiple Choice Problem}\label{PoT-MCP}

\ni Background Independence Postulate 9) (Uniqueness Specifications) for Background Independent schemes would then be that these are to be unambiguous.
[More precisely, there should only be physically meaningful ambiguities.]    

\mbox{ } 

\ni Whilst Sec \ref{Cl-PoT} exhibits multiplicity of classical candidate timefunctions, an ensuing Multiple Choice Problem is only relevant once one transcends to the quantum level.  
Canonical equivalence of classical formulations of a theory does not imply unitary equivalence of the quantizations of each \cite{Gotay00}.  
In other words, a single classical theory can lead to multiple inequivalent quantum theories. 
This follows from the Groenewold--Van Hove Theorem (see e.g. \cite{Gotay00}). 
This is a time issue foremost through different classical choices of time variables being capable of leading to inequivalent quantum theories.
N.B. that this can occur for finite theories rather than being the exclusive province of QFT's, diffusing a common misconception \cite{I93}.
\K \cite{Kuchar92} termed the Multiple Choice Problem an `embarrassment of riches'  in contrast to the Global PoT for timefunctions' being an `embarrassment of poverty'.

\subsection{Loop Quantum Gravity version} 

At the quantum level, this becomes different due to the access afforded by new representations and better-defined Functional Analysis.

\subsection{Deeper levels of quantum Background Independence} 

At the topological manifold level, topological field theories are an interesting offshoot of QFT.
On the GR side, yet again Wheeler's questions \cite{WheelerGRT, Wheeler64b} pointed to the interesting issue of incorporating {\it topology change in quantum GR}.
He envisaged this in terms of `spacetime foam' and of transition amplitudes between spatial manifolds of different topologies. 
Isham \cite{I89-Latt} went one level further down by considering quantization of topological spaces themselves.  
In fact, he has also considered quantizing even more general structures that are no longer based on equipped sets \cite{I03, I10-ToposRev}.

\subsection{Quantum threadings etc}

Isham's approach of quantizing additional structures applies likewise to the threading formulation, the characteristic formulation and 2 + 2 splits.

\mbox{ } 

\noindent {\bf Acknowledgements}: 
E.A. thanks close people, Julian Barbour, Sean Gryb, Jonathan Halliwell, Chris Isham, Tim Koslowski, Flavio Mercati and Hans Westman for discussions, 
Jeremy Butterfield, John Barrow, Marc Lachi$\grave{\me}$ze--Rey, Malcolm MacCallum, Don Page, Reza Tavakol and Paulo Vargas-Moniz for help with my career.

\vspace{10in}  


\end{document}